\documentclass[aps,prd, twocolumn,floatfix,preprintnumbers,superscriptaddress,showpacs]{revtex4-1}
\usepackage{fontenc}
\usepackage{graphicx}
\usepackage{amsmath,amsfonts,bm}
\usepackage{amssymb}
\usepackage{slashed}
\usepackage{color}

\begin{document}
                            \preprint{JLAB-THY-15-2145}
%


\title{Virtuality     and Transverse Momentum Dependence of \\ Pion Distribution Amplitude
}

\author{A. V. Radyushkin}

\affiliation{Physics Department, Old Dominion University, Norfolk,
             VA 23529, USA}
\affiliation{Thomas Jefferson National Accelerator Facility,
              Newport News, VA 23606, USA
}

\begin{abstract}

We  describe basics of a  
 new approach to transverse momentum dependence 
 in hard exclusive processes.  We develop it 
in application to  
the  transition process
 $\gamma^* \gamma \to \pi^0$  at the handbag level.
 Our  starting point   is  
coordinate representation for  matrix elements
 of operators  (in the simplest case, bilocal  ${\cal O} (0,z)$) 
 describing   a hadron with momentum $p$.
 Treated as  functions of $(pz)$ and $z^2$, they are  parametrized 
 through     {\it virtuality distribution  amplitudes} (VDA)   $\Phi (x, \sigma)$,  with  
 $x$ being  Fourier-conjugate to $(pz)$ and  $\sigma$  
  Laplace-conjugate  to $z^2$.  
 For   intervals with   $z^+=0$,  
 we   introduce the  {\it transverse momentum distribution amplitude}  (TMDA) 
 $\Psi (x, k_\perp)$,   and  write it 
 in terms of VDA   $\Phi (x, \sigma)$. 
The  results of covariant calculations,  
written in terms  of $\Phi (x, \sigma)$ are  converted   into expressions 
involving  $\Psi (x, k_\perp)$. 
   Starting with scalar toy models, we  extend the analysis  
   onto the case of spin-1/2 quarks  and QCD. 
 We propose  simple models for soft VDAs/TMDAs,
and  use them  for comparison of handbag results with experimental 
(BaBar and BELLE)  data 
on  the pion transition form factor. We also discuss  how one can generate 
high-$k_\perp$  tails from primordial soft distributions.


\end{abstract}

\pacs{11.10.-z,12.38.-t,13.60.Fz}



 \maketitle



   
\setcounter{equation}{0}   \section{Introduction}

Analysis  of effects due to  parton  transverse momentum  is 
an important direction in  modern  studies of hadronic structure. 
The main effort is to use  the transverse-momentum
dependence of inclusive processes, such as 
semi-inclusive deep inelastic scattering (SIDIS) and Drell-Yan (DY)
pair production, describing their cross sections 
in terms of  {\it transverse momentum dependent  distribution}  
   (TMDs)    $F(x,k_\perp)$ \cite{Mulders:1995dh}. 
   
   TMDs  are generalizations
   of the usual  ``longitudinal''  parton densities $f(x)$  that correspond
   to TMDs integrated over $k_\perp$.  
Information about the transverse momentum structure
is contained in the   $ k_\perp^{2n}$ moments of $F(x,k_\perp)$ for $n \geq 1$.
 Dealing with   DY and  SIDIS processes,  one encounters   
    ${\cal O}( \langle  k_\perp^{2n} \rangle )$ contributions that are not 
    suppressed by inverse 
    powers of $Q^2$, the high momentum probe.   Such effects are  combined  
    into one  function, TMD  $F(x,k_\perp)$,  
the use of which   in these cases
 is unavoidable.

    However, in light-cone dominated processes, such as deep inelastic scattering
     in the limit of  large      $Q^2$,  
the higher moments of $ k_\perp^{2}$
   	generate just 
 power corrections
    of $\langle k_\perp^{2n} \rangle /Q^{2n}$ type to the leading power behavior
    described by the ``collinear'' parton distribution $f(x)$.   
     Still, 
if one is interested 
 in the region of  moderately large $Q^2$'s, 
 the  transverse momentum  corrections may be 
rather  important even for a  light-cone dominated process.  
Then  one may want to explicitly 
 represent them as generated from a common TMD-type function.
 Effectively, this corresponds to a resummation 
 of such  power corrections.

A situation when accessible $Q^2$ are not large enough 
to secure the dominance of the asymptotically leading
collinear approximation, is very common in hard exclusive 
processes, such as pion and nucleon electromagnetic  form factors,
where the  overlap contributions of soft wave functions \cite{Drell:1969km}
$\psi (x, k_\perp)$ are sufficient   to  describe existing  data. 
These wave functions are apparent   analogs of TMDs 
in case of exclusive processes.  In particular, integrating 
$\psi (x, k_\perp)$ over $k_\perp$ gives \cite{Lepage:1979zb} the  distribution amplitude 
$\varphi (x)$ \cite{Radyushkin:1977gp}, a  basic object of the asymptotic perturbative QCD
analysis for  exclusive processes \cite{Radyushkin:1977gp,Chernyak:1977fk,Efremov:1978fi,Efremov:1979qk,Lepage:1979zb}. 
The latter is based on the operator product expansion (OPE) \cite{Wilson:1969zs,Brandt:1970kg},
with $x^n$ moments of $\varphi (x)$ related to matrix elements
of the lowest-twist operators. 
 
One may expect that  higher $ k_\perp^{2n}$ moments of $\psi (x,k_\perp)$
should correspond to  matrix elements
of  higher-twist operators of the OPE.
However, a  subtle point is that the standard OPE 
\cite{Wilson:1969zs,Brandt:1970kg}     is constructed within a  
 covariant 4-dimensional quantum field theory (QFT)  framework, while the wave 
functions  $\psi (x, k_\perp)$  \cite{Drell:1969km}  mentioned in the context of the overlap 
contributions are the objects of a  3-dimensional  light-front approach  \cite{Kogut:1969xa,Lepage:1980fj}.

Our goal in the present paper is to  develop the basics of 
a formalism (its outline  was given
in Ref. \cite{Radyushkin:2014vla})  that starts from  a  covariant 4-dimensional 
approach,  but describes the  
structure of hadrons in hard exclusive processes  in terms of  
  functions $\Psi (x, k_\perp)$ incorporating the  dependence 
on the  transverse momentum of its constituents.   
Just like in the light-front formalism, the organization of these 
 functions has the structure of the Fock state decomposition,
 i.e. each function is characterized by the number of constituents 
 involved. 
 
 The lowest, 2-body component   is described 
 by a function $\Psi (x, k_\perp)$ that depends on a 3-dimensional 
 variable $x, k_\perp$. 
To emphasize the distinction, we use ``transverse momentum
dependent distribution amplitude'' (TMDA) as 
the name for  the  function $\Psi (x, k_\perp)$    that 
appears  in our approach.  
By construction,  $\Psi (x, k_\perp)$ has  a direct connection
with the operators that appear in the OPE of a  covariant QFT.
As a specific application,  we   choose the  
    hard  exclusive   process  of  
$\gamma^* \gamma \to \pi^0$ transition  that 
  involves   just  one hadron, and thus has the  simplest  structure.

    The paper is organized as follows. 
Since the basic features  of our approach 
are not sensitive  to the spin of the particles, we begin 
 in  Sect.  \ref{Scalar_FF}  with   the discussion of   
 the structure of the $\gamma^* \gamma \to \pi^0$  transition form factor 
 at the handbag(i.e., 2-body)  level in a scalar model.  
With the  goal of  keeping  the closest contact with the OPE, 
 we  start with a general   analysis of  the handbag diagram 
 using the coordinate representation, in which 
the hadron structure  is described 
through the $z$-dependence of a matrix element
$ \langle p | \phi (0) \phi (z) \rangle \equiv \tilde  \chi_p (z) $
involving two parton fields $\phi$
($p$ being the hadron momentum).

 By Lorentz invariance,  
$\tilde \chi_p (z) $ depends on $z$ through two variables,
$(pz)$ and $z^2$. 
A double Fourier transform of $\tilde \chi_p (z) $ 
with  respect to $(pz)$ and $z^2$  gives
the 
{\it virtuality distribution amplitude} (VDA) $\Phi (x, \sigma)$,
the  basic object of our analysis. For any contributing Feynman diagram
the support of the VDA is restricted by  $0\leq x \leq 1$ 
and $0\leq \sigma \leq \infty$.  
The variable $x$ has the usual meaning 
of the fraction of the hadron momentum $p$ carried
by a parton,  while  the  variable $\sigma$ being conjugate 
to $z^2$ may be interpreted as a generalized virtuality. 

Projecting $\tilde \chi_p (z)$  on the light front $z_+=0$ 
results in the 
{\it impact parameter distribution amplitude} (IDA)
$\varphi (x, z_\perp)$, whose further Fourier transform 
leads to the {\it  transverse momentum dependent 
 distribution amplitude} (TMDA) $\Psi (x, k_\perp)$.
The properties of virtuality distributions, TMDAs and connections between  
 them and related functions are discussed in Sect.  \ref{TMDA}.   
A  key point  for  subsequent applications 
 is that $\Psi (x, k_\perp)$ has a 
simple expression in terms of VDA $\Phi (x, \sigma)$. 
This observation may be  used to rewrite the results 
of covariant calculations (initially given in terms  of VDA 
$\Phi (x, \sigma)$) through  TMDA $\Psi (x, k_\perp)$,
i.e., as a 3-dimensional integral over 
$x$ and $k_\perp$.  In this way we   derive  the expression 
for the transition form factor in terms of TMDA.

To emphasize a special role of the VDA representation,
in Sect.  \ref{handbag} 
we analyze the structure of the handbag amplitude in 
several other representations, namely, coordinate light-front variables,
Sudakov and IMF parameterizations for  the virtual momentum integration.
We show that  expressions for the form factor 
 in all these cases are much more complicated than the VDA form 
 (to which they are eventually equivalent), and one needs to resort 
 to approximations in order to get a compact
 formula.
 Continuing to discuss the relation between  the VDA approach
     and the method of operator product expansions, 
 in Sect. \ref{threebody} we outline the application of the VDA approach in  the three-body distribution 
      case.

  Modifications 
that appear for  spin-1/2 quarks and vector gluons are  considered in Sect. \ref{gaugeT}. 
In particular, we observe  that the basic  relations between the VDA-based distributions remain intact when 
matrix elements involve  
spin-1/2 quarks. 
Using  the parameterization
in terms of VDA 
$\Phi (x, \sigma )$, we calculate the handbag
 diagram  and express the result in terms of the TMDA 
 $\Psi (x, k_\perp)$. 
The change in the form factor formula reflects 
a more complicated structure  of the spin-1/2 hard propagator.  
Then 
we study the extension of our  results 
onto gauge theories.

 In Sect. \ref{TMDA_Models},  we formulate a few simple models for 
  soft  \mbox{TMDAs}, i.e. those 
 that decrease faster than any inverse power 
 of $1/k_\perp^2$ for large $k_\perp^2$.
In 
 Sect. \ref{FF_Models} we analyze  the results of using these
 models   to 
describe the data on the 
pion transition form factor.

Attaching perturbative propagators to the 
 soft TMDA, as  shown  in Sect.  \ref{Hard},  produces factors with 
 $1/k_\perp^2$ behavior. As a result, 
the quark-gluon interactions  in QCD  
generate a hard $\sim 1/k_\perp^2$ tail for \mbox{TMDAs.} 
The basic elements of generating hard tails from soft primordial 
TMDAs are illustrated in 
Sect. \ref{Hard_tail}.   Finally,  in Sect. \ref{Summary} we formulate 
our conclusions  and discuss directions of further applications 
of the VDA approach.



\setcounter{equation}{0}   \section{Transition form factor  in scalar model}
\label{Scalar_FF}

  \subsection{Handbag diagram   in coordinate representation}

\begin{figure}[h]
\centerline{\includegraphics[width=2.2in]{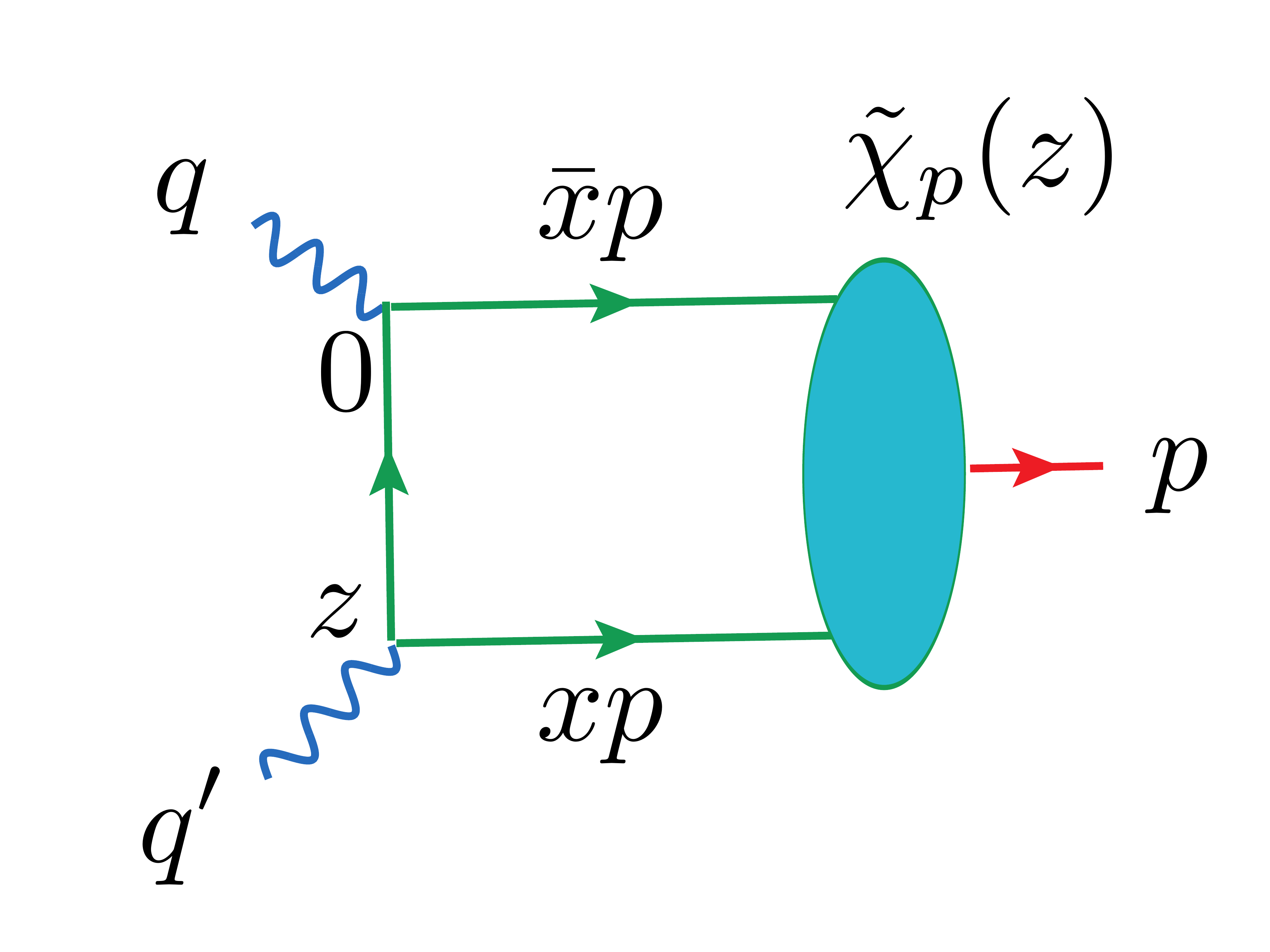}}
\caption{Handbag diagram in the coordinate  representation and parton momentum assignment.
\label{coord}}
\end{figure}

 Consider   a  general  handbag diagram for 
a scalar analog of the $\gamma^* \gamma \to \pi^0$ amplitude, 
see Fig. \ref{coord}, 
with the hadronic 
blob being a matrix element connecting  parton fields 
with the ``pion''.
In  
  the coordinate representation  we have 
  \begin{align}
T(q,p) = 
\int {d^4z}\, e^{-i(q'z)} \, D^c(z) \, 
\langle  p | \phi (0)  \phi(z)  | 0 \rangle  \  , 
\label{eq:Fscalar00}
\end{align}
where  $D^c(z) = i/4\pi^2 z^2$ is  the scalar massless propagator,
 $q'$ is 
the momentum of the initial real  ``photon'', $q'^2 =0$ 
given by $q'=p-q$, with 
$p$ being  the momentum of the final ``pion''  and 
$q$ is the momentum of the initial virtual  ``photon''
(\mbox{$q^2  \equiv - Q^2$}).

The pion structure is described by the 
matrix element 
$\langle  p | \phi (0)  \phi(z)  | 0 \rangle \equiv \tilde \chi_p (z) $
 of the bilocal operator.    To parametrize  it,  we   
start with a formal 
Taylor expansion 
 \begin{align}
  \phi(z) = \sum_{n=0}^{\infty} 
 \frac{1}{n!}z_{\mu_1} \ldots z_{\mu_n} \ 
 {\partial}^{\mu_1} \ldots
{\partial}^{\mu_n}   \phi(0) \ . 
\label{Taylor0}
\end{align}
Then,  information about the pion is 
contained in matrix elements
$\langle  p | \phi (0)   {\partial}^{\mu_1} \ldots
{\partial}^{\mu_n}  \phi (0)  | 0 \rangle $.
Due to Lorentz invariance, they may be written as
 \begin{align}
\langle  p | \phi (0)   {\partial}^{\mu_1} \ldots & 
{\partial}^{\mu_n}  \phi (0) | 0 \rangle = A_n^{(0)} i^n \,  p^{\mu_1} \ldots p^{\mu_n}
\nonumber \\ & + {\rm   terms \ containing} \  g^{\mu_i\mu_j}{\rm 's} \  .
\label{ln}
\end{align}
Utilizing the fact that the $\mu_k$ indices are symmetrized  in the Taylor expansion  
by the $z_{\mu_1} \ldots z_{\mu_n}$ tensor, we may use a more organized 
expression 
 \begin{align}
\langle  p | \phi (0)   (z\partial)^n \phi (0)  | 0 \rangle = i^n 
\sum_{l=0}^{[n/2]}  (z^2 \Lambda^2)^l (pz)^{n-2l}A_n^{(l)}  \ ,
\label{ln2}
\end{align}
with   information about the pion structure accumulated now in constants
$A_n^{(l)} $.  The momentum  scale $\Lambda$ was introduced 
to  secure that all $A_n^{(l)} $'s have the same dimension. 

Take the lowest, $l=0$ term. 
To perform summation over $n$, we may  treat the coefficients 
$A_n^{(0)}$ as moments of the {\it pion distribution amplitude} \cite{Radyushkin:1977gp} 
$ \varphi (x)$ 
\begin{align} 
A_{n}^{(0)} = \int_{0}^1 \varphi (x)\,   x^n  dx 
\label{An0}
\end{align}
(we want the quark at the $z$ vertex 
to carry  the momentum $xp$, see Fig.(\ref{coord})).
As a result,
\begin{align}
& \langle p |   \phi(0) \phi (z)|0 \rangle 
=  
\int_{0}^1  \varphi (x)\,  \,  e^{i  x (pz) } \, dx
 +  {\cal O} (z^2)  \  . 
\label{DDF}
\end{align}
For the amplitude $T(q,p)$ this gives
  \begin{align}
T(q,p) =&  \frac{i}{4\pi^2} \int_{0}^1 dx \,  \varphi (x) 
\nonumber \\ &  \times
\int \frac{d^4z}{z^2}\, e^{-i(q'z) +i x(pz)  }
 \biggl [ 1+   {\cal O} (z^2)  \biggr ] \,  . 
\label{eq:FscalarO}
\end{align}
The   term  given  by  $l=0$ part in Eq. (\ref{ln2})  produces 
 \begin{align}
T^{(0)} (q,p) = -  \int_{0}^1 \frac{\varphi(x) dx}{(q'-xp)^2 +i \epsilon}  \  .
\label{eq:Fscalar0a}
\end{align}
An extra $z^2$ factor in  the $l=1$ term  of (\ref{ln2}) cancels the 
$1/z^2$ singularity of $D^c (z)$, and results in a contribution 
 proportional to
 \begin{align}
\int {d^4z}\, e^{-i(q'z) + i  x (pz)  }
= (2 \pi)^4 \delta^4 (q'-xp) \ ,
\label{eq:Fscalar0b}
\end{align}
which apparently should be treated as zero, because $q'$ is not proportional
to $p$.  The same applies  to terms with higher powers $(z^2)^l$, which 
produce integrals proportional to $\Box^{2l-2} \delta^4 (q'-x p)$. 
If one would use  just  
a straightforward dimensional counting,  one would expect  
 that terms with higher powers of $z^2$
 result in contributions accompanied by powers of $1/(q'-xp)^2$,
but as we see,  
 they  produce terms that are ``invisible'' in the $1/(q'-xp)^2$ expansion.
 The actual power corrections appear   when  a ${\cal O} [(z^2)^l]$ 
term in  the matrix element
$  \langle p |   \phi(0) \phi (z)|0 \rangle $  is accompanied by some nonzero power
of 
$\ln z^2$.

Note also that since  $-(q'-xp)^2= xQ^2+ x \bar x p^2$,  the $1/(q'-xp)^2$ expansion
does not coincide with the $1/Q^2$ expansion when $p^2 \neq 0$.
The things  simplify when $p^2=0$. 
Then,  
for a massless propagator $D^c (z)$, a   power correction of $1/(Q^2)^{1+l}$ type can be  
obtained from  $(z^2)^l (\ln z^2)^{k\geq 1}$ terms only.

This simple  analysis leads us to  an important observation that, 
for a situation when $p^2=0$ and the matrix element 
$\langle  p | \phi (0)   (z\partial)^n \phi (0)  | 0 \rangle$ 
is given by a regular power expansion in $z^2$
(we shall say that one deals with a ``soft'' wave function in such a  case), the whole 
amplitude $T(q,p)$  with  a massless hard propagator $D^c (z)$
is {\it exactly}  given by the first term $T^{(0)}(q,p)$ only, namely, that 
 \begin{align}
T^{\rm soft} (q,p) |_{p^2=0} =   \int_{0}^1 \frac{\varphi(x) dx}{xQ^2}
\Biggl [1 + {\rm no \ powers \ of \ } 1/Q^2 \Biggr ]   \  ,
\label{eq:Fscalar000}
\end{align}
with no higher $1/Q^2$ corrections {\it under the $x$-integral.} 

On the other hand, if the matrix element has a logarithmic singularity  
$(z^2)^l \ln z^2$ with $l\geq 1$, the amplitude should have a power
correction with  $(1/Q^2)^l$ behavior.

These  observations  may be used  as a constraint  
(``OPE compatibility'') that  should  be required to hold  
 in schemes that add transverse momentum dependence
into the description of the pion structure. As we will see,
some previously used 
approximate  schemes that   look natural  otherwise,
are not ``OPE compatible''.

     \subsection{Introducing virtuality distributions}
   \label{VDA0}

 As mentioned already, parametrizing   matrix elements 
   of local operators resulting from     
 the Taylor expansion  (\ref{Taylor0}), 
one needs to deal with a set of parameters 
$A_n^{(l)}$ for each $n$, with $l$ being the number
of metric tensors $g^{\mu_i \mu_j}$ on the right hand side of 
Eq. (\ref{ln}),  or power of $z^2$ in   Eq. (\ref{ln2}).
Collecting together  terms with the same power of $(pz)$, we may write
\begin{align}
\langle  p | \phi (0)   \phi (z)  | 0 \rangle =
\sum_{l=0}^{\infty } \frac{1}{l!} \left (\frac{z^2 \Lambda^2}{4} \right )^l 
\sum_{N=0}^{\infty} i^N \,  \frac{(pz)^{N} }{ N!}  \, {B_N^{(l)}}   \ .
\label{lnB21}
\end{align}
By analogy with Eq. (\ref{An0}) which introduces  the pion distribution amplitude $\varphi (x)$  through the coefficients $A_{n}^{(0)}$, 
we define that 
 the coefficients $B_N^{(l)} $ are given by  
 double moments  of  a function of two variables $\Phi (x, \sigma)$, 
 which we call the {\it  virtuality distribution amplitude}  (VDA)  :
\begin{align}
B_N^{(l)}  
= & (-i)^l 
\int_{0}^{\infty} d \sigma \, \sigma^l e^{-\epsilon \sigma/4} \int_{0}^1 dx\, x ^N   \, 
 \Phi (x,\sigma) \,   \,  . \label{PhixsB02}
\end{align} 
Substituting this definition into Eq. (\ref{PhixsB02})  gives 
 \begin{align}
 \langle p |   \phi(0) \phi (z)|0 \rangle 
= & 
\int_{0}^{\infty} d \sigma \int_{0}^1 dx\,   \nonumber \\ & \times 
 \Phi (x,\sigma) \,  \,  e^{i x (pz) -i \sigma (z^2-i \epsilon)/4 } \,   . 
 \label{Phixs000}
\end{align} 

We have derived this  {\it VDA representation}  
from a formal Taylor expansion  of the matrix element 
$ \langle p |   \phi(0) \phi (z)|0 \rangle$ of the bilocal operator.
Such an expansion makes sense only if 
 the matrix elements of local operators 
$\langle  p | \phi (0)   (z\partial)^n \phi (0)  | 0 \rangle $
are finite. In such a case we say that one deals with 
a {\it soft wave function}.  However, Eq.  (\ref{Phixs000}) 
looks just like  a double Fourier transform
in  two variables   $(zp)$  and $z^2$. As such,  it  should hold
for a very wide range of functions, including the functions that {\it are not } 
 given by a convergent Taylor expansion in $z^2$.  

This observation suggests that the VDA representation
may be obtained under much weaker assumptions. 
One may also wonder why   we have imposed particular   limits 
of integration, namely, $0 \leq x \leq 1$ and $0\leq \sigma <\infty$
on these  Fourier integrals.
Obviously,  these limits do  not  appear  automatically  for any function 
of  $(pz)$ and $z^2$.  But 
  it can be shown (see below)  that 
any Feynman diagram
contributing to   $\langle p |   \phi(0) \phi (z)|0 \rangle$ 
has  the  VDA representation 
with exactly  these limits of integration.
Also, it does not matter if the Feynman diagram has   
 logarithmic singularities in $z^2$,  the VDA representation  
(\ref{Phixs000})  still holds, even though 
some of   
 the moments (\ref{PhixsB02}) diverge in that case.
It should be also noted that the   logarithmic singularities in $z^2$ 
come as $\ln ( 
z^2- i \epsilon)$,   
reflecting the causal structure  of Feynman diagrams. 

\subsection{VDA and $\alpha$-representation} 
\label{alpha}

Using the $\alpha$-representation and techniques outlined in
Refs. \cite{Radyushkin:1983wh,Radyushkin:1983ea,Radyushkin:1997ki}, 
it can be demonstrated  that the VDA representation  (\ref{Phixs000}) holds for 
any Feynman diagram contributing to  the relevant matrix  element.      \setcounter{paragraph}{0}

\subsubsection{Momentum space} 

Consider  the momentum representation
version of the matrix element 
\begin{align}
&  \int d^4z  e^{- i (kz)} \langle p |   \phi(0) \phi (z)|0 \rangle \equiv 
  (4\pi i )^2 \chi_p (k) \,  , 
 \label{Phixsmom}
\end{align} 
where $k$ is the momentum of the quark going from the ``$z$'' vertex,
and $\chi_p (k)$ is the Bethe-Salpeter  wave function.
Then, according to  Refs. \cite{Radyushkin:1983wh,Radyushkin:1983ea,Radyushkin:1997ki},
  the contribution of 
any Feynman diagram  ${\cal D}$ to 
   $\chi_p (k) $
can be represented as  
\begin{align}
\chi^{\cal D}_p&  (k)  =  i^{l} \, \frac{P({\rm c.c.})}{(4\pi i)^{Ld/2}}
\int_0^{\infty} \prod_{j=1}^l   d\alpha_{j} [D(\alpha)]^{-d/2}
\nonumber \\ & \times 
\exp \left \{ i k_1^2  \frac{A (\alpha)}{D(\alpha) } +i k_2^2
  \frac{ B (\alpha)}{D(\alpha) }
 \right \} 
\nonumber \\ & \times 
\exp \left \{ 
i  p^2  \frac{ C(\alpha) }{D (\alpha) }
- i  \sum_{j} \alpha_{j} (m_{j}^2- i\epsilon) \right \}  \ , 
\label{alphap}
\end{align}
where $k_1=k$, $k_2= p-k$, 
$d$ is the space-time dimension,  ${P({\rm c.c.})}$ is the relevant
 product of  coupling constants, 
 $L$ is the number of loops of the diagram,  and $l$
is the number  of its
internal lines.  For our purposes,
the most important property of this representation is that 
$A(\alpha), B(\alpha), C(\alpha), D(\alpha)$ 
are positive functions (sums of products) 
of the $\alpha_\sigma$-parameters 
of a  diagram.   
Thus, we have a general representation 
\begin{align}
\chi_p (k)  =& 
\int_0^1 dx
\int_{0}^{\infty} d \lambda \, 
e^{i \lambda \bar x k^2 + i \lambda  x (k-p)^2-\epsilon \lambda } 
F(x, \lambda\, ;\, p^2) \ , 
\label{alphap31}
\end{align}
where 
\begin{align}
  F(x, & \lambda  \, ;  \, p^2)=\sum_{\rm all  \ diag}   \, i^{l} \, \frac{P({\rm c.c.})}{(4\pi i)^{Ld/2}}
\int_0^{\infty} \prod_{j=1}^l   d\alpha_{j} [D(\alpha)]^{-d/2} 
\nonumber \\ & \times 
\delta \left (x - \frac{B (\alpha)}{A (\alpha)+ B (\alpha)} \right ) 
\delta \left ( \lambda - \frac{A (\alpha)+ B (\alpha)}{D(\alpha) } \right ) 
\nonumber \\ & \times 
\exp \left \{ 
i  p^2  \frac{ C(\alpha) }{D (\alpha) }  - i  \sum_{j} \alpha_{j} (m_{j}^2- i\epsilon) \right \}   \  .
\label{alphap3}
\end{align}
Eq.   (\ref{alphap31}) 
 may be also rewritten as 
\begin{align}
\chi_p (k)  =& 
\int_0^1 dx
\int_{0}^{\infty} d \lambda \, 
e^{i \lambda (k-xp)^2+i \lambda x \bar x p^2 -\epsilon \lambda } 
F(x, \lambda;p^2) \ . 
\label{alphap31}
\end{align}

\subsubsection{Coordinate space} 

Making Fourier transform  to the coordinate representation,
 we get  
\begin{align}
 \langle p |   \phi(0) \phi (z)|0 \rangle 
= & 
\int_{0}^{\infty} d \sigma \int_{0}^1 dx\, 
 \Phi (x,\sigma) \,  \,  e^{i x (pz) -i {\sigma} {(z^2-i \epsilon )/4}} \,  .   \label{Phixs0}
\end{align} 
The functions $F$ and $\Phi$ are related by 
\begin{align}
e^{i \lambda x  \bar x p^2 } 
F(x, \lambda;p^2) = \Phi (x, 1/\lambda) 
\label{FPhi}
\end{align}
(the VDA $\Phi$ also depends on $p^2$, i.e. in principle  it should be written as 
$\Phi (x, \sigma;p^2)$, 
but we will not indicate  this dependence explicitly, mainly because 
$p^2$  is fixed for a given matrix element).

\subsubsection{Important observations} 

Note that the momentum $p$ in (\ref{FPhi})  is the {\it actual } 
momentum that appears in the matrix element.
In this sense, a parton  in the VDA picture carries the
fraction $xp$ of the {\it  total} hadron momentum $p$,
not just the fraction $xp^+$ of its ``plus'' component $p^+$.
In fact, all our discussion so far was absolutely Lorentz covariant,  and 
there was no need to decompose momenta into 
any components, to project $p$  on  its ``plus'' part, etc.
We also emphasize  that 
there was  no need to assume that $p^2=0$. 

Another point is that the  representation (\ref{Phixs0})   has been  obtained 
 without any assumptions  about {\it regularity} of the $z^2 \to 0$ limit.
This means that  one can use the VDA   representation even in cases when 
a   formal Taylor expansion  
in $z^2$ does not exist because of singularities in  the $z^2 \to 0$ limit.
In other words, the  matrix  element   $\langle p |   \phi(0) \phi (z)|0 \rangle $ 
 may be non-analytic for $z^2=0$,
and still be given by a VDA representation. 

\subsection{Scalar handbag diagram in VDA representation}

Using the  VDA parametrization (\ref{Phixs000})  we  can take the $z$-integral
in Eq. (\ref{eq:Fscalar00}) to 
obtain 
 \begin{align}
T(q,p) = &
 \int_{0}^1 \frac{dx}{(q'-xp)^2 +i \epsilon} \, \int_{0}^{\infty}   d \sigma \, 
  { \Phi (x,\sigma)    } 
  \nonumber \\ &\times  
\left \{ 1- e^{i[{(q'-xp)^2 +i \epsilon} ]/  \sigma }  \right \} \  . 
\label{eq:Fscalar3000}
\end{align} 
The first term in the brackets does not depend on $\sigma$ and 
produces   the integral 
 \begin{align} 
\int_{0}^{\infty}  \Phi (x,\sigma) \, d \sigma  \equiv  \varphi  (x)   \  ,
\label{Phix00}
\end{align} 
 where $\varphi (x)$ is the distribution amplitude 
defined by Eq. (\ref{An0}). Indeed,  taking  $z^2=0$ in the VDA representation
  (\ref{Phixs000}) and comparing  the result with Eq. (\ref{DDF}), we see that 
  this  is formally the case. 

Of course,  this reasoning assumes  that the integral over $\sigma$ 
in (\ref{Phix00})  converges 
at  the upper limit, which happens when $\Phi (x,\sigma)$ 
decreases faster than $1/\sigma^{1+\epsilon}$ for large $\sigma$.
If $\Phi (x,\sigma) \sim 1/\sigma$ for large $\sigma$, then
the integral (\ref{Phix00}) logarithmically diverges, which 
corresponds to a $\ln z^2$ singularity for  the matrix element. 
However,  the $\sigma$-integral in Eq. (\ref{eq:Fscalar3000}) 
converges even in that case, because the sum of terms in the brackets
behaves like $1/\sigma$ for large $\sigma$.

As noted earlier, if  the matrix element  $\langle p |   \phi(0) \phi (z)|0 \rangle$ 
can be expanded in a Taylor series
in $z^2$, with finite coefficients,  the higher ($l\geq 1$) 
 terms $(z^2)^l$  of such an expansion 
  cancel the singularity  $1/z^2$ 
of the massless scalar propagator $D(z)$. As a result,
 these terms  produce terms proportional
to $\Box_{q'}^l$   derivatives of the $\delta^4 (q'-xp)$
function. Taken separately, each of these terms   is   
invisible in the $T(q,p)$ amplitude, simply  because $q'$ is not 
proportional to $p$.
Still,  an infinite sum of delta-function derivatives   in our case 
 produces a non-trivial function  given by the second term in Eq. 
 (\ref{eq:Fscalar3000}).    In  other words, 
the ``invisible''  contributions are combined in the second term
which, after integration over $\sigma$, 
results in  a nontrivial function of $(q'-xp)^2$. 
 
 The pion structure is now described by the VDA $ \Phi (x,\sigma) $,
 and  by just  modeling its $\sigma$-shape one can 
 study the impact of higher $l$ terms.
However, it is very instructive 
to give  an interpretation of these terms  
 using the concept  of parton transverse momentum.

  \setcounter{equation}{0}   \section{Transverse momentum distributions} 
\label{TMDA}

\subsection{Introducing  TMDA}

  To  bring in the  transverse momentum dependence, 
  we should decide, first,  which directions are ``transverse''. 
It is natural to  define that  the pion momentum $p$ 
has only   longitudinal components,
becoming  a purely ``plus'' vector in the $p^2=0$ limit. 
    Projecting the matrix element $ \langle p |   \phi(0) \phi (z)|0 \rangle$
    on the light front $z_+ =0$,  
  \begin{align}
 \langle p |   \phi(0) \phi (z)|0 \rangle  |_{z_+=0} 
   = 
 \int_{0}^1 dx \,  {\varphi } (x, z_\perp)
 \, e^{i x (pz_-) }  \  ,   \label{Phixbb}
\end{align} 
 we   introduce  the {\it impact parameter distribution amplitude} (IDA)
 $  {\varphi } (x, z_\perp)  $.  
 It is related to VDA by
 \begin{align}
 {\varphi } (x, z_\perp) = & 
\int_{0}^{\infty} {d \sigma }  \, 
 \Phi (x,\sigma) \,  \,  
 e^{ i \sigma (z_\perp ^2-i \epsilon ) /4} 
  \  .  \label{IDA} 
\end{align}

The next step is to  treat the  IDA function  
as a Fourier transform 
\begin{align}
{\varphi } (x, z_\perp) = \int  {\Psi}(x, k_\perp ) \, e^{i (k_\perp z_\perp)}
\,  {d^2 k_\perp }    \label{impaf} 
\end{align}
of the {\it transverse momentum dependent
distribution  amplitude}   (TMDA)  ${\Psi} (x, k_\perp )$.
 One can write the TMDA   in terms of VDA  as
\begin{align}
{\Psi} (x, k_\perp ) = & \frac{i }{\pi }
\int_{0}^{\infty} \frac{d \sigma }{\sigma} \, 
 \Phi (x,\sigma) \,  \,  
 e^{- i (k_\perp ^2-i \epsilon )/ \sigma} 
  \  .  \label{Phixs1tmd0} 
\end{align} 

The  moments of TMDA ${\Psi } (x, k_\perp )$ are 
 formally given by
 \begin{align}  
 \int  {\Psi } (x, k_\perp ) \, k_\perp^{2l} \, d^2 k_\perp  
=  \frac{l!}{i^l} 
 \int_{0}^{\infty} {\sigma}^l  \, 
 \Phi (x,\sigma) \, {d \sigma }  \ .
  \label{sigma_k} 
\end{align}
They are proportional to the $\sigma^l$ 
 moments of the VDA $\Phi (x,\sigma)$  and, 
hence,  finite for a soft VDA. 
This   means that a ``soft''  TMDA 
 ${\Psi } (x, k_\perp )$ should 
decrease  faster than any power of 
 $1/k_\perp^2$ for large $k_\perp$.

\subsection{Scalar handbag diagram in TMDA representation}

Using the TMDA/VDA relation (\ref{Phixs1tmd0}), one can  rewrite
Eq. (\ref{eq:Fscalar3000})   in terms of TMDA as 
 \begin{align}
 T(q,p) 
= &  
 -  \int_{0}^1   \frac{dx}{(q'-xp)^2}  
     \int_{k_\perp^2 \leq -(q'-xp)^2}  \, d^2 { k}_\perp 
 \,   \Psi (x, { k}_\perp   )  \ ,
 \label{Phixs0600}
 \end{align}
 which converts into 
 \begin{align}
 T(Q^2) 
= &  
   \int_{0}^1   \frac{dx}{xQ^2}   \int_{k_\perp^2 \leq {x} Q^2}    \Psi (x, { k}_\perp   ) 
 \, d^2 { k}_\perp 
 \label{Phixs060}
 \end{align}
 in the $p^2=0$ case.

 It should be  emphasized that no Taylor expansions 
 in $z^2$ have been  used in deriving (\ref{Phixs060}). 
 Still, if one deals with  a soft TMDA, 
 one can use Taylor expansion and 
  separate the $l=0$ term in Eq. (\ref{eq:Fscalar3000}).
  Then, incorporating the reduction relation 
 \begin{align}
\label{redrel}
\int  {\Psi} (x, k_\perp ) \, d^2 k_\perp =  \varphi (x) 
\end{align}
 one can   write 
  \begin{align}
 T(Q^2) 
= &  
   \int_{0}^1   \frac{dx}{xQ^2}  \left [ \varphi (x)  -   \int_{k_\perp^2 \geq {x} Q^2}    \Psi (x, { k}_\perp   )  
 \, d^2 { k}_\perp  \right ] 
      \  . 
 \label{Phixs06ht}
 \end{align}
As we have noted, a soft TMDA decreases  for large
$k_\perp^2$  faster than any inverse power of $k_\perp^2$.
As a result,  the second term in Eq. (\ref{Phixs060})
decreases  for large $Q^2$ 
 faster than any power of $1/Q^2$, 
 i.e. there are no $1/Q^2$ power corrections 
 to the $\varphi (x)$ term under the $x$-integral in Eq. (\ref{Phixs060}).
This means that 
 the 
VDA-based expression (\ref{Phixs060})  in case of a soft VDA
has an  OPE-compliant form of Eq.
(\ref{eq:Fscalar000}).

 Alternatively, if the matrix element 
 $ \langle p |   \phi(0) \phi (z)|0 \rangle $ 
 has a  logarithmic singularity $\ln z^2$ 
 starting with   $(z^2)^l $ power, the $\sigma$ moments  of $ \Phi (x,\sigma) $
 should diverge starting with $\sigma^l$, 
 and $k_\perp^2 $ moments  of ${\Psi } (x, k_\perp )$ should diverge 
 starting with $k_\perp^{2l} $. This  means  that ${\Psi } (x, k_\perp )$
 decreases as $(1/k_\perp^2)^{l+1} $  for large $k_\perp$, i.e., 
 $\Psi (x, { k}_\perp   )$ has a power-like ``hard tail''.  
 Then the second term in  Eq. (\ref{Phixs06ht})
produces a ${\cal O}((1/Q^2)^{l+1})$ contribution to  $T(Q^2)$.

 Thus, we see that the VDA-based  formula (\ref{Phixs060}) is  in full compliance 
 with the OPE  approach.

\subsection{Impact parameter representation} 

Substituting the expression of TMDA in terms of IDA
 \begin{align}
{\Psi}(x, k_\perp ) = \int \,  \frac{d^2 b_\perp }{(2 \pi)^2}  \, 
  {\varphi } (x, b_\perp)  \, e^{- i (k_\perp b_\perp)}
   \label{pafim} 
\end{align}
into the VDA-based  formula  (\ref{Phixs060}) for $T(Q^2)$ we obtain 
  \begin{align}
T(Q^2)
   {=}   \int_{0}^1 \frac{ dx}{\sqrt{x}Q }
 \int_0^\infty db \,
   \,J_1  (\sqrt{x}Q b) \, \varphi (x,b)\, .
\label{eq:Fscalar1230}
\end{align}
Note that we intentionally use here the notation $b_\perp$ for the impact parameter 
variable, to emphasize that it cannot be identified with the 
transverse part $z_\perp$ of the integration variable $z$ 
in  the original  coordinate representation
integral   (\ref{eq:Fscalar00}). 

Indeed,  recall that our procedure has started  with taking 
 integral over $d^4 z$ to obtain 
the result  expressed  by Eq.   (\ref{eq:Fscalar3000}) in terms of VDA $ \Phi (x,\sigma) $
which was transformed then  into  Eq. (\ref{Phixs060}) written in terms of TMDA 
${\Psi}(x, k_\perp )$. 
After this starting integration,   a connection 
of the final result 
with the $z_\perp$-integration has been   completely lost. 
Then 
 we have converted the TMDA result   (\ref{Phixs060})  into the expression (\ref{eq:Fscalar1230}) 
in  terms of IDA   ${\varphi } (x, b_\perp) $,
 in which $b_\perp$ is a new auxiliary variable.

\subsection{Twist decomposition} 

 \setcounter{paragraph}{0}

So far, we did not mention the concept of twist, since 
ordering contributions by 
$(z^2)^l$ power in Eq. (\ref{ln2}) was sufficient for our purposes.
But let us discuss now 
the twist expansion of the basic matrix element 
$\langle  p | \phi (0)  \phi(z)  | 0 \rangle $.

\subsubsection{Traceless combinations} 

The operators $ \phi(0){\partial}^{\mu_1} \ldots
{\partial}^{\mu_n}   \phi(0)$ do not correspond to an 
irreducible representation.
They are not traceless,  and that is why their parametrization
requires a set of numbers $A_n^{(l)}$ rather than just one number.  
To get matrix elements   corresponding to an   irreducible
representation   one has   to write  
the tensor $z_{\mu_1} \ldots z_{\mu_n}$
 as a sum of products of powers of $z^2$ and 
symmetric-traceless
combinations $\{\ldots z_{\mu_i } \ldots z_{\mu_j} \ldots \}$ 
satisfying the irreducibility condition \mbox{$g^{\mu_i  \mu_j} 
\{\ldots z_{\mu_i } \ldots z_{\mu_j} \ldots\}=0$}.  Using the notation
$
\{z\partial\}^{n} \equiv  \{z_{\mu_1} \ldots z_{\mu_n} \}  \, 
\partial^{\mu_1} \ldots \partial^{\mu_n} 
$ 
for products of traceless tensors, it is possible to derive \cite{Radyushkin:1983mj}
 \begin{align}
  \phi(z) = \sum_{l=0}^{\infty} 
 \left ( \frac{z^2 }{4}  \right )^l  \sum_{N=0}^{\infty} 
 \frac{N+1}{l!(N+l+1)!}  
\{z{\partial} \}^{N}
 ({\partial}^2)^l 
 \phi(0)  \  . 
 \label{twistD}
\end{align}
Now,  parametrizing  matrix elements of   traceless operators 
\begin{align} 
& \langle  p |\phi (0) \{  {\partial}^{\mu_1} \ldots
{\partial}^{\mu_N}  \}  ({\partial}^2)^l   \phi(0)  
| 0 \rangle \nonumber \\ &
= i^n 
C_{N}^{(l)}  \, \Lambda^{2l}    \{ 
p^{\mu_1} \ldots  \ldots p^{\mu_{N}}   \} 
\label{trC}
\end{align} 
one needs just  one number $C_{N}^{(l)} $ for each operator.
A usual way to make a projection 
on a traceless combination is to multiply Eq. (\ref{trC}) by a 
product $n^{\mu_1} \ldots  \ldots n^{\mu_{N}}$ built from an
auxiliary lightlike vector $n$. Since $n^2=0$,
one has a relation 
\begin{align} 
& \langle  p |\phi (0)  (n  \partial)^N ({\partial}^2)^l   \phi(0)  
| 0 \rangle 
= i^N
C_{N}^{(l)}  \, \Lambda^{2l}  (pn)^N
\label{trC2}
\end{align} 
involving ordinary scalar products $(n  \partial)$ and $(np)$.
Choosing $n$ to be in the ``minus'' direction, we may rewrite 
Eq. (\ref{trC2})
as 
\begin{align} 
& \langle  p |\phi (0)   \partial_+^N ({\partial}^2)^l   \phi(0)  
| 0 \rangle 
= i^N
C_{N}^{(l)}  \, \Lambda^{2l}    p_+^N \ ,
\label{trC21}
\end{align} 
with clear separation of derivatives $ \partial_+$ probing 
the longitudinal   structure of the hadron,  and contracted derivatives
${\partial}^2$  sensitive to distribution of quarks in virtuality. 
The operators containing powers of  ${\partial}^2$ have higher twist,
and  their contribution to the light-cone expansion
 is accompanied by powers of $z^2$.

\subsubsection{Twist expansion and target mass effects}

However, trying  to use the twist decomposition (\ref{twistD})  for getting 
 a closed expression
for $\langle  p | \phi (0)  \phi(z)  | 0 \rangle $ 
similar to  a VDA representation,
one needs to perform  a summation over  $N$
\begin{align}
\langle  p | \phi (0)   \phi (z)  | 0 \rangle = &
\sum_{l=0}^{\infty} 
 \left ( \frac{z^2 \Lambda^2}{4}  \right )^l  \sum_{N=0}^{\infty}  
 \frac{N+1}{l!(N+l+1)!} 
 \nonumber \\ & \times  i^N
\{zp \}^{N}
C_N^{(l)} 
\label{ln211}
\end{align}
that involves structures 
$
\{zp \}^N 
$
built from traceless combinations. 
It is possible to write them 
in  simple powers,
\begin{align} 
\{zp \}^{N}  = \frac{ [(zp)+r ]^{N+1} -[(zp)-r ]^{N+1} }{2^{N+1}  r} 
\  ,
\label{zpnR}
\end{align}
where 
$r=\sqrt{(zp)^2 -z^2 p^2 } $ (see, e.g., Ref. \cite{Radyushkin:1983mj}).
Since Eq.  (\ref{zpnR}) expresses $\{zp \}^{N}$ in terms of powers of
$[(zp) \pm r]$, treating $C_N^{(l)} $ coefficients as appropriately normalized 
$x^N\, \sigma^l$ moments of VDA  $\Phi (x, \sigma)$, one can explicitly 
perform summation over $N$ 
and  obtain formulas involving exponentials $e^{i x [(zp) \pm r]/2}$
instead of $e^{i x (zp)}$
(see  
Refs. \cite{Geyer:1999uq,Blumlein:1999sc} for formulas
 including also the spin-1/2 cases).  However,  further integration over $z$ 
is rather complicated because of the square root involved in $r$. 

Another way is to use  the inverse expansion 
\begin{align} 
\{zp \}^{N}  = (zp)^N - \frac14 {(N-1)} \, z^2 p^2 (zp)^{N-2} + \ldots \  .
\label{zpn}
\end{align}
After the  re-expansion of 
$\{zp \}^{N}$,  one would get   series  in powers of $(pz)$ and $z^2$, in which 
some  of $(z^2)^l$ terms are accompanied by $\Lambda^{2l }$ 
factors having a dynamical origin (virtuality of quarks) and some 
 $(z^2)^k$ terms  that are  accompanied 
by $(p^2)^k$   factors, which 
are purely kinematical (they come from the re-expansion 
of $\{zp \}^{N}$) and reflect nonzero mass of the hadron.

\subsubsection{VDA representation and target mass effects} 

Thus it looks  simpler to use Eq.  (\ref{zpn}), which would give target mass
corrections as a series in $p^2/Q^2$. 
In fact, the most simple way is to 
avoid the  twist  decomposition
altogether.   Note that  
the VDA representation, first,    
 is valid without approximations  and,   second,
 involves the actual 
hadron momentum $p$. This means that 
it is  sufficient     to  merely  treat $p$ ``as is'',
e.g. to use  $(q'-xp)^2 = -(x Q^2 + x \bar x p^2)$ for 
the combination present in our result 
(\ref{Phixs0600}) for the Compton amplitude.

Proceeding in this way,  one can   include,  if  necessary, the 
kinematical hadron  mass effects that are 
analogous to Nachtmann \cite{Nachtmann:1973mr,Georgi:1976ve} corrections 
in deep inelastic scattering. 
However, 
since our primary  goal  is  to   
 concentrate 
on  dynamical effects  (and also given the smallness of the pion mass)
we will   simplify the things by  just  taking   
 $p^2=0$, in which case 
 $\{zp \}^{N}  = (zp)^N$.  
  
 Still, the  discussion of the twist decomposition has an important outcome,
 namely,   the
 understanding that  higher  $l$  terms in Eq. (\ref{ln2})  correspond to 
 local operators with increasing powers of contracted derivatives $\partial^2$ that probe the 
 parton's   virtuality.   It is for this reason why   $\Phi (x, \sigma)$ is  referred  to
 as a virtuality distribution.

 \subsubsection{Equal virtualities}

There is a   kinematics in which the summation over 
spin $N$ is not necessary and only the $l$-sum remains.
Consider a situation when  both photons are virtual,
and moreover, have  equal virtualities, $q_1^2=q_2^2 \equiv - Q^2$.
For a lightlike $p$, we have 
in this case $(pq_1)=(pq_2) =0$. 
Choosing  $p$  in the ``plus'' direction, we conclude  that both 
$q_1$ and $q_2$ do not have  ``minus'' components, and 
their virtualities $q_i^2$ are given by the transverse component 
$q_\perp$ only, \mbox{$q_i^2=-q_\perp^2= -Q^2$.}  Similarly, 
$(q_1-xp)$  and $(q_2-\bar x p)$ do not have the minus component, and 
$(q_1-xp)^2 =(q_2-\bar x p)^2=-q_\perp^2= -Q^2$.
 Parametrizing  the bilocal matrix element, as usual, by (\ref{Phixs02}), 
 we obtain 
 \begin{align}
T(Q^2,Q^2) = &
 \frac{1}{Q^2}   \int_{0}^{\infty} d \sigma \,  \left \{ 1- e^{-[iQ^2 + \epsilon]/  \sigma }  \right \} \, 
  \nonumber \\ &\times   
   \int_{0}^1\, dx \, 
  { \Phi (x,\sigma)    }   
 \  . 
\label{eq:FQQ}
\end{align} 
In this result, VDA $\Phi (x,\sigma) $ enters 
only through the integrated distribution 
 \begin{align}
\Sigma (\sigma) =   \int_{0}^1\, dx \, 
   \Phi (x,\sigma)    \  ,
\label{eq:FQQ2}
\end{align} 
in terms of which we have
 \begin{align}
T(Q^2,Q^2) = &
 \frac{1}{Q^2}   \int_{0}^{\infty}  d \sigma \,  \left \{ 1- e^{-[iQ^2 + \epsilon]/  \sigma }  \right \}  
   { \Sigma (\sigma)    } 
\  . 
\label{eq:FQQ}
\end{align} 
In the OPE language, this means that operators with 
nontrivial, $N \geq 1$
 traceless combinations $\{ z \partial\}^N$ do not contribute,
simply because their matrix elements result in 
  $(pq)^N$ factors that vanish. 
As a result,  only the 
expansion in $\phi (\partial^2)^l \phi$ operators is left.

Switching to TMDA, we get 
 \begin{align}
 T(Q^2,Q^2) 
= &  
   \frac{1}{Q^2}   \int_{k_\perp^2 \leq  Q^2} d^2 { k}_\perp    \int_0^1 dx \,  \Psi (x, { k}_\perp   ) 
      \  . 
 \label{Phixs066}
 \end{align}
 Again, TMDA enters integrated over $x$.

\subsection{Basic relations for VDAs and TMDAs}

  \setcounter{paragraph}{0}

\subsubsection{Analytic continuation of TMDA} 

The  TMDA/VDA 
relation (\ref{Phixs1tmd0}) 
tells us that  ${\Psi} (x, k_\perp )$ is a function of $k_\perp ^2$. 
 In what follows, we will   also use the notation
 $\psi (x, k_\perp^2)\equiv \pi \Psi (x, k_\perp)$ emphasizing   that $\psi$ 
  is explicitly 
 a function of $k_\perp^2$.
 In fact, the relation
 \begin{align}
{\psi} (x, \kappa^2 ) = & {i }
\int_{0}^{\infty} \frac{d \sigma }{\sigma} \, 
 \Phi (x,\sigma) \,  \,  
 e^{- i (\kappa^2-i \epsilon )/ \sigma} 
  \label{psixs2tmd} 
\end{align} 
defines $\psi (x, \kappa^2)$ not only 
 for positive  $\kappa^2$, when it may be interpreted  in terms of 
 the transverse momentum 
 squared 
 $k_\perp ^2$, but also for negative values of $\kappa^2$, in which case it is 
understood  as a formal parameter.
 In other words, Eq. (\ref{psixs2tmd}) provides an analytic continuation 
 of  ${\Psi} (x, k_\perp )$ into the region of negative and complex $k_\perp ^2$.
Formally, the relation between $\psi (x, \kappa^2 ) $ and $ \Phi (x,\sigma)$
may be inverted:
 \begin{align}
\Phi (x,\sigma)= & \frac {1}{2 \pi i \sigma}
\int_{-\infty -i\epsilon}^{+\infty-i\epsilon} d \kappa^2  \, 
 e^{i \kappa^2/ \sigma}  \, {\psi} (x, \kappa^2 ) \ . 
  \label{psixs1inv} 
\end{align} 
In practice,   it is $ \Phi (x,\sigma)$   that  is  a  primary function: it  
is extracted from explicit expressions for  matrix elements
(or their models),
and then one obtains ${\psi} (x, \kappa^2 ) $ using \mbox{Eq. (\ref{psixs2tmd}).}

\subsubsection{VDA representation for the Bethe-Salpeter function} 

Sometimes it is convenient to use the momentum representation
version of the matrix element 
\begin{align}
&  \int d^4z  e^{- i (kz)} \langle p |   \phi(0) \phi (z)|0 \rangle \equiv 
 (4\pi i)^2  \chi_p (k) \,  , \label{Phixsmom}
\end{align} 
where $k$ is the momentum of the quark going from the ``$z$'' vertex,
and $\chi_p (k)$ is the Bethe-Salpeter  wave function.
In the VDA representation,
\begin{align}
  \chi_p (k)  =& 
\int_{0}^{\infty} \frac{d \sigma}{\sigma^2}  \int_{0}^1 dx\,  
 \Phi (x,\sigma) \,  \,  e^{i (k-xp)^2/\sigma -\epsilon/\sigma  }  \ .
 \label{Phiksmom2}
\end{align} 
Comparing  (\ref{psixs2tmd}) 
and (\ref{Phiksmom2})
we may  formally write
\begin{align}
  \chi_p (k)  =  \int_{0}^1 dx\,
  \left [\frac{\partial}{\partial k_\perp^2} {\psi} (x, k_\perp^2 )  \right ]_{k_\perp^2 = -
  (k-xp)^2} \  . 
 \label{Phichi}
\end{align} 
In the  regions, where $(k-xp)^2$ is positive, one should 
understand  ${\psi} (x, k_\perp^2 )$ through 
 the  analytic continuation specified by 
Eq. (\ref{psixs2tmd}).

Thus, the function $\chi_p (k)$ for all $k$ may be obtained 
from  the TMDA  ${\psi} (x, k_\perp^2 )$ and its analytic continuation 
into the region of  negative $k_\perp^2$.
We can also say that  the Bethe-Salpeter  wave function in the
coordinate representation 
\begin{align}
\tilde \chi _p (z) \equiv \langle p |   \phi(0) \phi (z)|0 \rangle 
  \label{Phizcoor}
\end{align}  
 is determined 
for all $z$ by an analytic continuation from 
its values on the light-front $z_+$=0. 

There is also a reduction relation from $\chi_p (k)$ to ${\psi} (x, k_\perp^2 )$.
Taking for simplicity $p^2=0$ (and normalization $k^2 =  k^+ k^- - k_\perp^2$), we have 
\begin{align}
 \int_{-\infty}^{\infty}  &  dk_-  \chi_p (k)  = - 2 \pi  i \,  \int_{0}^1 dx\,   \delta (k^+ - xp^+) \, {\psi} (x, k_\perp^2 ) \ .
 \label{Phiksmom21}
\end{align}

 \subsubsection{Bilocal function} 

In some cases,  it is convenient to use    an intermediate 
 distribution $B(x, z^2/4)$, the {\it bilocal  function}
 defined  through 
\begin{align}
 \langle p |   \phi(0) \phi (z)|0  \rangle 
  \equiv  &
 \int_{0}^1 dx \, B(x, z^2/4)
 \, e^{i x (pz) }   \ . \label{Phixb0}
\end{align} 
Note  that $B(x, z^2/4)$ describes  both  
positive and negative $z^2$, while IDA ${\varphi } (x, z_\perp)=B(x,-z_\perp^2/4)$
corresponds to negative $z^2$ only.

As we have  seen, for any  Feynman diagram
of perturbation theory  $B(x, z^2/4)$ 
 is a function of $z^2-i \epsilon$.
 If the pion is in the initial state, the matrix element
\begin{align}
 \langle 0 |   \phi(0) & \phi (z)|p \rangle 
=  
\int_{0}^{\infty} d \sigma \int_{0}^1 dx\,   \nonumber \\ & \times 
 \Phi (x,\sigma) \,  \,  e^{-i  x (pz) -i \sigma {(z^2-i \epsilon )}/{4}} \,  . \label{Phixs02}
\end{align} 
is still a function of $z^2-i \epsilon$. 
This property is essential in the definition of VDA  
$ \Phi (x,\sigma)$ through 
  \begin{align}
  B(x, \beta) 
  =  \int_{0}^{\infty}  \, e^{-i \beta \sigma} 
   \Phi (x,\sigma) \, d \sigma 
   \label{Bxbeta0}
\end{align} 
which has a form of a  Laplace-type representation.
Its  formal inversion gives
 \begin{align}
\Phi (x,\sigma)= & \frac {1}{2 \pi  }
\int_{-\infty}^{+\infty} d \beta  \, 
  \,  \,  
 e^{-i \beta \sigma}  \,B(x, \beta-i\epsilon) \ . 
  \label{Phiphi} 
\end{align} 
In particular, combining (\ref{Bxbeta0}) and (\ref{psixs2tmd}) gives
\begin{align}
B(x, z^2/4) = \frac1{\pi i} \int_{-\infty -i\epsilon}^{+\infty-i\epsilon} d \kappa^2  
\, K_0 (\sqrt{\kappa^2 z^2}) \,  \psi (x, \kappa^2 )  
\label{bipsi}
\end{align}
(for imaginary arguments, $K_0$ should be understood as the Hankel function $H_0^{(2)}$).

\subsubsection{Moments of TMDA}

The VDA/TMDA relation  (\ref{Phixs1tmd0}) is   quite general  in the sense that it  holds even 
if   the   the matrix element 
$\langle  p | \phi (0)  \phi(z)  | 0 \rangle$ of the bilocal operator 
is non-analytic in the   $z^2 \to $  limit. 
However, if this limit is regular (which happens for a soft VDA 
$\Phi (x,\sigma)$ that vanishes for large $\sigma$ faster than any power of 
$1/\sigma$), one can  
connect  the $\sigma$ 
moments of VDA $ \Phi (x,\sigma)$  
and  $k_\perp^2$ moments of  TMDA ${\Psi} (x, k_\perp )$, 
\begin{align}
\int  {\Psi } (x, k_\perp ) \, k_\perp^{2n} \, d^2 k_\perp =  
\frac{n!}{i^n} 
\int_{0}^{\infty} {\sigma}^n  \, 
 \Phi (x,\sigma) \, {d \sigma }  \  .
 \label{sigma_k}
\end{align}
This connection allows one to get the relation 
\begin{align}
B(x, \beta) = \int  \Psi (x, k_\perp ) \, J_0 (2k_\perp  \sqrt{-\beta})\, d^2 k_\perp \ .
\label{zipsi}
\end{align}
For 
 negative   $\beta=-z_\perp^2/4$, this formula may be 
 also  obtained by performing the angular 
 integration in 
 \mbox{Eq. (\ref{impaf})}, which means that Eq. (\ref{zipsi}) is valid even 
 if the TMDA  $ \Psi (x, k_\perp )$ is not soft. 
When  \mbox{$\beta $}    is positive  (then we can write
$\beta= |z|^2/4$), one may understand  Eq.(\ref{zipsi}) as  
\begin{align}
B(x, |z|^2/4)  = \int  \Psi (x, k_\perp ) \, I_0 (k_\perp |z|)\, d^2 k_\perp \ ,
\end{align}
 where $I_0$ 
is  the modified Bessel function.  This integral converges, e.g.,
for a Gaussian TMDA   $\Psi (x, k_\perp ) \sim e^{-k_\perp^2/\Lambda^2}$,
and the 
 basic
  bilocal  function $B(x, \beta)$ may be then expressed in terms of TMDA
$ \Psi (x, k_\perp )$ both for spacelike and timelike values of $\beta$.
For a TMDA with an  exponential $\sim e^{-\Lambda |k_\perp|}$ fall-off,
the integral diverges for $z^2 \geq \Lambda^2$, which reflects a
singularity of $B(x, z^2/4) $ for  time-like intervals with $z^2 = \Lambda^2$. 

  \setcounter{equation}{0}
 
\section{Scalar handbag diagram in terms of VDA and TMDA } 

\label{handbag}

\subsection{Reducing handbag to a 3-dimensional integral} 
\label{kminus} 

Our approach to get  the TMDA expression
 for the scalar handbag
diagram 
 \begin{align}
 T(q,p) 
= &  
 -  \int_{0}^1   \frac{dx}{(q'-xp)^2}  
     \int_{k_\perp^2 \leq -(q'-xp)^2}  \, d^2 { k}_\perp 
 \,   \Psi (x, { k}_\perp   )  \ ,
 \label{Phixs06000}
 \end{align}
by-passes    the standard   idea 
of starting with 
a   4-dimensional integral
\begin{align}
T(q,p)  =  \int  \frac{ \chi_p (k)} { (q'-k)^2}  \ d^4 k  \  , 
\label{eq:Fscalarmom0}
\end{align}
 decomposing  the integration momentum 
$k$ in the  light-front components \mbox{$k =\{k_+ ,  k_-, k_\perp\}$,} 
 with the ``plus'' direction given by $p$,
and then  trying  to integrate over $k_-$. 
An obvious difficulty of such an approach 
is that the $k$-dependence of $\chi_p (k)$ is not explicit.
 \begin{figure}[h]
 	\centerline{\includegraphics[width=2.5in]{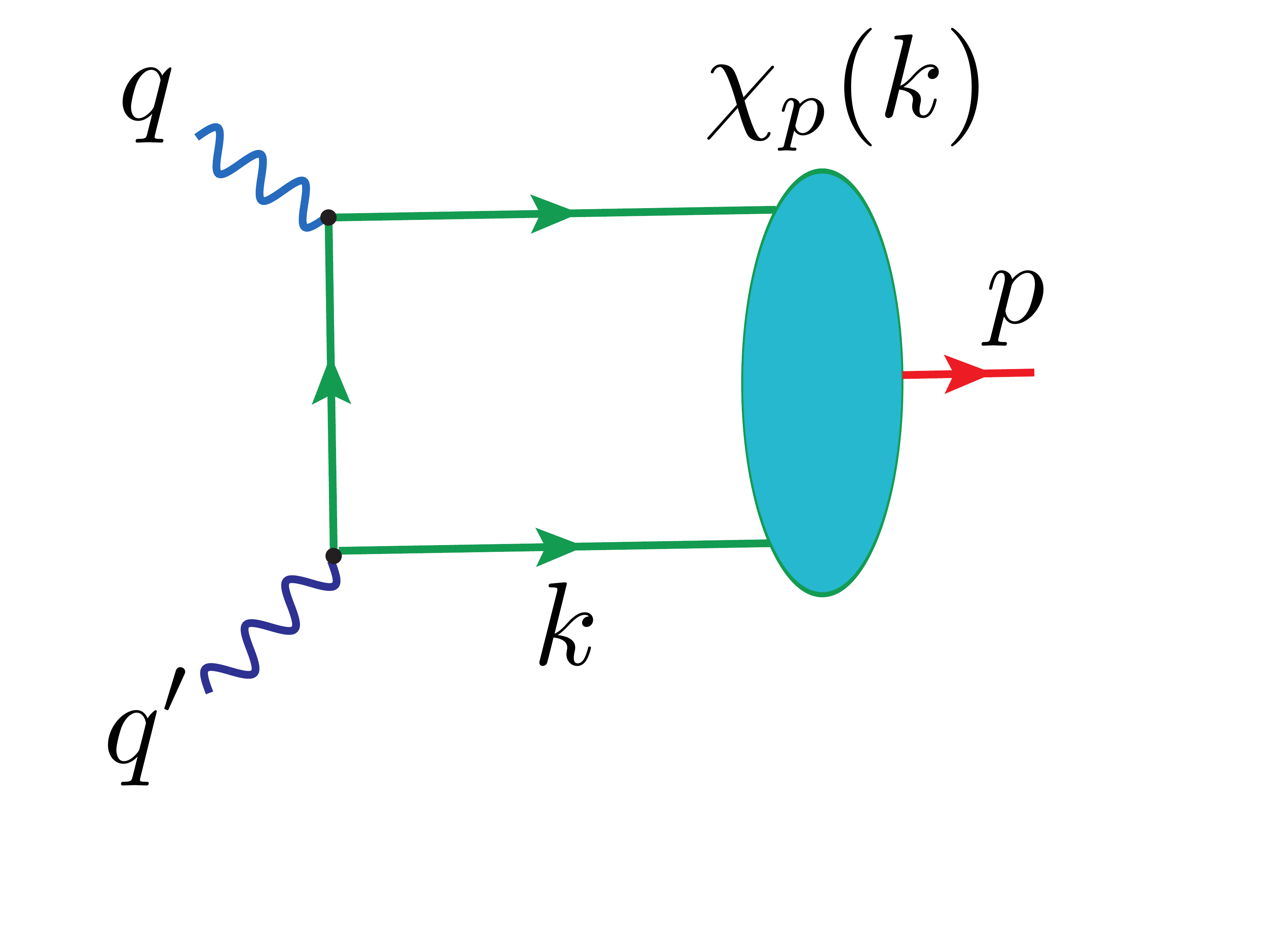}}
 	\vspace{-0.5cm}
 	\caption{Handbag diagram in momentum representation.
 		\label{sudak2}}
 \end{figure}
The usual way out  of this situation  is to use some  approximation
that eliminates the $k_-$-dependence of the hard propagator $1/(q'-k)^2$.
After that, one deals with the function $\chi_p (k)$ integrated   over $k_-$,
that  depends  on $k_+$ and $k_\perp$.
Below we  consider  two  approximations of this kind.

\subsubsection{Neglecting $k^2$ in hard propagator} 

Since  the $q'$ photon is  real,  ${q'}^2=0$,
we  deal with 
 \begin{align}
T(q,p)  = \int   \frac{ \chi_p (k)}{2(q' k)-k^2} \, d^4 k  \  . 
\label{eq:Fscmom0}
\end{align}
  As usual, for $p^2=0$ one may choose 
$p$ to define   the  plus direction and introduce $x$ 
through   $k_+ =xp$. Another light-like vector  ${q'}$ may be chosen to define
 the minus direction. Then $2(q' k) =2x (q'p) = xQ^2$,   and 
 if we neglect $k^2$ in the denominator we obtain 
  \begin{align}
T ^{(k^2\Rightarrow 0)} (q,p) |_{p^2=0} =   \int_0^1 dx \,  \frac{\varphi(x) }{xQ^2}  \  ,
\label{eq:Fscal000}
\end{align}
 where $\varphi (x)$ is the distribution amplitude 
  \begin{align}
\varphi(x) =  \int  \, d^4 k \,    \delta (x- k_+/p_+ ) \, \chi_p (k)   \  .
\label{eq:Fscmom1}
\end{align}
 Apparently, this formula gives the desired result  (\ref{eq:Fscalar000}) for   
 soft wave functions. However, since it cannot produce  any power  correction 
 in principle, it  cannot be correct for
 wave functions corresponding to 
 matrix elements that have logarithmic singularities  
 for $z^2=0$  in $l\geq 1$ terms: in such cases we {\it should have}
 $1/Q^2$ corrections. 
  
  Most importantly, neglecting virtuality $k^2$ one also neglects transverse 
 momentum effects altogether, while we want to keep track of them.

\subsubsection{Neglecting $k_-$  in hard propagator}  

To this end, we  write a more detailed 
  decomposition 
    \begin{align}
k= xp + k_- + k_\perp 
 \end{align}
which gives  $k^2= 2x (p k_- ) -k_\perp^2$,
and the approximation is to neglect $ 2x (p k_- )$,
 while keeping the $k_\perp^2$ part of $k^2$
in the propagator.
This gives
    \begin{align}
T ^{(k_-\Rightarrow 0)} (q,p) |_{p^2=0} =   \int dx \, \int d^2 k_\perp  
\frac{\Psi(x,k_\perp) }{xQ^2+k_\perp^2 }  \  ,
\label{eq:Fscal0kmin}
\end{align}
where $\Psi (x, k_\perp)$ is the  transverse momentum dependent 
distribution amplitude, 
  \begin{align}
\Psi (x,k_\perp) =  \int  \, d k_+  dk_-  \,    \delta (x- k_+/p_+ ) \,  \chi_p (k)   \  .
\label{eq:Fscmom11}
\end{align}

Now, the formula (\ref{eq:Fscal0kmin}) always generates a tower of 
$(k_\perp^2/Q^2)^n$ corrections, so it cannot  be  correct for soft wave functions.
Furthermore, since taking $k_-=0$ in the hard propagator is an approximation,
 it  cannot be absolutely correct for ``hard'' wave functions as well.
 
 To estimate the quality of this approximation, note that 
 if $k^2=0$, we have $2x (pk_-) = k_\perp^2$, i.e. the  term that 
 is neglected has the same magnitude as the one that is kept.
 Assuming that  $k^2$ has  some  average value of  $\Lambda^2$
 while $  k_\perp^2$  averages to $\langle   k_\perp^2 \rangle$,
 we conclude that $2x (pk_-)$ averages to 
 $\Lambda^2 +\langle   k_\perp^2 \rangle$
 which is not necessarily zero.  In other words,  neglecting 
 $2x (pk_-) $ is equivalent to assuming that a nonzero 
 vrituality comes entirely from parton's transverse momentum,
 which is a dynamical question, the answer to which is not clear {\it a priori}.

In general, the main  idea of this procedure is based on neglecting
$k_-$ in the hard subprocess amplitude, and 
thus  it creates an impression that the description of the hadron structure
in terms of two variables $x$ and $k_\perp^2$ may be only obtained as 
a result of some  approximation.

However, deriving our result (\ref{Phixs06000})  
we did not make any approximations.
Such an outcome became possible because we 
 were able to perform the 4-dimensional integration over $k$
using the  VDA representation which explicitly 
specified  the dependence of $\chi_p (k)$ on $k$.
So, let us study what  happens if we use 
 a framework that  involves a usual 
explicit decomposition of the integration variable
(4-momentum $k$ or 4-dimensional coordinate $z$) into light-front 
components. When  necessary, we will  also  incorporate the VDA representation
adjusted to such a decomposition.

 \subsection{Handbag diagram in  coordinate light-front variables } 
 
To begin with, we try  a Sudakov-type 
decomposition of the original coordinate space integral
 (\ref{eq:Fscalar00}).  
 Taking  $p$ in ``+'' direction, and $q'$ in ``--'' direction, and using  $z= \{z_+, z_-, z_\perp \}$, 
we have 
   \begin{align}
 \nonumber 
T(q,p)& =  -\frac{i}{2(2\pi)^2}  \int_{0}^1 dx
 \int d^2 z_\perp\,
\int_{-\infty}^\infty   dz_+
\int_{-\infty}^\infty   
dz_- 
\\ & \times
e^{-iq'_-z_++i x p_+z_- } \, 
 \frac{ B(x, z^2/4)}{z_+z_- - z_\perp^2 -i \epsilon}  \  . 
    \label{eq:FscalarI1} 
\end{align}

The integrand has an explicit  pole  at \mbox{$z_+= (z_\perp^2 +i \epsilon)/z_-$},
 which corresponds to \mbox{$z^2=0$.} 
One may wish   to  calculate the  integral over $z_+$   by taking 
 residue at this location. 
Whether this is possible, depends on the analyticity properties of 
$B(x, z^2/4)$. 

\setcounter{paragraph}{0}

\subsubsection{Soft wave function}  

Take first a soft wave function case when 
 $B(x, z^2/4)$ 
 is given by a $(z^2)^l$ Taylor expansion     with finite coefficients. Now, 
if one   treats  this  expansion term by term,   then one should take 
$z^2=0$ in $B(x, z^2/4)$, which amounts to keeping just   the lowest $l=0$ 
contribution. Since $q'_- >0$, the integral is nonzero for $z_- <0$ only,
i.e.,  the ``$0$'' vertex   corresponding to the virtual  photon is 
{\it later}  in the light-cone  ``time''   $z_-$
than the real photon  vertex located at $z$. 
 The result is 
   \begin{align}
   T^{\{ {\rm soft}\}} &  (q,p) =   \frac{1}{2\pi} \int_{0}^1 dx
 \int d^2 z_\perp\,
\int_{0}^\infty   \frac{dz_-}{z_-} \,  
\nonumber \\ & \times 
 e^{-iz_\perp^2q'_- /z_- +i x p_+z_- } \, 
 B^{\rm ( soft )} (x, 0) 
 \nonumber \\ & =
 T^{{\rm ( soft },l=0)} (q,p)
  \  . 
\label{eq:Fscalar121}
\end{align}

Taking    $z_-$ integral and using that $ B^{\rm ( soft )} (x, 0) =  \varphi (x)$,
we obtain the  representation
   \begin{align}
T^{{\rm ( soft })} (q,p)  {=} &  \int_{0}^1 dx
 \int z_\perp d z_\perp\,
 K_0 (z_\perp \sqrt{xQ^2}) \, 
  \varphi (x)  \   , 
\label{eq:Fscalar122}
\end{align}
in which the distribution amplitude  $ \varphi (x)$  has no $z_\perp$ dependence.
Then the integral over $z_\perp$ is trivial, with the result 
   \begin{align}
T^{{\rm ( soft })} (q,p)  {=} &  \int_{0}^1 \frac{dx}{xQ^2} 
\, 
  \varphi (x)  \  
\label{eq:Fscalarz22}
\end{align}
that agrees with Eq. (\ref{eq:Fscalar000}), as expected. 

Note that the  only thing that is needed 
from the 
$z_\perp$-dependent factor 
$ K_0 ( \sqrt{x z_\perp^2Q^2})$  for this agreement   is that its 
 integral over $z_\perp^2$ gives $1/xQ^2$. 
In other words, any function $Z(a)$  of $a= x z_\perp^2Q^2$ 
producing 1 after the \mbox{$0\leq a \leq \infty$}  integration,
would produce  Eq. (\ref{eq:Fscalarz22}).
Thus one has  little grounds to argue 
that  the specific  \mbox{$z_\perp$-dependence}  
of 
$ K_0 ( z_\perp \sqrt{x Q^2})$ should be present 
in general case when   the $z_\perp$  dependence is added to 
 the IDA

Note also that to get the $z_\perp$-dependent IDA $ \varphi (x, z_\perp^2)$ 
one should project    
the 
bilocal function $B(x,z^2)$,   onto the light-front $z_+=0$.
However, 
for a  residue taken at   
\mbox{$z_+= (z_\perp^2 +i \epsilon)/z_-$} this  is not the case
   when  $z_\perp^2 \neq 0$. 

\subsubsection{General case, preliminary steps and reproduction of VDA result}

 Furthermore, one cannot take the integral (\ref{eq:FscalarI1})
 by a simple residue if $B(x, z^2/4)$ has singularities, like $\ln z^2$,  in the 
 complex $z^2$ plane. 
To analyze 
a  general case,  we write   $B(x, z^2/4)/z^2$   in terms of VDA, 
   \begin{align}
   \nonumber   
T(q,p) = & \frac{1}{2(2\pi)^2}  \int_{0}^1 dx
 \int d^2 z_\perp\,
\int_{-\infty}^\infty   dz_+
\\ & \times
\int_{-\infty}^\infty   
dz_- e^{-iq'_-z_+ +i x p_+z_- } \, 
\int_{0}^{\infty} \sigma \, d \sigma  \int_{0}^1 d \beta
  \nonumber \\ & \times 
 \Phi (x,\beta \sigma) \,  \,  e^{-i \sigma {(z_+z_- - z_\perp^2-i \epsilon )}/{4}} \, 
  \  .    \label{eq:FscalarI2}
\end{align}
 Integrating over $z_+$ produces 
   \begin{align}
    \nonumber    
T(q,p) = & \frac{1}{4\pi}  \int_{0}^1 dx
 \int d^2 z_\perp\, 
\int_{0}^{\infty} \sigma \, d \sigma
\int_{0}^1 d \beta\,  \Phi (x,\beta \sigma) \,
\\ & \times   \,  e^{i \sigma ( z_\perp^2+i \epsilon )/4} \,
\int_{0}^\infty   
dz_- e^{i x p_+z_- } \, 
 \delta ( q_-'+ \sigma z_-)
  \  . \label{eq:FscalarI3}
\end{align}
Since $q'_->0$, we  have  $z_-<0$, just like when we  took a   residue.  
   Integrating over $z_-$ and changing    $\beta \sigma \to \sigma$ 
gives 
   \begin{align}
 \nonumber  
T(q,p) = & \frac{1}{4\pi}  \int_{0}^1 dx\,
 \int d^2 z_\perp\, \int_{0}^{\infty} d \sigma \,  \Phi (x, \sigma) \,
\\ & \times
 \, 
\int_{0}^1 \frac{d \beta}{\beta}
  \,  e^{i \sigma ( z_\perp^2+i \epsilon )/4\beta} \, 
  e^{-i \beta x Q^2/\sigma} 
  \  .    \label{eq:FscalarI7}
\end{align}

It is rather easy to convert  this expression into  the  VDA formula  for $T(q,p)$. Indeed,  
  integrating over $z_\perp$ results in 
 \begin{align}
   \label{eq:FscalarI6}
T(q,p) = & i  \int_{0}^1 dx
 \, 
\int_{0}^{\infty} \frac{d \sigma}{\sigma} \, \Phi (x, \sigma) 
\int_{0}^1 d \beta \, 
 \,  e^{-i \beta x Q^2/\sigma} 
  \  ,
\end{align}
which,  after the $\beta$ integration, gives 
 \begin{align}
T(q,p) = &
 \int_{0}^1 \frac{dx}{xQ^2} \, \int_{0}^{\infty}   d \sigma \, 
  { \Phi (x,\sigma)    }  
\left \{ 1- e^{-i xQ^2 /  \sigma }  \right \} 
\label{eq:Fscalar3om}
\end{align} 
that  coincides with Eq. (\ref{eq:Fscalar3000})
in our case of $p^2=0$.

\subsubsection{General case, keeping $z_\perp$ dependence}

However, if we want to keep  $z_\perp$ variable, we need 
to take the  integrals over $\sigma$ or $\beta$ instead. 
  Changing $\beta = \sigma \rho$ in (\ref{eq:FscalarI7}) gives 
   \begin{align}
\nonumber  
T(q,p) = & \frac{1}{4\pi}  \int_{0}^1 dx\,
 \int_{0}^{\infty} d \sigma \,  \Phi (x, \sigma) \,  \int d^2 z_\perp\,
\\ & \times
 \, 
\int_{0}^{1/\sigma}  \frac{d \rho}{\rho}
  \,  e^{i ( z_\perp^2+i \epsilon )/4\rho} \, 
  e^{-i \rho x Q^2} 
  \  .    \label{eq:FscalarI8}
\end{align}
At this stage, it is instructive to return to 
Eq. (\ref{eq:FscalarI2})  and  combine there the integrations over 
$z_+$ and $z_-$    into one 2-dimensional integration
over $z_\|$   
   \begin{align}
   \nonumber   
T(q,p) &=  \frac{1}{2(2\pi)^2}  \int_{0}^1 dx
\int_{0}^{\infty} \sigma \, d \sigma  \int_{0}^1 d \beta \, 
 \Phi (x,\beta \sigma) \,
  \nonumber \\ & \times  
 \int d^2 z_\perp d^2 z_\| \, e^{i \tilde q z_\| } \, 
 \,  e^{-i \sigma {(z_\|^2- z_\perp^2-i \epsilon )}/{4}} \, 
  \  ,  \label{eq:FscalarI21}
\end{align}
where $\tilde q \equiv q' -xp$  has  longitudinal components only.
Now  it is clear that the factor $ e^{-i \rho x Q^2}  $ in Eq.(\ref{eq:FscalarI8})
comes 
 from the $d^2 z_\|$ integration. Using the fact that 
$\tilde q$ is space-like, $\tilde q^2 = -xQ^2$, we can
represent
 \begin{align}
 e^{-i \rho x Q^2}  = \int \,  \frac{d^2 \zeta_\perp }{4 \pi  i \rho}  \, 
 \, e^{i  \zeta_\perp^2/4\rho - i (\kappa_\perp  \zeta_\perp)} \ , 
   \label{eim} 
\end{align}
where $\kappa_\perp$  is 
a two-dimensional vector  
satisfying \mbox{$\kappa^2_\perp =-\tilde q^2 = xQ^2$}. This gives 
  \begin{align}
 \nonumber  
T(q,p) &=  \frac{1}{(4\pi)^2i}  \int_{0}^1 dx\,
 \int_{0}^{\infty} d \sigma \,  \Phi (x, \sigma) \,
 \int d^2 z_\perp\, 
  \int \,  {d^2  \zeta_\perp }
 \, 
  \nonumber \\ & \times  
  \int_{0}^{1/\sigma}  
 \frac{d \rho}{\rho^2}
  \,  e^{i ( z_\perp^2+ \zeta_\perp^2+i \epsilon )/4\rho} \, 
  e^{-i  (\kappa_\perp  \zeta_\perp) } 
  \  .   \label{eq:FscalarI81}
\end{align}
Integrating  over $\rho$ and   then  over $\sigma$  to switch to  IDA gives 
 \begin{align}
   \nonumber   
T(q,p) = & \frac{1}{(2\pi)^2}  \int_{0}^1 dx\,
 \int d^2 z_\perp\, 
  \int \,  {d^2  \zeta_\perp } \\ & \times
 \, 
  \,  \frac{\varphi (x,  z_\perp^2+ \zeta_\perp^2)} {z_\perp^2+ \zeta_\perp^2}\, 
  e^{-i  (\kappa_\perp  \zeta_\perp) } 
  \  .  \label{eq:FscalarI83}
\end{align}
 Integrating  over the angle between $\kappa_\perp$ and
$\zeta_\perp$  we have  
 \begin{align}
 \nonumber   
T(q,p) = & \frac{1}{2}  \int_{0}^1 dx\,
 \int_0^\infty  dz_\perp^2 
  \int_0^\infty  \,  {d  \zeta_\perp^2 } \\ & \times
 \, 
  \,  \frac{\varphi (x,  z_\perp^2+ \zeta_\perp^2)} {z_\perp^2+ \zeta_\perp^2}\, 
J_0 \left ( \sqrt{xQ^2  \zeta_\perp^2} \right )
  \  .  \label{eq:FscalarI84}
\end{align}
Thus, the total impact parameter variable 
\mbox{$b^2 \equiv z_\perp^2+ \zeta_\perp^2$} of the IDA $\varphi (x, b^2)$
comes from the transverse  $z_\perp$ part  of the original $d^4z$ integration,
and from an additional  term $\zeta_\perp^2$ reflecting the result of the 
integration over the 
longitudinal  $z_\|$ 
part of      $z$. 
 In other words, the  additional term   $ \zeta_\perp^2$  
 is associated with the virtuality $xQ^2$ of the 
hard quark propagator. 

\subsubsection{Approximate expression} 

Due to weighting of $ \zeta_\perp^2$ by the Bessel function 
$J_0 \left ( \sqrt{xQ^2  \zeta_\perp^2} \right )$ that rapidly decreases
with $ \zeta_\perp^2$ for large $xQ^2$, one may estimate $ \zeta_\perp^2 \sim 1/xQ^2$ 
in this formula. 
Neglecting $ \zeta_\perp^2$ in the argument of IDA 
(but keeping it in the $1/(z_\perp^2+ \zeta_\perp^2)$ factor) 
 and using 
  \begin{align}
   \label{eq:FscalarI831}
  \int_0^\infty \, 
  \,  \frac{ {\zeta_\perp  d  \zeta_\perp }} {z_\perp^2+ \zeta_\perp^2}\, 
  J_0 \left ( \sqrt{x Q^2  \zeta_\perp^2} \right )
   = 
   K_0 \left ( |z_\perp| \sqrt{x Q_\perp^2 } \right ) 
  \  , 
\end{align}
we obtain the expression 
   \begin{align}
T(q,p){=} &  \int_{0}^1 dx
 \int z_\perp d z_\perp\,
 K_0 (z_\perp \sqrt{xQ^2}) \, 
 \varphi (x, z_\perp^2) + \ldots \  . 
\label{eq:FscalarI222}
\end{align}
Its  explicit part  coincides with a conjecture 
\begin{align}
	T(q,p) \stackrel{?}{=} &  \int_{0}^1 dx
	\int z_\perp d z_\perp\,
	K_0 (z_\perp \sqrt{xQ^2}) \, 
	\varphi (x, z_\perp^2) \  ,
	\label{eq:Fscalarz22a}
	\end{align}
that is used 
	as an impact parameter representation for the pion transition form factor
 in many papers (see, e.g., Ref. \cite{Jakob:1994hd}). 
	However, Eq. (\ref{eq:Fscalarz22a})  is not OPE compliant 
	for a soft wave function, since 
	a $(z_\perp^2)^l$ term from  the expansion of $\varphi (x, z_\perp^2)$ would 
	produce  
	an unwanted  tower of 
	$(1/xQ^2)^{l+1}$ power corrections under the \mbox{$x$-integral.}
As we  see now,  Eq. (\ref{eq:Fscalarz22a})  may be treated as an approximation 
only, based on the assumption that   $\zeta_\perp^2  \ll z_\perp^2$ in the argument 
of the IDA $\varphi (x,  z_\perp^2+ \zeta_\perp^2)$.

In fact, due to a rapid decrease of the modified Bessel 
function $K_0 (y)$ with increasing $y$,
the essential values of $z_\perp^2$ in  Eq. (\ref{eq:FscalarI222}) 
are  restricted to  $z_\perp^2 \sim 1/xQ^2$,  which is of 
the same size as  those for $\zeta_\perp^2$ that were  neglected compared to $z_\perp^2$
in the argument of IDA.  Thus, the correctness  of the approximation leading to 
Eq. (\ref{eq:FscalarI222})  is very questionable. 

The momentum space equivalent of  Eq. (\ref{eq:FscalarI222})
is the formula  (\ref{eq:Fscal0kmin}) obtained by neglecting 
the minus component $k_-$ in the hard part, but keeping $k_\perp^2$.
Thus, our coordinate space considerations suggests that the 
neglected $k_-$ effects may be  comparable in magnitude to those
caused by $k_\perp^2$.

\subsubsection{Comparison with exact  result} 

To proceed without approximations, 
we 
change \mbox{$z_\perp^2+ \zeta_\perp^2 = b^2$},    $ \zeta_\perp^2 = \gamma b^2$
in 
 Eq. (\ref{eq:FscalarI84})   to get 
 \begin{align}
 \nonumber    
T(q,p) = & \frac{1}{2}  \int_{0}^1 dx\,
 \int_0^\infty  d b^2 
  \int_0^1 d \gamma \\ & \times
 \, 
  \,  {\varphi (x,  b^2)}  
J_0 \left (2 \sqrt{\gamma \, xQ^2 b^2} \right )
  \  .  \label{eq:FscalarI85}
\end{align}
Integrating over $\gamma$ gives the expression  
 \begin{align}
   \label{eq:FscalarI85}
T(q,p) = &  \int_{0}^1 \frac{dx} { \sqrt{ x} Q}\,
 \int_0^\infty  db 
 \, 
  \,  \varphi (x,  b^2)
{J_1 \left (2 \sqrt{ x} Q b  \right )}
\end{align}
 that coincides with  Eq. (\ref{eq:Fscalar1230}), the VDA formula written in the impact parameter representation.

This discussion shows that the impact parameter $b$ in
the VDA/IDA formula (\ref{eq:FscalarI85})  differs from
the transverse distance $z_\perp$ in the original coordinate space integral
(\ref{eq:FscalarI1}) 
for $T$, namely, $b^2 = z_\perp^2 +\zeta_\perp^2$ with $ \zeta_\perp^2 ={\cal O} (1/xQ^2)$.
However, the essential $z_\perp^2$ are also of the order of $1/xQ^2$.

  \subsection{Calculation in momentum representation}

The VDA result  (\ref{Phixs060})  for  the handbag diagram 
can  also be obtained using the momentum representation for the VDA.
 The integral now reads 
     \begin{align}
T(q,p) & =    \int_{0}^1 dx
\int \frac{d^4k}{(q'-k)^2}
\nonumber \\ & \times  \int_0^\infty d \alpha \, e^{i\alpha (k-xp)^2} 
 \, \Phi (x, 1/\alpha) \  . 
\label{eq:Fscalarmom}
\end{align}
  Integrating over $k$ gives
    \begin{align}
T(q,p) = &  \int_{0}^1 dx
  \int_0^\infty d \alpha \, \int  \frac{d \alpha_1}{(\alpha+\alpha_1)^2}    \nonumber \\ &  \times 
e^{i \alpha_1  \alpha \tilde q^2/(\alpha+\alpha_1)} \, 
 \, \Phi (x, 1/\alpha) \ . 
\label{eq:Fscalarmom2}
\end{align}
 Switching to $\sigma_1 =1/\alpha_1$, $\sigma =1/\alpha$ and 
integrating over $\sigma_1$  gives the same VDA result
 as in  Eq. (\ref{eq:Fscalar3om}), which 
 may be converted into 
  \begin{align}
 T(Q^2,p^2) 
= &  -
   \int_{0}^1   \frac{dx}{\tilde q^2}  
  \int_{\kappa_\perp^2 \leq -\tilde q^2}  \, d^2 { \kappa}_\perp 
 \,   \Psi (x, { \kappa}_\perp   )  \ .
 \label{Phixs060a}
 \end{align}
Note that 
 the transverse momentum variable ${ \kappa}_\perp $ here  has formally 
 no direct connection with the momentum $k$ of the starting integral
 (\ref{eq:Fscalarmom}).

Alternatively, one may  wish to  choose  
 a particular decomposition of $k$, say, the Sudakov parametrization, 
 in which $k$ is split  into plus, minus and transverse
 components,  and perform integration over the minus component,
 trying to get an expression in terms of   the Sudakov  transverse 
 momentum $k_\perp$.  
 
 \setcounter{paragraph}{0}

 \subsubsection {Sudakov representation}  
 
 Switching for simplicity to $p^2=0$ case and
 using parametrization 
    \begin{align}
 k= \xi p + \eta q' + k_\perp \ ,
 \label{kSud2} 
  \end{align}
gives 
     \begin{align}
T(Q^2) = &  \int_{0}^1 dx   \int_0^\infty d \alpha  \, F(x, \alpha)  \int_0^\infty d \alpha_1  
   \nonumber \\ &  \times 
\frac{Q^2}{2} \int_{-\infty}^\infty  {d\xi \, d\eta } \int d^2 k_\perp 
\, e^{i\alpha (\xi -x)\eta Q^2} \,
   \nonumber \\ &  \times   e^{i\alpha_1 (\eta-1)\xi  Q^2} \, e^{-i(\alpha+\alpha_1)  k_\perp^2} \ .
\label{eq:FscalarSud}
\end{align}
 The minus component of $k$  is proportional  to $\eta$. Integrating  over 
it gives
        \begin{align}
T(q,p) & =  \int_{-\infty}^\infty  {d\xi }    \int d^2 k_\perp  \int_0^\infty {d \lambda}\int_0^1  d \beta
 \, \int_{0}^1 dx   \, F(x,  \beta  \lambda)  
   \nonumber \\ &  \times 
\, 
   e^{-i \bar \beta \lambda \xi  Q^2}\, \delta (\xi  -  \beta x) \, e^{-i \lambda  k_\perp^2} 
     \  .
\label{eq:FscalarSud4}
\end{align}

Thus, the Sudakov variable  $\xi =\beta x$ is smaller than the VDA variable
$x$.  Integrating over $\xi$ results in 
        \begin{align}
T(q,p) & =     \int d^2 k_\perp  \int_0^\infty {d \lambda}\int_0^1  d \beta
 \, \int_{0}^1 dx   \, F(x,  \beta  \lambda)  
   \nonumber \\ &  \times 
\, 
   e^{-i  \lambda \beta \bar \beta x Q^2}\,e^{-i \lambda  k_\perp^2} 
     \  .
\label{eq:FscalarSud4add}
\end{align}
Using integral over $\lambda$ to introduce TMDA gives 
        \begin{align}
T(Q^2) & =   -  \int_0^\infty  d k_\perp^2 
 \, \int_{0}^1 dx   \,   \int_0^1  d \beta \, 
    \nonumber \\ &  \times   \frac{\partial} {\partial k_\perp^2}
\psi \left (x,   \frac{k_\perp^2}{\beta }  + \bar \beta xQ^2 \right )  
     \  .
\label{eq:FscalarSud5add}
\end{align}
Our intention is  to keep $k_\perp$, but let us see   first what happens 
if we   integrate   over   it. Then 
     \begin{align}
T(q,p) & =  
   \int_{0}^1  {dx }  \,  \int_0^1 {d \beta}   \,  
\psi (x,\bar \beta  x Q^2)  
     \  ,
\label{eq:FscalarSud6add}
\end{align}
which leads to  the VDA result  
    \begin{align}
T(q,p) & =  
    \int_{0}^1  \frac{dx } { x Q^2}  \int_0^{xQ^2}  {d \kappa^2}  \, 
\psi (x,\kappa^2 )  
     \  . 
\label{eq:FscalarSud7add}
\end{align}

To keep  the original Sudakov 
 variable  $k_\perp$, we will try 
to  integrate over $\beta$   in Eq. (\ref{eq:FscalarSud5add}). 
Using 
         \begin{align}
 -  \frac{\partial} {\partial k_\perp^2}
\psi (x,  &  k_\perp^2/\beta  + \bar \beta xQ^2)  =
  \frac{\beta}{ k_\perp^2+ \beta^2  xQ^2} \, 
     \nonumber \\ &  \times 
     \frac{\partial} {\partial \beta} \,
      \psi  \left (x,   \frac{k_\perp^2}{\beta}   + \bar \beta xQ^2 \right ) 
\label{eq:FscalarSud8add}
\end{align}
 we obtain a rather long and complicated expression 
    \begin{align}
 T(q,p)  = &  \int_{0}^1 dx  \,  \int_0^\infty  d k_\perp^2 
 \Biggl [  \frac{\psi (x, k_\perp^2)}{xQ^2 +k_\perp^2}  
 \nonumber \\ &
 +
 \,   \int_0^1  d \beta
 \frac{\beta^2 x Q^2- k_\perp^2}{ (k_\perp^2+ \beta^2  xQ^2)^2 }
  \, \psi  \left (x,   \frac{k_\perp^2}{\beta}   + \bar \beta xQ^2 \right )  \, \Biggr  ]
     \  .
\label{eq:FscalarSud5add}
\end{align}
Only the first term here is rather simple 
      \begin{align}
T^{(1)} (q,p)  = &   
 \int_{0}^1 dx    \int d k_\perp^2 
   \frac{ \psi \left  (x, { k_\perp^2 } \right  )  }{xQ^2 +k_\perp^2} \ ,
   \label{eq:FscalarSud1fin}
\end{align}
and gives  the   expression used in many papers
(see, e.g.  Ref. \cite{Jakob:1994hd})  based on the Sudakov parametrization.
Note, however, that in our derivation 
it involves the VDA  variable $x$ rather than the Sudakov variable $\xi$. 

As we discussed in Sec. \ref{kminus}, one can get such a formula  
by  neglecting the minus component of $k$ in the hard propagator.
We have  also emphasized  that  this  expression is 
 not ``OPE compliant''. In particular, 
 unlike the VDA approach expression  (\ref{eq:FscalarSud7add}),  
it   generates a  tower of $k_\perp^2/Q^2$ corrections which should be absent for 
soft wave functions $\psi \left  (x, { k_\perp^2 } \right  )$. 
Still, it correctly reproduces in this case the leading power term   
 (\ref{eq:Fscalar000}). Also, for hard tails, when   $\psi^{\rm hard}  (x, { k_\perp^2 } )
 \sim (\ln k_\perp^2)^m/k_\perp^2$, it correctly reproduces 
 the leading part of the resulting $(\ln Q^2)^{m+1}/Q^2$ contribution to $T(Q^2)$.
 These observations  justify, to some extent,   the  
use of  Eq. (\ref{eq:FscalarSud1fin}).  
Nevertheless,  
 it  is just an approximation,
 while  Eq. (\ref{eq:FscalarSud7add})  is an exact result.
 
The difference between  them is   given  by    
the second integral in  Eq. (\ref{eq:FscalarSud5add}).
As one can see,     it has a rather complicated form and contains TMDA  in which 
    the   transverse momentum
    argument ${ k_\perp^2 }  $ is   rescaled by $1/\beta$ factor 
    and then  shifted 
by 
    a $\bar \beta$ fraction of the hard virtuality $x Q^2$.
We can restore in this term  the Sudakov variable $\xi = \beta x$  
to see that the first argument of TMDA here differs from 
 the Sudakov variable $\xi$ by the $1/\beta$ factor.
 
  In fact, since  the $\beta$ integration is present  in the second  term
  of (\ref{eq:FscalarSud1fin}),
 we did not reach our goal of reducing the $d^4 k$ integral
 to integration over just the plus momentum fraction and transverse momentum. 
 The only  way we see to get rid of the $\beta$ integral here 
 is to return to the VDA result (\ref{eq:FscalarSud7add}).

\subsubsection {Comparison in the  impact parameter representation}

The Sudakov variable $k_\perp$ is Fourier-conjugate to
the transverse coordinate $z_\perp$ of the virtual photon vertex 
in Eq.  (\ref{eq:FscalarI1}), so we may write  
 \begin{align}
{\Psi}(x, k_\perp ) = \int \,  \frac{d^2 z_\perp }{(2 \pi)^2}  \, 
  {\varphi } (x, z_\perp)  \, e^{- i (k_\perp z_\perp)} \ ,
   \label{pafim2} 
\end{align}
and present  the approximation (\ref{eq:FscalarSud1fin})  in the  impact parameters space as 
   \begin{align}
T^{(1)}(Q^2)
   {=}   \int_{0}^1 { dx}
 \int_0^\infty  
d  z_\perp  z_\perp\,
 K_0 (z_\perp \sqrt{xQ^2}) \, 
 \varphi (x, z_\perp^2)  \  , 
\label{eq:Fscalar1237}
\end{align}
which coincides with the term explicitly written in Eq. 
(\ref{eq:FscalarI222}).  It is instructive to compare this result with 
the impact parameter version (\ref{eq:Fscalar1230}) of the VDA formula, 
   \begin{align}
T(Q^2)
   {=}   \int_{0}^1 \frac{ dx}{\sqrt{x}Q }
 \int_0^\infty db \,
   \,J_1  (\sqrt{x}Q b) \, \varphi (x,b) \  .
\label{eq:FscalarI23}
\end{align}

 When $\varphi (x,b) = \varphi (x)$ (which corresponds 
 to  the leading  approximation of  keeping only the $l=0$ term 
in Eq. (\ref{ln2})), 
both formulas give the same $1/xQ^2$ result after integration over 
the impact parameter.

However, the two formulas have a different attitude with respect to 
 further  terms
of the $(b^2)^l$ expansion of a soft $\varphi (x,b)$.
Indeed,  
the integrals of $(b^2)^l$ with $K_0  (\sqrt{x}Q b)$ converge for all
powers $l$ (producing unwanted  $(1/Q^2)^{l+1}$  power corrections), 
while $J_1  (\sqrt{x}Q b)$ integrals diverge 
starting with $l=1$. Because of
oscillating nature of the $J_1$ Bessel function,
a  proper regularization  would set all these integrals to 
 zero (or derivatives of $\delta (xQ^2)$, to be more precise), 
 which corresponds to having no $1/Q^2$ corrections
under the $x$-integral.

Another evident difference between the two expressions 
is that  $J_1  (\sqrt{x}Q b)/b$ is finite  for $b=0$ while 
$K_0  (\sqrt{x}Q b)$  has a logarithmic singularity there.
Moreover, if $\Psi (x, k_\perp)$ is finite for $ k_\perp=0$, the 
exact  formula (\ref{Phixs06000}) gives a 
 finite value  for $T(Q^2)$  in the $Q^2\to 0$ limit, namely, 
  \begin{align}
T (Q^2\to 0) = 
 \pi  \int_{0}^1  \Psi (x, k_\perp =0) \, dx  \  . 
\label{eq:Fscalar3htQ0}
\end{align} 
On the other hand, Eq. (\ref{eq:FscalarSud1fin}) 
produces a logarithmically divergent result even if $\Psi (x, k_\perp=0)$
is finite.  We now see that this well-known deficiency of Eq. (\ref{eq:FscalarSud1fin}) 
(see, e.g., \cite{Musatov:1997pu}) 
is just the result of approximations (equivalent to taking $k_-=0$ in the hard 
propagator) used in its derivation.

 \subsubsection{Calculation in the  IMF variables}
 
 Another possibility is to write the $d^4k$ integral in 
  Eq. (\ref{eq:Fscalarmom}) using  a frame where $p$ defines the  
``plus''  direction, but 
the $q_+=0$. 
Defining the ``minus'' direction by a lightlike vector $n$, we have
\begin{align} 
  q' &= p+ Q^2 \, n - q_\bot  \  ,  \\ 
  k & = \xi p + \eta Q^2 n + k_\bot 
   \  ,
   \label{qprime2}
  \end{align}
  where $n^2=0$ and $2 (pn)=1$.
  Then $k^2= \xi \eta Q^2 -  k_\bot ^2$ and 
  \begin{align} 
  (q'-k)^2  & =  (1-\xi)(1- \eta) Q^2 -  (k_\bot + q_\bot )^2 
   \ . 
   \label{qprime2}
  \end{align}
 This gives 
     \begin{align}
T(q,p) &=   \int_{0}^1 dx   \int_0^\infty d \alpha  \, F(x, \alpha)  \int_0^\infty d \alpha_1  
   \nonumber \\ &  \times \frac{Q^2}{2} 
\int_{-\infty}^\infty  {d\xi \, d\eta } \int d^2 k_\perp 
\, e^{i\alpha (\xi -x)\eta Q^2} \,
   \nonumber \\ &  \times   e^{i\alpha_1 (\xi -1) (\eta -1)  Q^2} \, 
   e^{-i\alpha  k_\perp^2 - i\alpha_1(k_\perp +q_\perp)^2} \ .
\label{eq:FscalarIMF}
\end{align}
Performing integration over $\eta$, i.e.,  the minus component of $k$,  and 
changing $\alpha + \alpha_1 =\lambda$, $\alpha = \beta \lambda$, we obtain 
     \begin{align}
T(q,p) & =  \int d^2 k_\perp  \int_{0}^1 dx   \int_0^\infty d \lambda 
   \nonumber \\ &  \times  \, \int_0^1 {d \beta } 
F(x, \beta \lambda)  
\int_0^1  {d\xi } 
\, 
   \delta [\bar \xi  -\beta \bar x]    \nonumber \\ &  \times  
      e^{i \lambda   ( \bar \xi -\beta)  \bar \beta Q^2 -i \lambda ( k_\perp + \bar \beta  q_\perp)^2} 
    \  .
\label{eq:FscalarIMF3}
\end{align}
Thus, the IMF variable  $\bar \xi =\beta \bar x $ is smaller than the VDA variable
$\bar x$. i.e., $\xi_{\rm IMF}$ is larger than $x$.  
However, integrating over $\xi$   and shifting integration variable $k_\perp$  gives 
      \begin{align}
T(q,p) & =  \int d^2 k_\perp  \int_{0}^1 dx   \int_0^\infty d \lambda  \, F(x, \beta \lambda)  
   \nonumber \\ &  \times \int_0^1 d \beta \, 
   e^{-i \lambda  k_\perp ^2 -i  \lambda  
\beta \bar \beta xQ^2} \, 
    \  ,
\label{eq:FscalarIMFVf}
\end{align}
which coincides with the expression (\ref{eq:FscalarSud4add})  obtained using Sudakov variables, and further steps are the same.

\subsubsection{Summary} 

 Thus, our examination did not reveal any
  advantages of using explicit decomposition of the integration momentum $k$ 
  in either Sudakov or IMF variables. 
  To the contrary, their use results in the  expression 
   (\ref{eq:FscalarSud5add}) 
that is  much more complicated  than the  VDA formula 
 (\ref{eq:FscalarSud7add}).
Moreover, in the exact expression
(\ref{eq:FscalarSud1fin}) 
we did not reach  the goal of 
converting $d^4 k$ into $d\xi d^2 k_\perp$:
it   contains an extra   integration,
which we managed to get rid of only by making approximations.
Furthermore,  the approximate expression (\ref{eq:FscalarSud1fin}) 
is  not satisfactory since it  contains  $k_\perp$-dependent terms  in 
 hard factors that produce towers of $k_\perp^2/Q^2$ corrections.
 As a result, it is not OPE compatible for soft TMDAs.

  Note that trying to keep the Sudakov or IMF transverse momentum 
  variable, we have integrated over the plus-momentum variable $\xi$
  of these representations, thus using the VDA variable $x$ in further  expressions.
  As we have seen, $\xi$ differs from $x$ in both  cases
  (though in opposite directions: $\xi_{\rm Sud} \leq x$, while $\xi_{\rm IMF} \geq x$).
   If, instead,  we would try  to keep the $\xi$ variables by
   integrating over $x$, we would get much more lengthy expressions,
   with the $\beta$ integration variable now entering both arguments
   of TMDA, making further  simplifications virtually impossible.
   This is another argument  in favor of using the VDA-based  variables and
   formulas.


 \subsection{Three-body  contributions}
 \label{threebody}

Using VDA   we take into account the contributions 
of higher twist operators of $\phi \ldots (\partial^2)^n \phi (0)$ type.
By equations  of motion, like $  \partial^2 \phi  = g \chi \phi$
in the case of $g \phi^2 \chi$ interaction of  quarks $\phi$
with  gluons $\chi$, these operators may be converted into
multi-body operators like $\phi \ldots \chi^n(0) \phi (0)$.
A distinctive feature of these operators is that 
the gluon field $\chi (0)$ is taken at the same point 
as the quark field $\phi (0)$, so we still deal with 
an effectively bilocal operator. However,  
one may also wish to  include configurations with three (or  more)
partons participating in the short-distance subprocess,
which are described by multi-body 
operators with the gluon fields taken at locations
different from those of the photon vertices.

\begin{figure}[h]
\centerline{\includegraphics[width=2.2in]{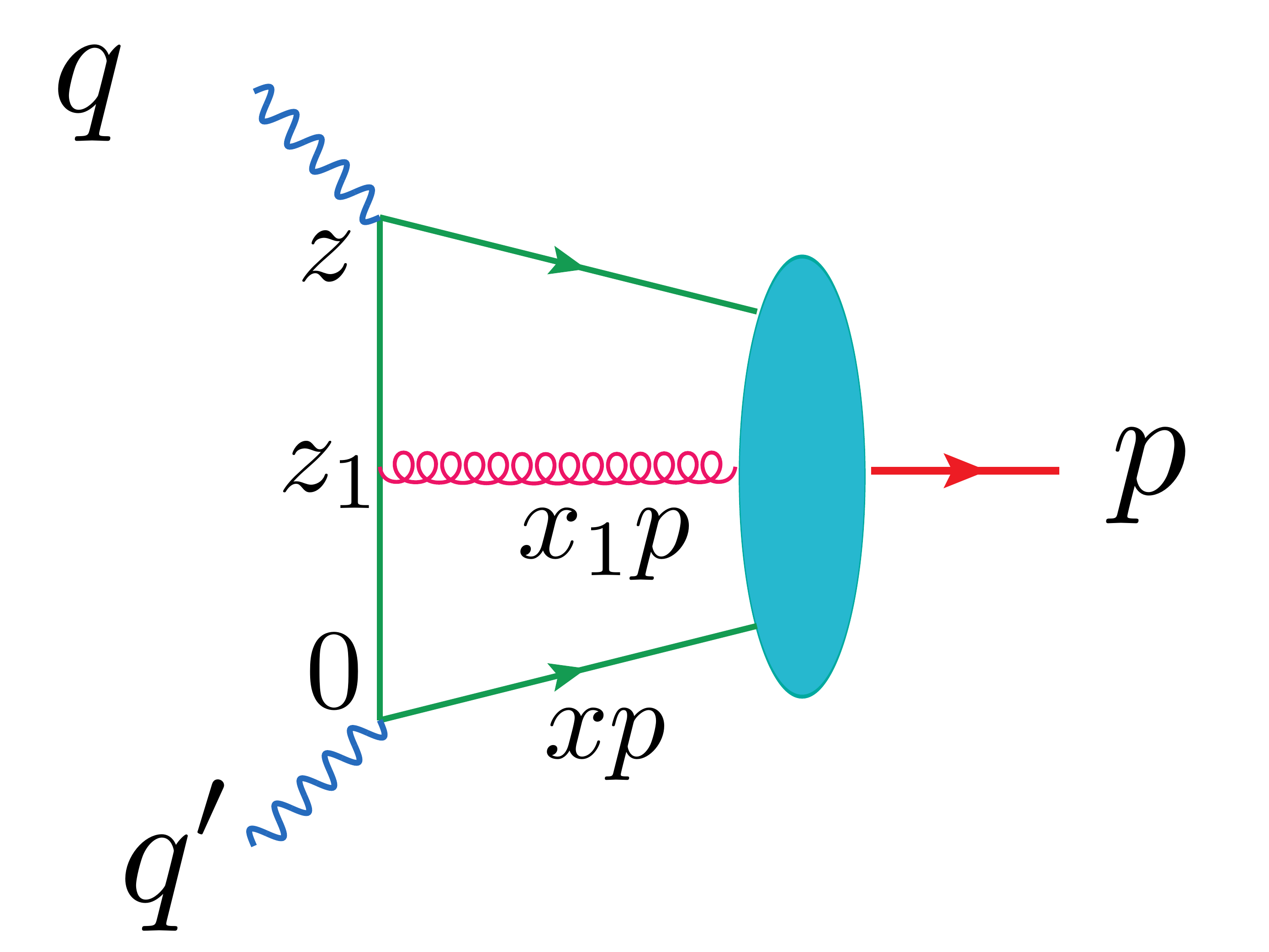}}
\caption{Scalar diagram  with a  gluon insertion.
\label{3body}}
\end{figure}

Take a contribution with one gluon insertion (see \mbox{Fig. \ref{3body}).} Then 
  \begin{align}
T_3(q,p) = &
\int {d^4z}\, e^{-i(qz)} \, \int d^4z_1 \,   D^c(z-z_1) \,  D^c(z_1)  
\nonumber \\ & \times 
\langle  p | \phi (z) \, \chi (z_1) \, \phi(0)  | 0 \rangle  \  .
\label{eq:Fscalar30}
\end{align}

The tri-local matrix element depends in general  on three
intervals: $z^2, z_1^2,(z-z_1)^2$ and two scalar products  $(pz), (pz_1)$.  
Neglecting virtuality-related dependence on  {\it all}  intervals,
we can parametrize
 \begin{align}
 \langle p |   \phi(z)  \, \chi (z_1) \, &  \phi (0)|0 \rangle 
= 
\int_0^1  \, dx \int_0^{\bar x}  dx_1 \,  f(x, x_1)\,  \nonumber \\ & \times 
 e^{i x_1 (pz_1) +i (\bar x-x_1) (pz)} 
 +  {\cal O} (z_i^2) \   .
\label{DDF3}
\end{align}
In this parametrization, the ``gluon''  has an (outgoing) momentum $x_1p$,
and ``quark''  at $0$ carries  momentum $x p$.
The quark  at $z$ has  momentum $(1-x-x_1)p$.
The spectral property $x\geq 0, x_1\geq 0, x+x_1 \leq 1$
(we will denote this region as $\Omega$)  
can be proven for any Feynman diagram contributing to $f(x,x_1)$,
see Refs. \cite{Radyushkin:1983wh,Radyushkin:1983ea}.

Taking $z_1=z$, we get an effectively  bilocal operator
$ \phi(0)  \, \chi (z) \,  \phi (z)$. 
Incorporating the equation of motion 
$\partial^2 \phi = g\chi \phi$, 
we get a reduction relation 
 \begin{align}
 g \int_0^{\bar x}   \,  f(x, x_1)\,  dx_1
 =\Lambda^2 \varphi_1 (x) 
\label{RedF3}
\end{align}
connecting 
$f(x, x_1)$ with the distribution amplitude
$\varphi_1 (x) $ corresponding to 
$\phi \partial^2 \phi$ operators
in Eq. (\ref{trC}). This connection just states 
that average parton virtuality  $\Lambda^2$ is proportional
to the average strength of the gluonic field $\chi$
inside the hadron.

Since these two contributions 
are governed by the same 
scale, one may wonder if we 
should consider them together.
However, there is an essential difference between the two. The virtuality correction
is ``invisible'' when taken on its  own,
and contributes to a nontrivial function
of $Q^2$ only after summation 
through VDA  with all 
other ``invisible'' contributions.
On the other hand, the diagram with a gluon insertion is an explicit power correction 
 to the handbag term. 
Its contribution is given by   the  parton formula
 \begin{align}
T_3 (q,p)   =&  \int_{\Omega}  \frac{f (x,x_1) dx \, dx_1}{[q'- xp]^2 
[q'-(x+x_1) p]^2 }   + {\cal O} (1/Q^6) \nonumber \\ & =\frac1{Q^4}
 \int_{\Omega}  \frac{f (x,x_1)  \, dx \, dx_1 }{x ( x+x_1)}
  + {\cal O} (1/Q^6)
\label{eq:3F3scalar}
\end{align}
from which it is evident that  the 3-parton term  is suppressed by $1/Q^2$ compared to the 
handbag diagram.  This  outcome is a consequnce of the fact 
that the 3-parton amplitude is less singular 
  \begin{align}
 \int d^4z_1  \,   D^c(z-z_1) \,   D^c(z_1)
\sim \ln (z^2)  
\label{eq:3Fscalarint}
\end{align}
on the light cone than the handbag diagram, which has $1/z^2$ singularity.

 One may also wish to  improve the precision and  include the virtuality  effects  by keeping $\chi (z_1)$
 in the above integral. Combining the denominators $1/z_1^2$ and $1/(z-z_1)^2$
 through Feynman parameter $u$ and shifting the integration variable
 $z_1 \to z_1 + uz$ , we arrive at (cf. Ref. \cite{Balitsky:1987bk}) 
    \begin{align}
 \int_0^1 du \int \frac{\chi (uz +z_1) \, d^4z_1 }{[z_1^2 + u (1-u) z^2]^2} \ .
\label{eq:3Fscalar3}
\end{align}
Note that $\chi (uz +z_1)$ is integrated over $z_1$ with a function that
depends on $z_1$ through $z_1^2$ only. Hence, if we expand $\chi(uz+z_1) $
around the  point  $uz$ using analog of Eq.(\ref{Taylor0}), all terms containing 
traceless combination  $\{z_1 \partial\}^N$   with $N\geq 1$ give zero
after integration over $z_1$, so we can use 
 \begin{align}
  \chi (uz+ & z_1) = \sum_{l=0}^{\infty} 
 \left ( \frac{z_1^2 }{4}  \right )^l 
 \frac{ ({\partial}^2)^l 
 \chi(uz) }{l!(l+1)!}  
 + {\rm traceless} \  \{z_1 \partial\}^N 
 \nonumber \\ &  \equiv 
 \int_{0}^{\infty} d \sigma_1 
\tilde  \chi (uz,\sigma_1) \,  \,  e^{-i \sigma_1 {(z_1^2-i \epsilon )}/{4}} \, 
+ \ldots \ 
 . 
 \label{3phiuz}
\end{align}
The $\sigma_1$-dependence of the field $\tilde  \chi (uz,\sigma_1)$ takes care 
of the effects due to the virtuality (off-shellness) of the scalar gluon  field $\chi$.
 Thus, 
 we end up  with the integral
  \begin{align}
&T_3(q,p) = 
 \int_0^1 du 
\int {d^4z}\, e^{-i(qz)} \, 
 \nonumber \\ &  \times 
  \int_{0}^{\infty} d \sigma_1 \,
  \langle  p | \phi (z) \,  \tilde \chi (uz,\sigma_1)\,  \phi(0)  | 0 \rangle  
\nonumber  
 \\ & \times 
 \int  \frac{d^4z_1}{[z_1^2 + u (1-u) z^2]^2}
  e^{-i \sigma_1 {(z_1^2-i \epsilon )}/{4}} 
 \ . 
\end{align} 
 The $d^4z_1$ integral
 can be calculated in terms of incomplete gamma function.
 The result depends on the combination
 $u\bar u z^2 \sigma_1$ and has a logarithmic 
dependence on it  for small $z^2$.  
One can in principle  keep this dependence, but  as the  first step
we may neglect   the  gluon virtuality effects given by 
 $ (z_1^2 \partial^2)^l \phi$  in Eq. (\ref{3phiuz}), 
and  look at  the virtuality effects due to non-zero 
value of  $z^2$.
Then one deals with 
\begin{align}
 T_3(q,p) =&   \int_0^1 du 
\int {d^4z}\, e^{-i(qz)} \, \ln (u \bar u z^2 M^2)  \, \nonumber \\ &  \times
\langle  p | \phi (z) \, \chi (uz) \, \phi(0)  | 0 \rangle  \,
\  .  \label{eq:Fscalar3u}
\end{align}

One can see that all  the  fields involved  in the  matrix element here 
are located on the straight line connecting $0$ and $z$,
i.e. we deal with a {\it ``string'' operator}  \cite{Balitsky:1987bk}. 
With respect to $z$,  the matrix element is a function of $(pz)$ and $z^2$
(note   that it has been 
never assumed that $z$ is on the light cone, so we can take $z^2 \neq 0$).
Then we can represent  
this matrix element  using a VDA-type parametrization like 
\begin{align}
  \langle p |   \phi(z) &  \,   \tilde \chi (uz) \,  \phi (0)|0 \rangle 
= \int_{0}^{\infty} d \sigma   
 \,  
 \int_{\Omega}  \, dx\, dx_1 \,  \Phi (x, x_1, \sigma)\, 
  \nonumber \\ & \times  e^{i x_1u (pz) +i(\bar x- x_1) (pz)-i \sigma (z^2-i \epsilon )/4} \,  . \label{3Phixs0}
\end{align} 
In the present paper,  we will not proceed further 
with the analysis of the 3-body and higher Fock components,
leaving it to future investigations.
However, some elements  of the technique used  above
are helpful in the analysis of gauge theories.

\setcounter{equation}{0}   \section{QCD case}
\label{gaugeT}

The justification for spending time on  scalar   models 
is that the same construction may be built in QCD.

\subsection{Spin-1/2 quarks}

\label{spin1/2}

A realistic case is when quarks have spin 1/2.  Then the 
handbag  diagram for the pion transition form factor 
 is  given by 
 \begin{align}
\int {d^4z}\, e^{-i(q'z)} 
  \langle p |   \bar \psi(0)  & \gamma^\nu \, S^c (-z) \, \gamma^\mu \,  \psi (z)|0 \rangle 
  \nonumber \\ &
=i \epsilon^{\mu \nu \alpha \beta} p_\alpha q_\beta F(Q^2) \ , 
\label{fermiC}
\end{align}
where $S^c (z) = \slashed z/ 2\pi^2 (z^2)^2$ 
is the  propagator for a massless fermion.
Writing the antisymmetric   part of $\gamma^\nu \, \slashed z \, \gamma^\mu $
as $i  z_\beta \epsilon^{\mu \nu \alpha \beta} \gamma_5 \gamma_\alpha $
we 
parametrize  the resulting matrix element as 
\begin{align}
  \langle &p |    \bar \psi(0)   \,\gamma_5 \gamma_\alpha   \,  \psi (z)|0 \rangle  
 = i  
\int_{0}^{\infty} d \sigma \int_{0}^1 dx\,  \nonumber \\ & \times 
 [ p_\alpha \Phi (x,\sigma) +z_\alpha \Phi_2 (x,\sigma)] \,  \,  e^{i x (pz) -i \sigma {(z^2-i \epsilon )}/{4}}  \  .  \label{Phixspin}
\end{align} 
Here 
we have two VDAs that are associated with the two
possibilities $p_\alpha$ and $z_\alpha$ for a vector carrying
the Lorentz index $\alpha$.
 In fact,  the  term proportional
to $z_\alpha$ does  not contribute to the  $\gamma^* \gamma \to \pi^0$ amplitude
at the handbag level, since it  disappears 
after 
convolution with  $z_\beta\epsilon^{\mu \nu \alpha \beta} $. 

All  the formulas relating    VDA with TMDA and IDA that
 were   derived in the  scalar case are valid   without changes.
However, the spinor propagator has a different functional form.
Using  it, we obtain 
 \begin{align}
F(Q^2) = &
\int_{0}^{\infty}   d \sigma  \int_{0}^1 { \Phi (x,\sigma)    } \frac{dx}{xQ^2} \, 
   \nonumber \\ &\times  
\left \{ 1+ \frac{i\sigma}{xQ^2} \left [ 1- e^{-[ixQ^2 + \epsilon]/  \sigma } \right ] \right \} \  . 
\label{eq:Fspinor3}
\end{align} 
The first two terms are given 
by two lowest moments of VDA $ \Phi (x,\sigma)$.
Assuming that they exist,  we may write  
 \begin{align}
F(Q^2) = 
&  \int_{0}^1 {    } \frac{dx}{xQ^2} \, 
 \, 
\left \{ \varphi (x) -  \varphi_1 (x) \frac{\Lambda^2}{xQ^2}  
\right.  \nonumber \\ & \left. 
 - \int_{0}^{\infty}   d \sigma  \, 
 \Phi (x,\sigma)\frac{i\sigma}{xQ^2}  e^{-[ixQ^2 + \epsilon]/ \sigma } \right \} \  . 
\label{eq:Fspinor33}
\end{align} 

For large $Q^2$,  Eq.  (\ref{eq:Fspinor33})  shows a  correction 
with a power-like  $\Lambda^2/Q^4$   behavior that 
corresponds to the \mbox{twist-4}  $\bar \psi \gamma_5 \gamma_\alpha 
D^2 \psi $ operator. Though it is accompanied by 
a $z^2$ factor,  the latter does not completely 
cancel the $1/z^4$ singularity
of the  spinor propagator,    and 
as a result this contribution has a ``visible'' $1/Q^4$ behavior.
The remaining term corresponds to 
contributions ``invisible'' in the OPE.

Note  that, taken separately,
the \mbox{twist-4}  contribution results in a very singular 
$\varphi_1 (x)/x^2$ integral.  If the pion ``daughter''  DA $\varphi_1 (x)$ 
does not vanish like $x^{1+\alpha}$
with a positive $\alpha$ in the end-point region,
the purely twist-4 contribution diverges.
However, it is easy to check that when  the ``invisible'' 
 terms are added,  the $x$-integral in 
 the original Eq.
 (\ref{eq:Fspinor3})
 converges. 
Indeed,  writing it  in terms of TMDA 
  \begin{align}
 F(Q^2) 
= &  
   \int_{0}^1   \frac{dx}{xQ^2}   \int_{0}^{xQ^2}  \frac{d k_\perp^2}{ xQ^2} 
    \int_{{k'}_\perp^2 \leq  k_\perp^2}   \Psi (x, { k'}_\perp   ) 
 \, d^2  k'_\perp
      \  , 
 \label{Fspinor40}
 \end{align}
  and then in terms of IDA 
    \begin{align}
F(Q^2)
   {=}  2  \int_{0}^1 \frac{dx}{xQ^2}
 \int_0^\infty  \frac{ db}{ b  }\,
   \,J_2  (\sqrt{x}Q b) \, \varphi (x,b)\, ,
\label{eq:Fspinor123}
\end{align}
 we see that  $J_2  (\sqrt{x}Q b)/x \to {\rm const}$   for small $x$,  
 so that $x$-integral converges  if $\varphi (x,b) \sim x^{-1+\alpha} $
 with  however small positive 
 $\alpha$, i.e.  $\varphi (x,b)$ may be even singular for 
 small $x$.  Taking the  $Q^2\to 0$ limit under the integral, we get
   \begin{align}
F(Q^2=0)
  {=}  \frac14   \int_{0}^1 {dx}
 \int_0^\infty  b \, { db}  \, 
  \varphi (x,b)\, ,
\label{eq:Fspinor123b}
\end{align}
 assuming that the $b$-integral is finite. This is the case if 
$\Psi (x, k_\perp)$ is finite for $ k_\perp=0$. Then we have  
  \begin{align}
F (Q^2\to 0) = 
 \frac{\pi }{2}   \int_{0}^1  \Psi (x, k_\perp =0) \, dx  \  . 
\label{eq:Fspinor3htQ0}
\end{align} 
This result also follows  directly from Eq.  (\ref{Fspinor40}) if
 $\Psi (x, k_\perp=0)$ is finite.
 It also coincides with the IMF  light-front approach result  of 
  Ref.\cite{Lepage:1980fj}. 

In fact, Eq. (\ref{Fspinor40})  can  be re-written in the form involving just one 
transverse momentum integration 
   \begin{align}
 F(Q^2) 
= &  
   \int_{0}^1   \frac{dx}{xQ^2}  
       \int_{{k_\perp}^2 \leq  xQ^2}   \Psi (x,  k_\perp   )  \left [ 1-  \frac{ k_\perp^2}{ xQ^2} \right ]
 \, d^2  k_\perp
 \label{Fspinor40a}
 \end{align}
and  explicitly showing  the twist-4 correction  given by the $k_\perp^2$ moment  of TMDA 
 $ \Psi (x,  k_\perp   )$.

\subsection{Adding a gluon }

\begin{figure}[h]
   \centerline{\includegraphics[width=2.5in]{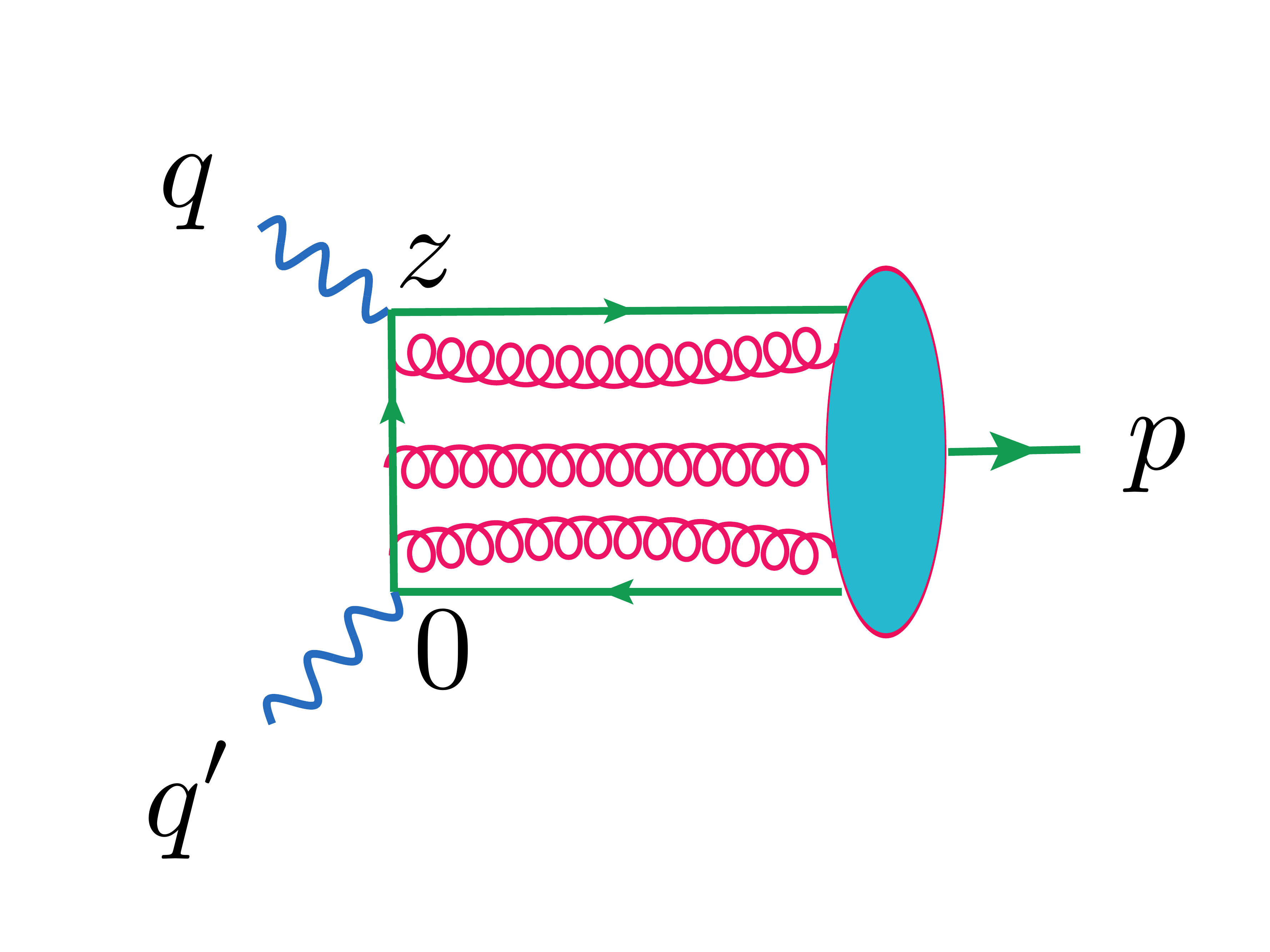}}
        \vspace{-0.5cm}
   \caption{Structure of the  handbag contribution in QCD.
   \label{QCD}}
   \end{figure}

In gauge theories, the handbag  contribution in 
a covariant gauge should be complemented 
by   diagrams corresponding to operators 
$\bar \psi (z)  \ldots \slashed A(z_i) \ldots  \psi (0)$
containing twist-0 gluonic field $A_{\mu_i}  (z_i)$ inserted 
into the fermion line between the points $z$ and $0$
(see Fig.\,\ref{QCD}).

Consider an insertion of the gluon field $A^\alpha (z_1)$ into the quark propagator
connecting $z$ and $0$ vertices. 
Taking the gluon with momentum $k$ we have 
 \begin{align} 
 \int     d^4z_1   \, S^c(z-z_1)  \gamma^\alpha S^c(z_1) e^{i(kz_1)}   
    \label{loop1}
 \end{align}
or
 \begin{align} 
 \int     d^4z_1 &   \, \int_0^\infty \sigma_1 \, d\sigma_1 
  \int_0^\infty \sigma_2 \, d \sigma_2  (\slashed z - \slashed z_1)  \gamma^\alpha 
  \slashed z_1 
    \nonumber \\ &  \times e^{-i \sigma_1 (z-z_1)^2/4-i  \sigma_2 z_1^2/4}
 e^{i(kz_1)}   
    \label{loop2}
 \end{align}
Integrating over  $z_1$ and changing $\sigma_1+\sigma_2 = \sigma$,
$\sigma_1 = t \sigma$ gives
  \begin{align} 
& \int_0^\infty\sigma {d\sigma }  \int_0^1 d t\, t \bar t 
    \exp \left \{  -i  \, t \bar t  \frac{z^2}{4} \sigma+
    i t (kz) 
  + i \frac{k^2}{\sigma}  \right \} 
  \nonumber \\ &  \times  
 \left [  \left  (\bar t \slashed z - 2 \slashed k / \sigma
  \right )  \gamma^\alpha  
   (t   \slashed z +2 \slashed k / \sigma) -4i\gamma^\alpha/{ \sigma}   \right ] \  .
    \label{loop31}
 \end{align}
  Changing further $\sigma \to \sigma/ t \bar t$  produces the integral
   \begin{align} 
 \int_0^\infty\sigma {d\sigma }  \int_0^1 d t &
    \exp \left \{ -i \frac{z^2}{4} \sigma 
    +
    i t (kz) +i t \bar t \frac{k^2 }{ \sigma}
    \right \}   \nonumber \\ &  \times  
 \left [   {\rm Trace}  \right ] \ , 
    \label{loop6n}
 \end{align}
 with the 
  trace given by 
   \begin{align} 
&  
 \left [   {\rm Trace}  \right ] = 
   2 z^\alpha \slashed z    - \gamma^\alpha \,( z^2 + {4i}\sigma ) \nonumber \\ &  +
   2  k_\beta [\bar t \slashed z \gamma^\alpha \gamma^\beta -t \gamma^\beta \gamma^\alpha \slashed z  ]/\sigma
    -4 t \bar t \slashed k   \gamma^\alpha \slashed k /\sigma^2   \  .
    \label{loop7n}
 \end{align}
 Writing 
    \begin{align} 
e^{ i t \bar t k^2 / \sigma} = 1+ \left [ e^{ i t \bar t k^2 / \sigma} -1 \right ] \  ,
   \label{loop8}
 \end{align}
 we can integrate over $\sigma$ the term corresponding to  ``1''.
 In particular, 
   \begin{align} 
 \int_0^\infty\sigma {d\sigma } & 
    \exp \left \{ -i \frac{z^2}{4} \sigma 
    \right \}   
   =- \frac{16}{(z^2)^2}  \  ,
    \label{loop9}
 \end{align}
 while 
    \begin{align} 
 \int_0^\infty\sigma {d\sigma } & 
    \exp \left \{ -i \frac{z^2}{4} \sigma 
    \right \}   
 \left [      z^2 + \frac{4i}{\sigma}  \right ] = 
  0   
   \label{loop9a}
 \end{align}
 and 
   \begin{align} 
 \int_0^\infty  {d\sigma } & 
    \exp \left \{ -i \frac{z^2}{4} \sigma 
    \right \}   
 =- \frac{4i}{z^2}    \ .
    \label{loop10}
 \end{align}
 Thus, the leading  singularity $1/(z^2)^2$ is accompanied by the same 
 gamma-matrix  factor $\slashed z$ as in the original quark propagator.
 The factor $z^\alpha$ means that the field $A^\alpha$ appears  as $(zA)$,
 while the exponential $e^{it (kz)}$ shows that  the field is taken 
 at the running argument $tz$, with integral over $t$ from $0$ to $1$.
 This means that  it  corresponds to the linear in $A$ part
    \begin{align} 
ig  \int_0^1 dt \, z_\alpha A^\alpha (tz)     \label{loop10n}
 \end{align}
 of the  gauge link 
  \begin{align}
{ E}(0,z; A) \equiv P \exp{ \left [ ig \,   \int_0^1dt \,  z_\alpha\, A^\alpha (t z) 
 \right ] }  \  .
 \label{straightE}
\end{align}
 (Our derivation given above is inspired by that given in Ref. \cite{Balitsky:1987bk}). 
 
 Furthermore,  terms corresponding to the sub-leading singularity $1/z^2$  are 
 proportional to $k$, i.e. the derivative of the gluon field, which in fact combines  into 
 the field-strength tensor $F_{\alpha \beta}$.
  Indeed,  if one takes 
 the $\alpha \leftrightarrow  \beta$  symmetric 
 part of  the $\gamma$-matrix  terms in $ k_\beta  \gamma^\alpha \gamma^\beta $ and 
  $k_\beta   \gamma^\beta \gamma^\alpha$, it 
  gives $k^\alpha$ which results in a vanishing contribution since $(kA)=0$,
  and only terms proportional to   $k_\beta \sigma^{\alpha \beta}$,  
 i.e.,
  $k_\beta A_\alpha-k_\alpha A_\beta  $
  remain, which corresponds to the field-strength tensor $F_{\alpha \beta}$.
 
 One can also take the $\sigma$ integral in its  original form:
   \begin{align} 
 \int_0^\infty   \sigma {d\sigma }  &
    \exp \left \{ -i \frac{z^2}{4} \sigma + i t \bar t \frac{k^2 }{ \sigma}
    \right \}     \nonumber \\ &
   =- \frac{8 t \bar t k^2}{z^2}  K_2 (\sqrt{ t \bar t k^2 z^2}) 
    \nonumber \\ & 
  = - \frac{16}{(z^2)^2}  
   + \frac{4 t \bar t k^2}{z^2} +\ldots \ , 
    \label{loop9o}
 \end{align}
 while 
    \begin{align} 
 \int_0^\infty\sigma {d\sigma }  &
    \exp \left \{ -i \frac{z^2}{4} \sigma + i t \bar t \frac{k^2 }{ \sigma}
    \right \}   
 \left [   z^2 + \frac{4i}{\sigma} \right ] 
 \nonumber \\ & = - 8 t \bar t k^2 K_0 (\sqrt{ t \bar t k^2 z^2})  
  \nonumber \\ & =  
 {\cal O} (k^2 \ln (k^2 z^2)) \ , 
      \label{loop9a0}
 \end{align}
 and 
   \begin{align} 
 \int_0^\infty  {d\sigma } &
    \exp \left \{ -i \frac{z^2}{4} \sigma + i t \bar t \frac{k^2 }{ \sigma}
    \right \}    \nonumber \\ & =- 4 i \sqrt{\frac{ t \bar t k^2}{z^2}}
     K_1 (\sqrt{ t \bar t k^2 z^2})    \nonumber \\ & 
 =- \frac{4i}{z^2}    +  {\cal O} (k^2 \ln (k^2 z^2))  \ , 
    \label{loop10}
 \end{align}
 which produces the results discussed above, 
 provided that one neglects ${\cal O} (k^2/z^2)$ and 
 $ {\cal O} (k^2 \ln (k^2 z^2))$  terms.
  
  \subsection{Quark propagator in external gluon field} 

To see  how the ${ E}$-factor  (\ref{straightE}) emerges
to all orders in the external field $gA$ \cite{Efremov:1978fi,Efremov:1978xm}, we 
observe  that 
the sum of gluon  insertions is equivalent
to substituting the free propagator $S^c (z_1-z_2)$ by
a propagator ${\cal S}^c (z_1, z_2;  A)$ of a  fermion 
in an external gluonic  field $A$. 
  This propagator 
 satisfies the Dirac equation 
\begin{align}
i \left [ \frac{\slashed \partial}{\partial z_1 } - ig \slashed A (z_1)  \right ]{\cal S}^c (z_1,z_2;  A) = - \delta^4 (z_1-z_2)  \ .
\end{align}
Looking  for a solution of this equation in the form
\begin{align}
{\cal S}^c (z_1,z_2;  A) =  E (z_1,z_2; A)  {\mathfrak S}^c_{\rm FS }
(z_1,z_2;A) 
\label{FSProp}
\end{align}
involving the straight-line exponential (\ref{straightE}),
one can see that  the factor $ {\mathfrak S}^c_{\rm FS }$ should satisfy the Dirac 
equation 
\begin{align}
i \left [ \frac{\slashed \partial}{\partial z_1} - ig \slashed {\mathfrak A} (z_1)  \right ]{\mathfrak  S}^c 
_{\rm FS }(z_1, z_2;  A) = - \delta^4 (z_1-z_2)  \ ,
\label{Dirac}
\end{align}
with the field  \cite{Efremov:1978fi,Efremov:1978xm} 
\begin{align}
 {\mathfrak A}^\mu  (z; z_1) = (z-z_1)_\nu 
 \int_0^1 \,  s \,  G^{\mu \nu} (z_1+s (z-z_1)) \, ds
 \label{FSA}
\end{align}
being  the 
vector potential in the Fock-Schwinger (FS) 
gauge  \cite{Fock:1937aa,Schwinger:1951nm},
\begin{align}
\label{FS}
(z-z_1)_\mu  {\mathfrak A}^\mu  (z,z_1) = 0 \ .
\end{align}
Here,  $z$  denotes   an arbitrary 
position in space while  $z_1$
specifies the ``fixed point'' of the 
gauge and in our case 
 refers  to  an end-point in the Compton amplitude.

Since the field-strength tensor  $G^{\mu \nu} $
has twist equal to (at least)  1, the insertion 
of this field into the free propagator results 
in power $(\Lambda^2Q^2)^l$ corrections 
to the Compton amplitude. Thus, we can write
\begin{align}
{\cal S}^c (0,z;  A) =  E (0,z; A) \left \{ { S}_c (z) 
+ {\cal O} (G) \right \} \  . 
\label{extF}
\end{align}

  \subsubsection{Bilocal operator in gauge theories}
  
Keeping the first term in Eq. (\ref{extF}) we need to deal with 
 the gauge-invariant bilocal operator
\begin{align}
{\cal O}^\alpha  (0,z; A) \equiv \bar \psi (0) \,
 \gamma_5 \gamma^\alpha \,  { E} (0,z; A) \psi (z)  \  .
\end{align}
Its  important property  is that the   Taylor expansion for  
 ${\cal O}^\alpha  (0,z; A) $   has 
the same structure 
as that for the original $\bar \psi (0)  \gamma_5 \gamma^\alpha \psi (z)$ 
operator, with the only change that
one should use covariant derivatives 
\mbox{$D^\mu =\partial^\mu  - ig A^\mu$}  instead of the
ordinary  $\partial^\mu $ ones:
\begin{align}
  { E} (0,z; A)\,  \psi (z) = \sum_{n=0}^{\infty} 
 \frac{1}{n!} (z D)^n  \psi (0) \ . 
\label{Taylorpsi}
\end{align}
This result follows  from the relation 
  \begin{align}
 \frac{d}{d z_\alpha} E(0,z;A) 
 =  { E} (0,z; A) [D^\alpha  + ig {\mathfrak A}^\alpha  (z,0)]     
 \label{Edz}
\end{align}
and the property $z_\alpha {\mathfrak A}^\alpha  (z,0)=0$ of the Fock-Schwinger 
field ${\mathfrak A}^\alpha  (z,0)$. 
As we have seen in Sect. \ref{VDA0}, the Taylor expansion 
for a matrix element may be written in the form of 
the  VDA parametrization 
\begin{align}
   \langle p |  \, & {\cal O}^\alpha  (0,z; A)  |0 \rangle  
  = i 
\int_{0}^{\infty} d \sigma \int_{0}^1 dx\,  
\bigl [ p_\alpha \Phi (x,\sigma) \nonumber \\ & 
+z_\alpha \Phi_z (x,\sigma) \bigr ] \, 
 \,  e^{i  x (pz) -i \sigma {(z^2-i \epsilon )}/{4}} \ . 
 \label{OPhixspin0}
\end{align} 

Again,  since 
the matrix element $\langle p |  \,  {\cal O}^\alpha  (0,z; A)  |0 \rangle $
is just a function of $(pz)$ and $z^2$, 
one can treat  the  VDA representation 
as a particular  case of  the double  Fourier
representation with specific constraints on the limits of 
$x$ and $\sigma$ integration.
These limits, in turn,  reflect only 
the positivity of the functions 
$A, B, C, D$  in the exponential of the $\alpha$-representation
(\ref{alphap}),  which are completely determined 
by the denominators of the momentum space 
propagators of the relevant Feynman  diagram
and are the same in any theory.  
This observation justifies the  use  of 
the VDA representation (\ref{OPhixspin0}) 
in QCD  in general case.

  \subsubsection{Multilocal operators }

Insertions of the nonzero-twist  FS 
field ${\mathfrak A}^\mu$  
 result in matrix elements  of $\bar \psi (0) \ldots G(sz) \ldots \psi (z)$
operators, which should be parametrized in terms of  trilocal, etc. VDAs.
These terms are analogous to  
$\phi (0) \ldots \chi (z_1) \ldots \phi (z)$, etc. operators of 
the scalar model. 
The technology of  how to work with insertions
of the Fock-Schwinger field ${\mathfrak A}^\mu$ 
is well-developed,
see e.g. Refs.  \cite{Nikolaev:1981ff,Nikolaev:1982rq,Smilga:1982wx,Balitsky:1987bk}. 
 For the Compton amplitude, the contribution due  to a single  
insertion 
of ${\mathfrak A}^\mu$  was  calculated by Balitsky and Braun 
 \cite{Balitsky:1987bk} and shown to produce a $1/Q^2$ correction to the 
 leading term.

\setcounter{equation}{0}   \section{Models of  soft VDAs}
\label{TMDA_Models}

\subsection{ Explicit models of  soft transverse momentum dependence }

Let us now  discuss  some explicit  models  of the 
$k_\perp$ dependence of  soft TMDAs $\Psi (x, k_\perp)$.
In general, they are   functions
of two independent variables $x$ and $k_\perp^2$.
But it makes sense to start with  a  simple case of     factorized models
  \begin{align}
\Psi (x, k_\perp) = \varphi (x) \, \psi (k_\perp^2)  \  ,
\end{align} 
in which   $x$-dependence and  $k_\perp$-dependence appear in  separate factors.
Since relations between VDAs, TMDAs and IDAs are the same in 
scalar and  spinor cases, we will refer  for simplicity to scalar operators.

\subsubsection{Gaussian model}

It is  popular to assume  a Gaussian dependence on
$k_\perp$, 
  \begin{align}
\Psi_G (x, k_\perp) = \frac{\varphi (x)}{\pi \Lambda^2}  e^{-k_\perp^2/\Lambda^2} \ . 
\label{Gaussian}
\end{align} 
In the impact parameter space, one gets IDA
  \begin{align}
\varphi_G (x, z_\perp) ={\varphi (x)} e^{-z_\perp^2 \Lambda^2/4} 
\label{GaussianI}
\end{align} 
that also  has a Gaussian dependence on $z_\perp$.
Writing 
\begin{align}
 \varphi_G (x, z_\perp)=&\frac{ {\varphi (x)} }{2\pi  }
\int_{-\infty}^{\infty} \frac{ i  \, d \sigma }{\sigma +i \Lambda^2} \, 
 \,  
 e^{- i z_\perp ^2 \sigma/4} 
 \  ,  \label{gausssigma} 
\end{align} 
we see that the integral here involves both positive and
negative $\sigma$, i.e. formally  $\varphi_G (x, z_\perp)$ 
cannot be written in the VDA representation 
(\ref{IDA}).  This is a consequence 
of the fact that $\psi_G (x, \kappa_\perp^2)$,
the analytic continuation of $\Psi_G (x, k_\perp)$
into the time-like region of momenta,
has an exponential  increase for large negative  $\kappa^2$. 

However, the transition form factor 
for  space-like virtual photons involves only the integral over 
positive $\kappa^2$, i.e. it is not 
sensitive to the  behavior of   $\psi_G (x, \kappa_\perp^2)$ 
for negative $\kappa^2$. One can see that   the 
form factor formula in terms of TMDA 
(\ref{Fspinor40})  shows no peculiarities in case of the Gaussian ansatz.
For this reason, 
 we will  use  this model  because of its calculational simplicity.

\subsubsection{Simple non-Gaussian models} 

 One may still  argue that a Gaussian fall-off for large $z_\perp$ 
is too fast.  In particular, 
 propagators $D^c (z,m)$ of massive particles have an exponential 
 $e^{-m|z|}$ decrease for spacelike intervals $z^2$.

To build models for TMDAs that  are more  closely resembling 
perturbative propagators, we 
 recall that the propagator of a scalar particle 
with mass $m$ may be written as 
\begin{align} 
D^c(z,m) = &  \frac1{(4 \pi)^2}  \int_0^{\infty} e^{-i \sigma z^2/4 - i  (m^2 - i\epsilon)/\sigma  }  
 {d \sigma}  \  .
\label{alpharD}
\end{align}
The mass term assures that the propagator falls off exponentially $\sim e^{-|z| m}$
for large  space-like 
distances. 
At small intervals $z^2$, however,  the free particle propagator 
has $1/z^2$ singularity while we want the soft part of 
$\langle p| \phi (0) \phi (z) |0 \rangle$ 
to be finite at $z=0$. The simplest way is to add a constant term $(-4/\Lambda^2)$ to $z^2$ 
in the VDA representation  (\ref{Phixs0}). So, we take


\begin{align} 
\Phi (x, \sigma) = \frac{\varphi (x)}{ p(\Lambda,m)} e^{i \sigma /\Lambda^2  - i  m^2 /\sigma -\epsilon \sigma }  
\label{alpharD}
\end{align}
as a model for VDA. 
The sign of the $\Lambda^2$ term is fixed 
from the requirement that $(4/\Lambda^2-z^2)^{-1}$ should not 
have singularities for space-like $z^2$.  The normalization factor  $p(\Lambda,m)$
is given by
\begin{align} 
{ p(\Lambda,m)} =&  \int_0^\infty e^{i \sigma /\Lambda^2  - i  m^2 /\sigma   -\epsilon \sigma }  d \sigma \nonumber \\ & =
2i {m}{\Lambda} K_1 (2 m/ \Lambda)  \ .
\label{alpharD}
\end{align}

\paragraph{$m=0$ model.} 

To begin with, let us take  $m=0$, i.e.
\begin{align} 
\Phi (x, \sigma) = {\varphi (x)} \,  \frac{e^{i \sigma /\Lambda^2  
  -\epsilon \sigma} }{ i \Lambda^2 } \ .
\label{alpharD0}
\end{align}

The bilocal matrix element in this case is given by
\begin{align}
& \langle p |   \phi(0) \phi (z)|0 \rangle 
=  
\, \frac{1}{1-z^2 \Lambda^2 /4}
\int_{0}^1 dx\, {\varphi (x)} \,  e^{i x (pz)}  \,  ,
\label{Phixs}
\end{align} 
which  corresponds to 
\begin{align}
{\varphi } (x, z_\perp) =
\, \frac{\varphi (x)}{1+ z_\perp^2   \Lambda^2/4 }  
\label{impaxs}
\end{align} 
for  IDA.  
One can see that the $z_\perp^2$ term of the 
$z_\perp$ expansion  of  ${\varphi } (x, z_\perp)$
in this model was adjusted to coincide with that of the exponential model,
so that    $\Lambda^2$ has the same 
meaning of  the  scale of $\phi \partial^2 \phi$  operator.
 The TMDA for this  Ansatz is given by
\begin{align}
& {\Psi} (x, k_\perp ) = 2 \varphi (x)\,  \frac{K_0 ( 2 k_\perp / \Lambda) }{ \pi  \Lambda^2 } 
\  .
\end{align}
It has a logarithmic singularity for small $k_\perp$
that reflects a slow $\sim 1/z_\perp^2$ fall-off  of  ${\varphi} (x, z_\perp)$ 
for large  $z_\perp$. 
The integrated TMDA that enters the form factor formula
(\ref{Fspinor40})  is given by 
  \begin{align}
    \int_{{k'}_\perp^2 \leq  k_\perp^2}   \Psi (x, { k'}_\perp   ) 
 \, d^2  k'_\perp = 
 \varphi (x)  
  \left [ 1- \frac{2k_\perp }{ \Lambda}  K_1 ( 2 k_\perp / \Lambda) \right ]
      \  . 
 \label{Fspinor4}
 \end{align}
It is also possible to   calculate explicitly the next $k_\perp$ integral 
 involved there, see Eq. (\ref{eq:Fscalar3htmod12}) below.
  
  For negative $k_\perp^2=-\kappa^2 - i \epsilon$, the model gives
  \begin{align}
 {\psi} (x, & k_\perp^2=-\kappa^2 - i \epsilon ) =  \frac{ 2 \varphi (x)}{  \Lambda^2 }  
 {K_0 ( -2i  \kappa / \Lambda) }  \nonumber \\ & = \pi
  \frac{ \varphi (x)}{ \Lambda^2 } 
  \left [ -N_0 ( 2\kappa / \Lambda)+iJ_0 ( 2\kappa / \Lambda) \right  ]
\  ,
\end{align}
i.e.  the analytic continuations of TMDA into the timelike region
in this  case generates an imaginary part. 

\paragraph{$m\neq 0$ model.} 

The model with nonzero mass-like term 
\begin{align} 
\Phi_m (x, \sigma) = \varphi (x) \frac{e^{i \sigma /\Lambda^2  - i  m^2 /\sigma  } }{ 2i {m}{\Lambda} K_1 ( 2 m/ \Lambda)  }  
\label{alpharD2}
\end{align}
corresponds to the  function 
\begin{align}
& \Psi_m(x, k_\perp ) = \varphi (x)\,  \frac{K_0 \left (2 \sqrt{ k_\perp^2 +m^2 } / \Lambda \right ) }{ \pi  m \Lambda
 K_1 (2 m/ \Lambda) }  \  .
 \label{psim}
\end{align}
 that  is finite for $k_\perp =0$ in accordance with the  fact that
the impact parameter distribution amplitude in this case,  
\begin{align}
{\varphi}_m (x, z_\perp)
=  {\varphi (x)}   \,  \frac{ K_1 \left ( m\sqrt{4/ \Lambda^2 + z_\perp^2} \, \right ) }{
 K_1 (2 m/ \Lambda) \, \sqrt{1 +\Lambda^2   z_\perp^2/4 }  }    \ , 
\end{align} 
 has an exponential $\sim e^{-m |z_\perp|}$
fall-off for  large $z_\perp$.

 \subsection{Modeling transition form factor by soft term}
\label{FF_Models} 

Let us now use these models to calculate  the modification
of the 
contribution to the transition form factor modified by 
higher twist terms absorbed  into  a 2-body TMDA $\Psi (x,k_\perp)$.

\subsubsection{Gaussian model} 

For the Gaussian model (\ref{Gaussian}), we have
 \begin{align}
F_G (Q^2) = &
 \int_{0}^1  \frac{dx}{xQ^2} \, 
{\varphi (x)} \, 
\int_{0}^{xQ^2}  \frac{d k_\perp^2}{ xQ^2} 
 \left [ 1-  e^{-k_\perp^2/\Lambda^2} \right ] \  , 
\label{FGauss}
\end{align} 
which gives 
 \begin{align}
F_G (Q^2) = &
 \int_{0}^1 \frac{dx}{xQ^2}   \, 
{\varphi (x)} \, 
 \left [ 1-  \frac{\Lambda^2}{xQ^2}  \left (1 -   e^{-xQ^2/\Lambda^2} \right ) \right ]  \  .
\label{FGauss2}
\end{align} 
If the pion DA $\varphi$ vanishes 
as any positive power $x^\alpha$ for \mbox{ $x \to 0$,}
the $x$-integral for the purely twist-2 contribution
converges. For large $Q^2$,  Eq. (\ref{FGauss2}) 
 displays  the power-like twist-4 contribution and the 
 term that  corresponds  to 
contributions of ``invisible''   operators with twist 6 
and higher. 
As we discussed,  inclusion of 
virtuality corrections improves the convergence 
in the small-$x$ region. In particular,   we obtain 
 \begin{align}
F_G (Q^2=  0) 
=  \frac{ f_\pi } { 2 \Lambda^2}   \  ,
\label{FGauss4}
\end{align} 
where we have used  the normalization condition 
 \begin{align}
 \int_{0}^1  \varphi (x) \, dx  =f_\pi  \  . 
\label{eq:phinorm}
\end{align} 
Restoring the overall normalization
    \begin{align}
 F_{\gamma \gamma^* \to \pi^0}  (Q^2) 
 =  \frac{ \sqrt{2}} { 3}  F(Q^2)   .
 \label{Ftra}
 \end{align} 
we find that our interpolation  
into small-$Q^2$ region gives 
   \begin{align}
  F^G_{\gamma \gamma^* \to \pi^0} (Q^2=0) 
 =   \frac{  s_0  } { 6\Lambda^2} F_{\rm anomaly}    \ , 
 \label{Fpihard0}
 \end{align} 
where $s_0 = 4 \pi^2 f_\pi^2 \approx 0.67\,$GeV$^2$, and 
   \begin{align}
F_{\rm anomaly}  
 =  \frac{ \sqrt{2}f_\pi   } { s_0}    
 \label{Fanomaly}
 \end{align} 
is the value of   $F_{\gamma \gamma^* \to \pi^0} (Q^2=0) $ 
given by the axial anomaly.  If we take $\Lambda^2=0.2\,$GeV$^2$, the coefficient 
   $ s_0 /6\Lambda^2$ is about 0.53,
   which is close to  the value 1/2  that was argued in Ref. \cite{Lepage:1980fj} 
  to be  an exact result for the $\bar q q$ contribution 
  in the light-front approach. In our approach, interpolation to  $Q^2=0$ 
is model-dependent. As shown below, the $m=0$ non-Gaussian model
gives a divergent result as $Q^2\to 0$.

\subsubsection{$m=0$ model} 

Using the  non-Gaussian model with $m=0$  (\ref{alpharD0})  gives 
 \begin{align}
F (Q^2) = &
 \int_{0}^1  \frac{dx}{xQ^2} \, 
{\varphi (x)} \, 
\left [ 1-    \frac{\Lambda^2}{xQ^2}  + 2  
K_2 (2 \sqrt{x }Q/\Lambda) 
 \right ] 
\ . 
\label{eq:Fscalar3htmod12}
\end{align} 
 Recall  that the size of the twist-4 term 
is determined by the magnitude of   the matrix element of the 
$\bar \psi \gamma_5 \gamma_\alpha 
D^2 \psi $ operator. The fact that the twist-4 contribution in the expression
above looks identical to that in  the Gaussian model 
(\ref{FGauss2})   means that we use definitions 
in which  the average parton virtuality in 
the two models  is measured in the same $\Lambda$ units.

Extracting  the  $Q^2 \to 0$ limit from   
Eq.  (\ref{eq:Fscalar3htmod12}), we  observe that it contains 
 logarithmically singular $\ln (\Lambda/Q)$ terms:
 \begin{align}
F (Q^2) = 
 \int_{0}^1  \frac{dx}{\Lambda^2} \, 
{\varphi (x)} \, 
& \left \{\ln \left (\frac{ \Lambda}{\sqrt{x}Q} \right )  +\frac34 -\gamma_{\rm E}  \right.
    \nonumber \\   &  \left.
+{\cal O} \left (\frac{xQ^2}{\Lambda^2} \right ) \right \}  \ . 
\label{eq:Fspi3htmod120}
\end{align} 
As established  earlier, if 
 the value of TMDA $ \Psi (x, k_\perp )$ at zero transverse momentum
 is finite, 
the $Q^2 \to 0$ limit of $F(Q^2)$ is given by
Eq. (\ref{eq:Fscalar3htQ0}) that  involves  \mbox{$ \Psi (x, k_\perp =0)$. }
In the $m=0$ model, $ \Psi (x, k_\perp )$ is proportional to $K_0 (2 k_\perp / \Lambda) $ 
and  is logarithmically divergent as  $k_\perp \to 0$.
Hence,  a  formal small-$Q^2$ expansion
of \mbox{Eq.  (\ref{eq:Fspinor123})}  leading to Eq. (\ref{eq:Fspinor123b}) 
 is not applicable in this case.

\subsubsection{$m\neq 0$ model} 

Turning  to the $m \neq 0$ model, we have 
 \begin{align}
F &(Q^2) = 
 \int_{0}^1  \frac{dx}{xQ^2} \, 
{\varphi (x)} \, 
\left \{1-  \frac1{ xQ ^2 } \frac{ \Lambda m K_2 (2m/\Lambda)}{ K_1 (2m/\Lambda) }  \right. 
\nonumber \\ & \left. +  
 \frac{ \Lambda  (x Q^2 +m^2) K_2 (2{\sqrt{x Q^2 +m^2}/\Lambda})}
 { xQ ^2 m  K_1 (2m/\Lambda)   }
 \right \} \ . 
\label{eq:Fspinor3htm}
\end{align}
Again,   the twist-4 power correction is  explicitly displayed here.
Note that the average quark virtuality understood as the 
scale that appears in the matrix element of the 
$\bar \psi \gamma_5 \gamma_\alpha 
D^2 \psi $ operator now depends on the interplay of the 
confinement scale  $\Lambda$ and mass-type scale $m$.

Using this expression or 
$
 \Psi_m (x, k_\perp =0)$ from Eq. (\ref{psim})  we obtain 
 \begin{align}
F_m (Q^2=  0) 
=  f_\pi  \frac{K_0 \left (2 m / \Lambda \right ) }{2  m \Lambda
 K_1 ( 2 m/ \Lambda) }  \  .
\label{eq:Fscalar3ht2Q0}
\end{align} 
Thus, we see  again that
 the $m=0$ limit is logarithmically divergent. Writing 
  \begin{align}
F_m  (Q^2=  0) 
=    \frac{f_\pi }{    \Lambda^2} \left [ 
\ln \left (\frac{ \Lambda}{m } \right )- \gamma_{\rm E}  + {\cal O} (m^2/\Lambda^2)  \right ] \  ,
\label{eq:Fscalar3ht2Q0l2}
\end{align} 
we may say that the size  of  $F (Q^2=  0) $ is basically set by the confinement
 scale $\Lambda$, with a coefficient logarithmically dependent on the
 ratio of  $\Lambda$ and the mass-type  scale $m$.
 
 \subsection{Comparing the  data with soft models} 
 \label{Data}
 
 In QCD, the  twist-2  approximation 
 for    $F (Q^2)$  in the leading (zeroth) order in $\alpha_s$
 is 
  \begin{align}
F^{\rm LOpQCD} (Q^2)= & 
 \int_{0}^1  \frac{dx}{xQ^2} \, 
{\varphi (x)}  \ . 
\label{pQCD}
\end{align} 
 Thus, in the asymptotic region,  the value of 
 $$I (Q^2) \equiv \frac1{f_\pi}  Q^2 F (Q^2)$$
 taken  from the data 
 gives  information about the shape of the pion DA.  
 In particular,  for DAs of  $\varphi_r (x) \sim (x \bar x)^r$ type, one has  
 $I_r=1+2/r$,
i.e.   \mbox{$I^{\rm as}(Q^2)  = 3 $} 
 for the ``asymptotic'' wave function $\varphi^{\rm as} (x)  =  6 f_\pi x\bar x$.

\begin{figure}[h]
\centerline{\includegraphics[width=2.8in]{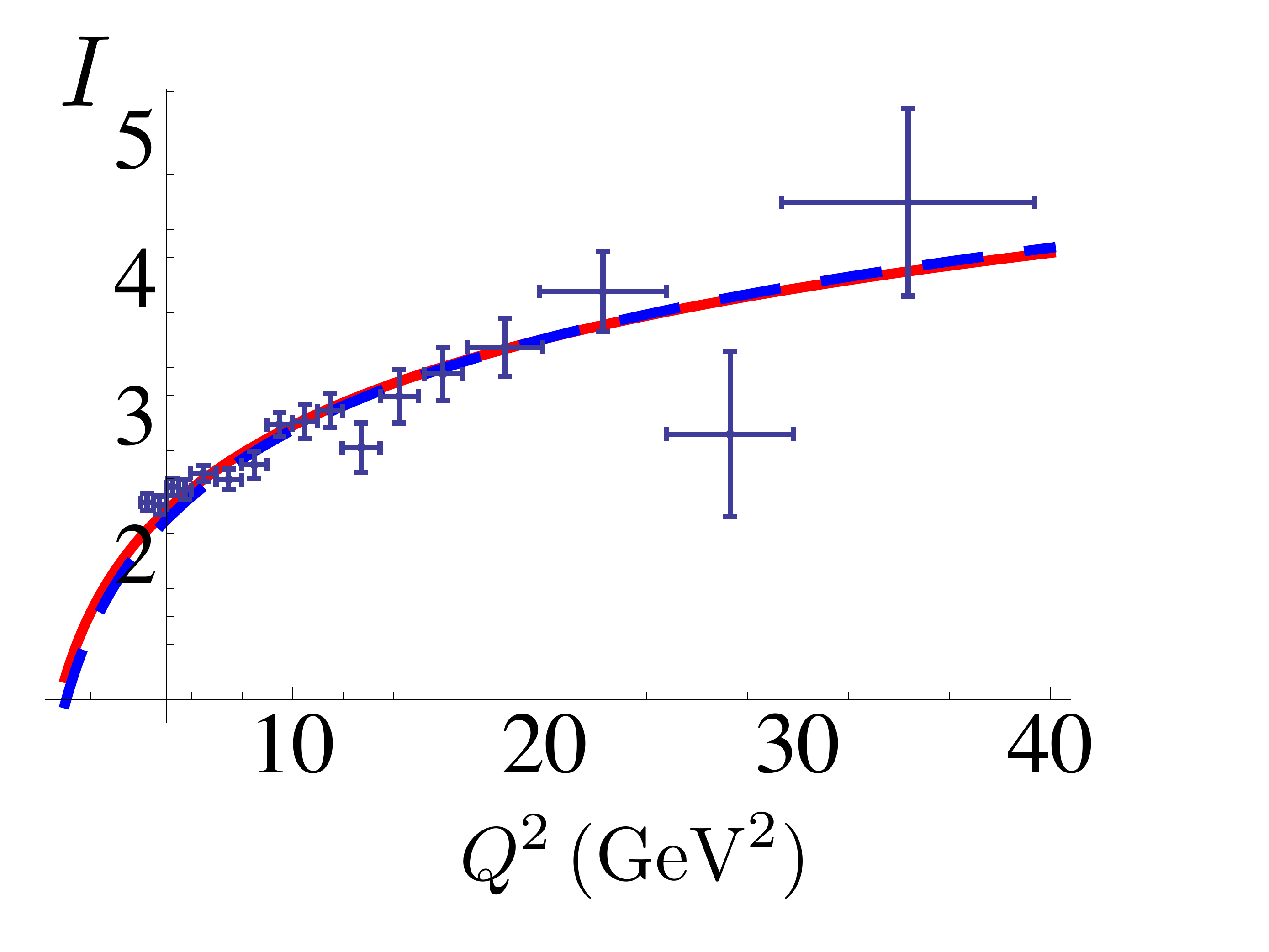}}
\vspace{-0.4cm}
\caption{{\sc BaBar}  data compared to model curves  described in the text.
\label{babar}}
\end{figure}

\begin{figure}[h]
\centerline{\includegraphics[width=3.1in]{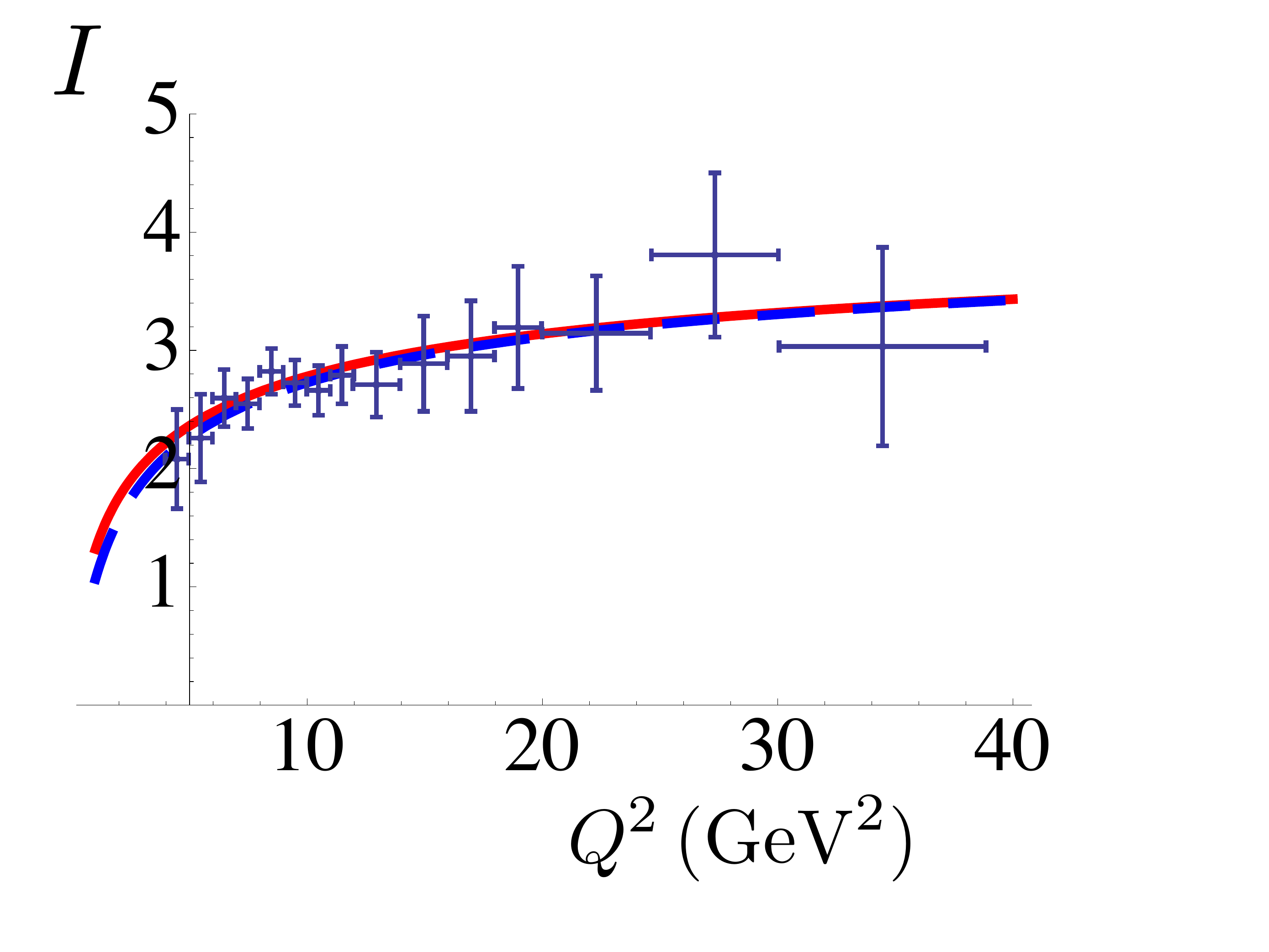}}
\caption{BELLE data compared to model curves 
described in the  text.
\label{belle}}
\end{figure}

 The most  recent data \cite{Aubert:2009mc,Uehara:2012ag}
 still show a $Q^2$ variation of   $I(Q^2)$ (see Figs. \ref{babar}, \ref{belle}),
 especially in case of {\sc BaBar}  data \cite{Aubert:2009mc} which  contain several 
 points with $I(Q^2)$ values well above 3.  It was argued \cite{Radyushkin:2009zg,Polyakov:2009je} 
 that BaBar data indicate that the pion DA is close to a flat function $\varphi^{\rm flat}  (x) = f_\pi $.
 The latter corresponds to $r=0$, and pQCD gives $I^{\rm flat}=\infty$. 
 As shown in Ref. \cite{Radyushkin:2009zg}, inclusion of
  transverse momentum dependence
 of the pion wave function in the  light-front formula  
 of Ref.  \cite{Lepage:1980fj} (see also \cite{Musatov:1997pu})
   eliminates the divergence at $x=0$, and
 one 
 can produce a curve that fits the {\sc BaBar}  data. 
 Similar curves may be obtained within the VDA approach described in the present  paper.

In Fig.  \ref{babar}, we compare {\sc BaBar}  data with model curves 
corresponding to  flat DA  $\varphi (x) =f_\pi $ and two types of 
transverse momentum distributions. First, we take the 
Gaussian model 
of Eq.   (\ref{FGauss2}). 
For large $Q^2$, it gives (for a flat distribution)
 \begin{align}
F_G^{\rm flat}  (Q^2) = &\frac{ f_\pi } { Q^2}\left [\ln \frac{Q^2}{\Lambda^2}
-1 + \gamma_E  + \frac{\Lambda^2}{Q^2}  +\ldots 
\right ]
   \  .
\label{FGauss2flas}
\end{align} 
The logarithmic divergence of the pQCD formula converts here
into a logarithmic increase of the $Q^2 F (Q^2)$ combination.

A curve closely following the data is obtained for 
a value of $\Lambda^2=0.35\,$GeV$^2$  which is 
larger than the standard estimate $\delta^2=0.2\,$GeV$^2$ 
\cite{Novikov:1983jt} for the 
matrix element of the $\bar \psi \gamma_5 \gamma_\alpha D^2 \psi$ operator.
However, the higher-order pQCD corrections are known 
\cite{Li:1992nu}   to 
shrink the $z_\perp$ width  of the IDA $\varphi (x, z_\perp)$,
effectively increasing  the observed $\Lambda^2$ compared to the 
primordial value of  $\Lambda^2$.
One should also take into account 
the correction due  to the 3-body TMDA 
corresponding to $\bar \psi G \psi$  operator generated 
by the insertion of the Fock-Schwinger field.
Its magnitude is governed by the same scale $\Lambda^2$  that 
appears  in the $\bar \psi D^2 \psi$ operator. 
We plan to include this term in future studies.
Our goal now is just to show that 
the VDA approach results in curves that are able 
to easily fit  the data  over a wide range of $Q^2$ values.

For illustration, we also take the non-Gaussian $m=0$  model of Eq. (\ref{eq:Fscalar3htmod12}),
 to  check  what  happens 
 in  case of unrealistically slow  $\sim 1/z_\perp^2$ decrease 
  for large $z_\perp$.   Still, 
if we take  a   larger value of 
$\Lambda^2=0.6\,$GeV$^2$, this model  
 produces practically the same curve 
as the $\Lambda^2=0.35\,$GeV$^2$ Gaussian model. 
The explanation is that in this case we have 
 \begin{align}
F_{m=0}^{\rm flat}  (Q^2) = &\frac{ f_\pi } { Q^2}\left [\ln \frac{Q^2}{\Lambda^2}
-1 + 2\gamma_E  + \frac{\Lambda^2}{Q^2}  +\ldots 
\right ]
\label{F2flas}
\end{align} 
for large $Q^2$, which, compared to Eq. (\ref{FGauss2flas}),
amounts to adding the $e^{\gamma_E} \approx 1.8$
factor in the  argument of $\ln (Q^2/\Lambda^2)$.

 Data from BELLE \cite{Uehara:2012ag} give lower values for $I$,
suggesting a non-flat DA. In Fig.  \ref{belle},
 we show the curves corresponding to  
\mbox{$\varphi (x) \sim f_\pi (x\bar x)^{0.4}$} DA.
If  we take the  
Gaussian model   (\ref{FGauss2}),  a good eye-ball fit to data is  produced
if we take 
$\Lambda^2=0.3\,$GeV$^2$.
Practically the same curve is obtained in  the   non-Gaussian
 \mbox{$m=0$}   model  
of Eq. (\ref{eq:Fscalar3htmod12})  for 
 $\Lambda^2=0.4\,$GeV$^2$. Again, a VDA-based analysis  of the higher-order 
 Sudakov effects \cite{Li:1992nu}   is needed to extract 
the value of $\Lambda$ in the primordial TMDA.
Note also that the curve is still well below the  pQCD value
$I_{0.4} =4.5$ for this DA. 

\section{Modeling hard tail} 
\label{Hard}

\setcounter{equation}{0} 

\subsection{Vertex models} 
\label{Vertex}

\subsubsection
{Hard vertex and two propagators}

\begin{figure}[h]
   	\centerline{\includegraphics[width=1.6in]{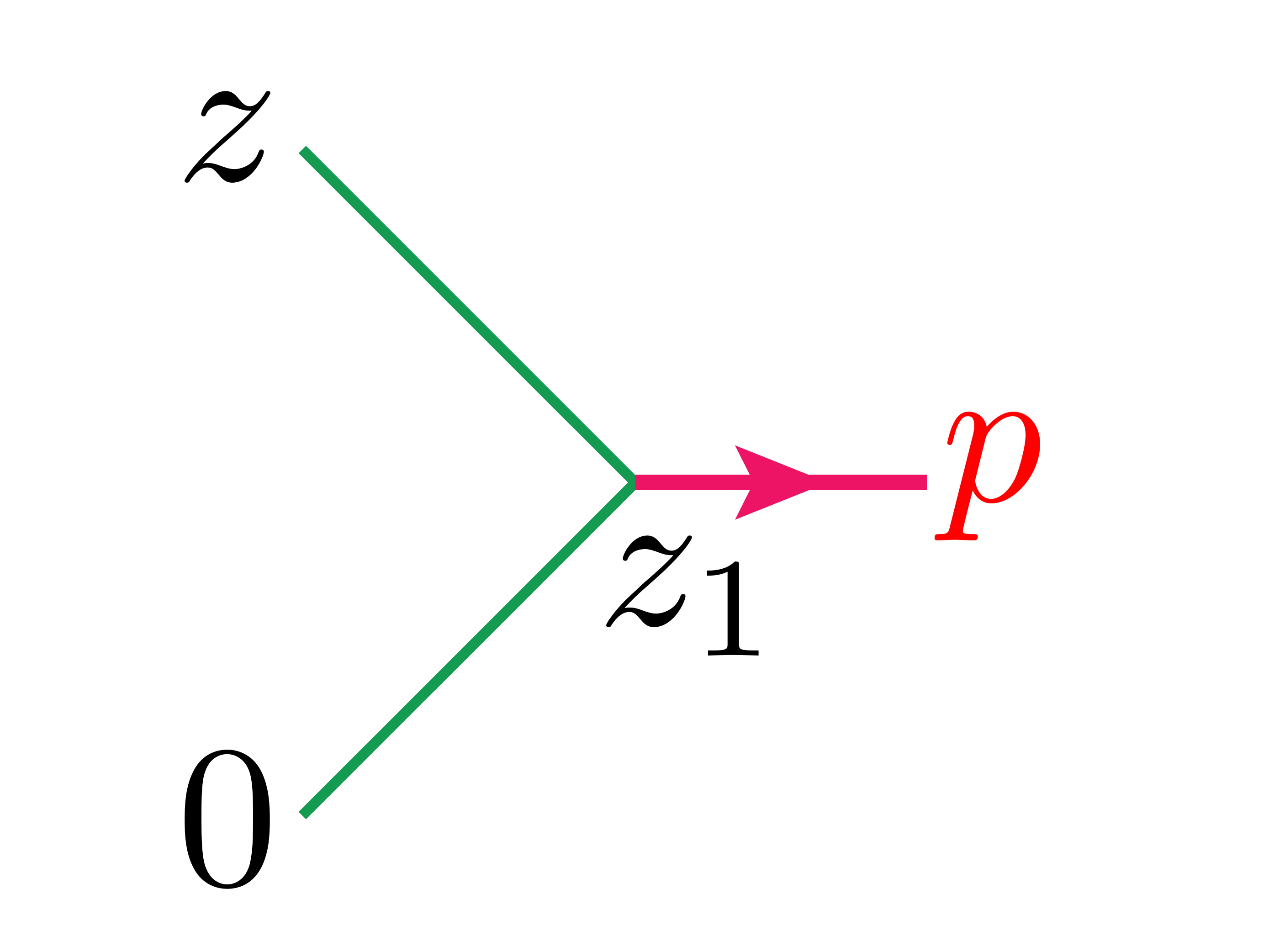}}
   	\caption{Modeling VDA by a local current source.
   		\label{current}}
   \end{figure}

The simplest explicit example of a VDA-like object based on Feynman diagrams 
 is given by a toy model 
in which the  matrix element of the bilocal operator 
is given  by a  graph  consisting of 
two  perturbative propagators $D^c(z_1,m)$ and \mbox{$D^c (z-z_1,m)$}
($m$ being the relevant mass) 
 joined at the   point $z_1$ in which the external 
momentum $p$ enters (see Fig. \ref{current}).  
To  derive the relevant VDA, let us    use  the momentum representation.
Then  we have 
\begin{align} 
& \chi (k,p)  =
\frac{1  }{(k^2-m^2) [ (p-k)^2-m^2] } 
\\ &      \equiv \int_{0}^{\infty} d \lambda 
\int_{0}^1 dx\,  
F (x,\lambda) \,  \,  e^{i  \lambda \bar x k^2 + i   \lambda x (k-p)^2
	-i  \lambda m^2 -\epsilon  \lambda} \, ,  \nonumber
\end{align}
which gives the relation 
\begin{align}
& \, e^{i  \lambda x^2 p^2}  F_2 (x,  \lambda) 
=   \lambda  {e^{i  \lambda x p^2 -i  \lambda m^2}  }\,  . 
\label{Phi2momf}
\end{align} 
 Using $F (x,  \lambda)= \Phi (x, 1/ \lambda)$,  
we find the VDA for  this case 
  \begin{align}
\Phi  (x, \sigma) =   \frac1{\sigma} \, e^{i (x \bar x p^2 -m^2)/\sigma }  \  ,
\label{Phi2}
\end{align} 
which  yields  the following   result
  \begin{align}
\Psi  (x, k_\perp) =   \frac1{\pi} \, \frac1{k_\perp^2+m^2-x \bar x p^2  }  
\end{align} 
 for the   TMDA.  In the  large transverse momentum region,
 it has a ``hard'' power-like $\sim 1/k_\perp^2$ behavior. 
   In the impact parameter space, we have 
 $$\varphi  (x, z_\perp) = 2 K_0 (mz_\perp) \ , $$ a function that has a logarithmic
 singularity for $z_\perp =0$.  This feature explains why,
  formally   integrating $\Psi  (x, k_\perp)$  over  $d^2 k_\perp$  to 
  produce DA, one faces  in this case a logarithmic divergence.

  \subsubsection
{Soft vertex and propagators}  

\begin{figure}[h]
       \vspace{-3mm}
   \centerline{\includegraphics[width=2in]{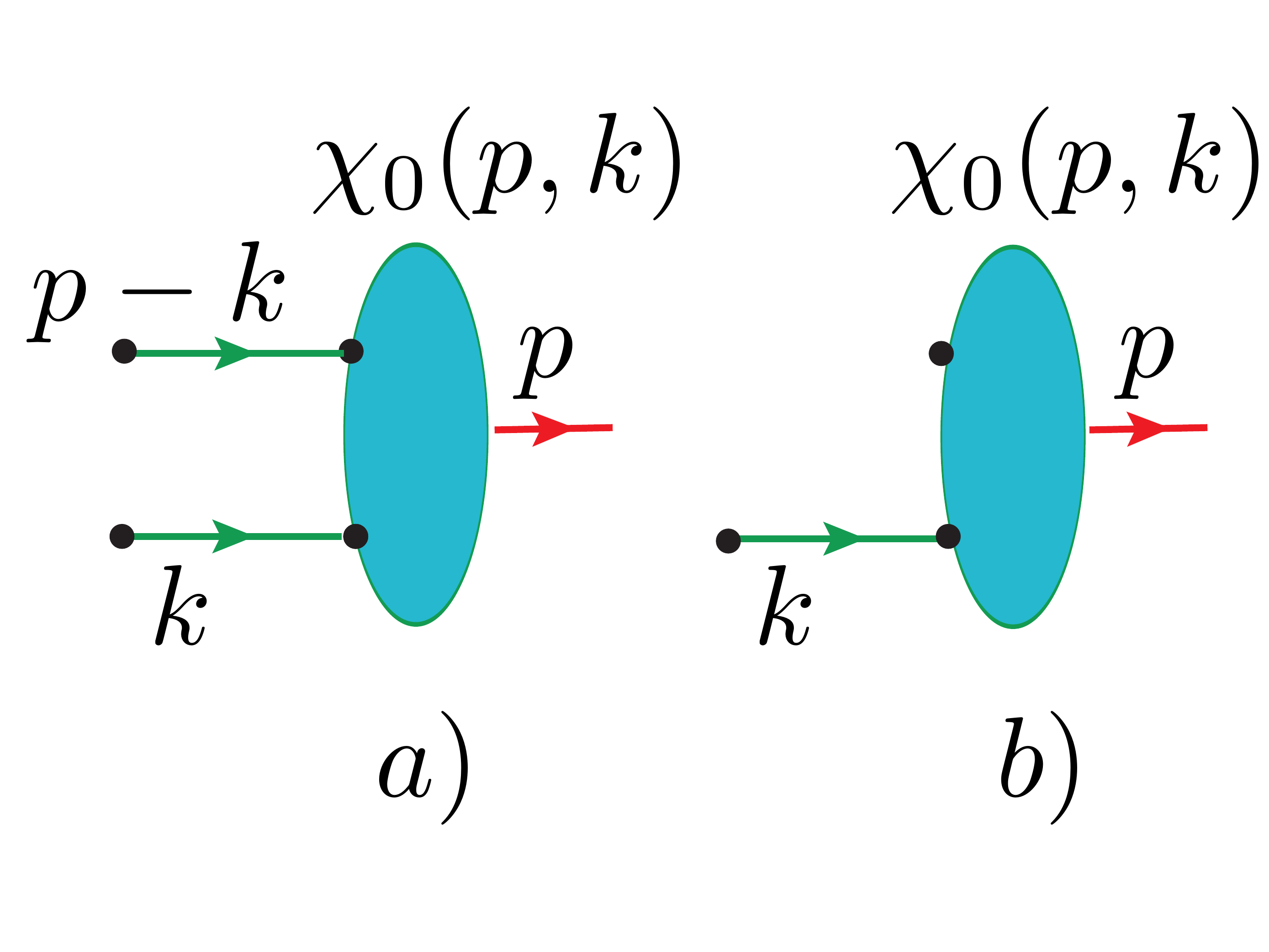}}
   \vspace{-5mm}
   \caption{Attaching propagators to a soft vertex.
   \label{B0}}
   \end{figure}

Instead of a point current,
  we  can use a soft vertex, i.e.   consider a model
  (see Fig. \ref{B0}a) 
  \begin{align}
&  \chi_2 (k,p) = 
  \frac{\chi_0 (k,p)  }{k^2  (p-k)^2 } \ , 
\label{chi2mom}
\end{align} 
 where,   for simplicity, we took massless propagators.
  Also, we will take $p^2=0$ in this model.  To proceed, we  write the soft vertex 
  in the VDA representation as  
\begin{align}
& \chi_0 (k,p) = 
\int_{0}^{\infty} d \alpha \int_{0}^1 dy\,  
 F_0(y,\alpha) \,  \,  e^{i \alpha (k-yp)^2-\epsilon \alpha} \,  . \label{Phixsmom}
\end{align}

 \paragraph{One perturbative propagator.}  Consider first the case 
  when just one propagator is added to the soft vertex   (see Fig. \ref{B0}b), 
   \begin{align}
&  \chi_1 (k,p) = -
  \frac{\chi_0 (k,p)  }{k^2  } \ . 
\label{chi1mom}
\end{align} 
 Then
    \begin{align}
&  \chi_1 (k,p) =  i
 \int_{0}^1 dy
\int_{0}^{\infty} d \alpha  \int_{0}^{\infty} d\alpha_1 \,  
  \nonumber \\ &  \times 
 F_0(y,\alpha) \,  \,  e^{i \alpha (k-yp)^2 + i \alpha_1 k^2}
   \nonumber \\ &= i \int_{0}^1 dy\,
   \int_{0}^{\infty} \lambda d  \lambda \int_{0}^1 d \beta
    e^{i \lambda (k-\beta y p)^2 }  F_0(y,\beta \lambda) 
    \,
 \ , 
\label{chi1mom2}
\end{align} 
 which gives 
    \begin{align}
&  F_1 (x,\alpha) =  i
 \alpha \int_{0}^1 dy\,  
\int_{0}^1 d \beta \,  \,  \delta (x-\beta y ) \, 
 F_0(y, \beta \alpha) 
    \,
\label{F1mom2}
\end{align} 
or
   \begin{align}
&  F_1 (x,\alpha) =  i
 \alpha \int_{x}^1 \frac{dy}{y} \,  
 F_0(y, \alpha x/y) 
    \, .
\label{F1mom21}
\end{align} 
For the TMDA we have 
    \begin{align}
&  \psi_1 (x,k_\perp^2) = - \frac{\partial}{\partial k_\perp^2} 
\int_{x}^1 \frac{dy}{y} \,  
 \psi_0(y, y k_\perp^2/x ) \,  
    \, . 
\label{F1mom2a}
\end{align}

To illustrate the impact of adding a perturbative 
leg, let us take   the  simplest  model when 
the soft distribution has no $x$-dependence (i.e. is ``flat''),  
 $ \psi_0(y, k_\perp^2 )  =\psi_0^F (y, k_\perp^2 ) \equiv \psi_0( k_\perp^2 ) $.
 Despite the flatness of the soft distribution,  we obtain a function
    \begin{align}
&  \psi_1^F (x,k_\perp^2) = \frac{\psi_0 (k_\perp^2)
-\psi_0 (k_\perp^2/x)} {k_\perp^2}
\label{F1mom3}
\end{align} 
with a nontrivial $x$-profile.
For  large $k_\perp^2$, and a fast-decreasing soft function $\psi_0 (k_\perp^2)$,
one can use a naive approximation 
  $\psi_1^F (x,k_\perp^2) \approx
 \psi_0 (k_\perp^2)/k_\perp^2$ for some  range of $x$-values 
 that are not very close to $1$.
But eventually $\psi_1^F (x,k_\perp^2)$
vanishes for $x=1$, i.e. the parton corresponding 
to the ``hard''  $1/k^2$  propagator cannot carry the 
whole momentum of the hadron,
even though the soft vertex allowed this,
and the purely soft $(p-k)$ parton still can carry 
the $\bar x=1$ fraction.

 \paragraph{Two perturbative propagators.} 
 Switching now to the soft vertex model with two perturbative
 propagators  attached 
(\ref{chi2mom}), we obtain
    \begin{align}
  F_1 (x,\alpha) =  & - 
 \alpha^2 \int_{0}^1 {dy}  \int_{0}^1 d \beta   
\int_{0}^{1-\beta}  d \beta_2 \, 
 F_0(y, \beta \alpha) \,   \nonumber \\ &
 \times   \delta (x-(\beta_2 + \beta y) )
 \nonumber \\ = &
-  \alpha^2 \int_{0}^1 {dy}  \int_{0}^{V_0(x,y)}  F_0(y, \beta \alpha)  d \beta   
    \, ,
\label{F1mom4}
\end{align} 
where 
   \begin{align}
V_0 (x, y) = \frac{x}{y} \theta(x<y) +  \frac{ \bar x}{\bar y }  \theta(x>y) \  .
\end{align} 
One may recognize in this function a part of the ERBL evolution kernel
\cite{Efremov:1979qk,Lepage:1979zb}; we will turn to this point later. 
For TMDA in this case we have 
 \begin{align}
 & \psi_2 (x, k_\perp^2) =  \left ( \frac{\partial}{\partial k_\perp^2} \right )^2
\int_{0}^1 {dy}\,  \int_{0}^{V_0(x,y)} d \beta \, 
 \psi_0(y, k_\perp^2/\beta ) \,  
  \nonumber \\ = & - \frac{1}{k_\perp^2} \, 
  \frac{\partial}{\partial k_\perp^2} \int_{0}^1 {dy}\, V_0(x,y)  \, 
  \psi_0(y, k_\perp^2/V_0(x,y) )
    \, . 
\label{F1mom2b}
\end{align} 
For illustration, taking again a flat model
$ \psi_0(y, k_\perp^2 )   = \psi_0( k_\perp^2 ) $, 
we obtain
   \begin{align}
&  \psi_2^F (x,k_\perp^2) = \frac{
\psi_0 (k_\perp^2) - 
x \psi_0 (k_\perp^2/x) - \bar x \psi_0 (k_\perp^2/\bar x) 
} {k_\perp^4} \  .
\label{F2mom3a}
\end{align} 
Thus, for  large $k_\perp^2$ and a fast-decreasing soft function $\psi_0 (k_\perp^2)$,
one can use a naive approximation 
  $\psi_2^F (x,k_\perp^2) \approx
 \psi_0 (k_\perp^2)/k_\perp^4$ for some  range of $x$-values 
 that are not very close to 0 or $1$.
However,  for non-zero $k_\perp^2$, the function 
$\psi_1^F (x,k_\perp^2)$
vanishes both for $x=0$ and $x=1$, i.e. 
neither   parton  can carry the 
whole momentum of the hadron. 

For small $k^2_\perp$, we have 
   \begin{align}
&  \psi_2^F (x,k_\perp^2\to 0) =   - \frac{1}{k_\perp^2}
  \,
 \psi_0^{'}  (k_\perp^2) + \ldots 
\  , 
\label{F2mom31}
\end{align} 
i.e.,  
the function 
$\psi_2^F (x,k_\perp^2) $ behaves like 
$\psi_0^{'}  (0)/k^2_\perp$   for small $k^2_\perp$,
showing no dependence on $x$ except for narrow 
regions of $x$ close to 0 or 1.

 \subsubsection{Equations of motion}

 A situation, when one $\chi$  function  differs from another 
 one by a perturbative propagator is encountered when
 one considers 
 equations  of motion. In particular,
 we have $  \partial^2 \phi  = g \chi \phi$
in the case of $g \phi^2 \chi$ interaction of  quarks $\phi$
with  gluons $\chi$.  This imposes a relation 
\begin{align}
\partial^2
 \langle p |   \phi(0) \phi (z)|0 \rangle 
&= 
 \langle p |   \phi(0)  g \chi (z)  \phi (z)|0 \rangle 
 \label{vdad20}
\end{align} 
between 2-body and 3-body matrix elements,
which should be also satisfied by their VDA representations. 
In  the momentum representation, 
the equation of motion imposes the restriction 
   \begin{align}
- k^2  \chi  (k,p) = 
  {\chi_1 (k,p)  }\ , 
\label{chi1moma}
\end{align} 
connecting the 2-body amplitude $ \chi  (k,p) $ 
and the  reduced 
3-body amplitude $ \chi _1 (k,p) $  in which the gluon field $\chi$  is  
located at the same point with  one of the $\phi$ fields. Hence, $ \chi  (k,p) = -
  {\chi_1 (k,p)  }/ k^2 $, and we can 
  use   Eq. (\ref{F1mom21})  to write a relation 
    \begin{align}
  \Phi  (x,\sigma) = & \frac{i}{\sigma} 
  \int_{x}^1 \frac{dy}{y} \,  
 \Phi_1 (y, \sigma \, y/x )     
\label{F1mom21a}
\end{align}   
between the 2-body VDA  $ \Phi  (x,\sigma)$ corresponding to   $\chi  (k,p)$
and the reduced 3-body VDA 
 $\Phi_1 (x, \sigma) $    corresponding to    $\chi_1 (k,p)$.

However, for the purposes of  VDA model-building,  Eq. (\ref{F1mom21a})
is not convenient  since the basic 2-body VDA  $\Phi (x,\sigma)$  
looks like generated from the  3-body VDA $\Phi_1(x,\sigma)$
describing the matrix element 
$ \langle p |   \phi(z_1)  g \chi (z_3)  \phi (z_2)|0 \rangle $
in the $z_3 \to z_1$ limit.  

We would rather prefer  to 
start with  some model form  for  the  2-body  VDA $\Phi (x, \sigma)$ 
and then  treat Eq. (\ref{F1mom21a}) as a constraint on the  full 3-body VDA.
Thus, we need 
 an inverse  relation in which $\Phi_1(x,\sigma)$  
is written in terms of $\Phi (x, \sigma)$.
To this end,   let us  apply  
 $\partial^2$ to the VDA parametrization  (\ref{Phixs0}) of the 2-body 
matrix element:
\begin{align}
\partial^2
 \langle p |   \phi(0) \phi (z)|0 \rangle 
&= 
\int_{0}^{\infty} d \sigma \int_{0}^1 dx\, 
 \Phi (x,\sigma) \,  \,  e^{i x (pz) -i \sigma z^2/4} 
 \nonumber \\ &\times \left [ x (pz) \sigma - \sigma^2 z^2/4 -2i \sigma \right ]
 \label{vdad21}
\end{align} 
(we take $p^2=0$ and $z^2 \to z^2 -i \epsilon $ is implied here and below).
Integrating  the first term by parts over $x$  assuming \mbox{$\Phi (0,\sigma) = \Phi (1,\sigma) =0$}, while  the second one over $\sigma$ 
 assuming 
$\sigma^2 \Phi (x,\sigma)e^{-\epsilon \sigma}|_{ \sigma \to \infty} =0$,   \mbox{$\sigma^2 \Phi (x,\sigma)|_{\sigma=0}=0$,} 
 we obtain \begin{align}
\partial^2 
 \langle p |   \phi(0) \phi (z)|0 \rangle 
= & i
\int_{0}^{\infty} d \sigma \int_{0}^1 dx\,     \,  e^{i x (pz) -i \sigma z^2/4}
 \nonumber \\ &\times   \left [x \frac{\partial}{\partial x}  
 +\sigma  \frac{\partial}{\partial \sigma} \right 
 ]   \sigma \Phi (x,\sigma) 
 \label{vdad25}
\end{align}

 Parametrizing $ \langle p |   \phi(0) g \chi (z) \phi (z)|0 \rangle $
by a VDA $ \Phi_1 (x, \sigma)$ gives 
\begin{align}
 \Phi_1 (x, \sigma) =
i \left [x \frac{\partial}{\partial x}  
 +\sigma  \frac{\partial}{\partial \sigma} \right 
 ]   \sigma \Phi  (x,\sigma)   \ . 
 \label{vdad26}
\end{align}
As a check,  using   Eq. (\ref{vdad26}) in the right-hand side of Eq. (\ref{F1mom21a} ),
one can verify that
the outcome is  indeed  given by $ \Phi_1  (x,\sigma) $. 
Now, Eq. (\ref{vdad26})   can be  converted    into a relation
     \begin{align}
\psi_1&  (x, k_\perp^2) =   k_\perp^2 \psi  (x, k_\perp^2   ) -
   x \frac{\partial}{\partial x}
 \int_{k_\perp^2}^\infty   \, d  \kappa^2  \, 
   \psi  (x, \kappa^2   )
\label{vdad27}
\end{align} 
between TMDAs. As one might expect, the action of $\partial^2$
on the 2-body function has resulted in a term containing
the $ k_\perp^2 $ factor.
For the  distribution amplitude 
    \begin{align}
\psi_1 (x) \equiv & \int_0^\infty  dk_\perp^2 \psi_1  (x, k_\perp^2) 
\end{align} 
 this gives
     \begin{align}
\psi_1 (x)
 =  
\left [1 -     x \frac{\partial}{\partial x} \right ] \varphi^{(1)}  (x) \ , 
\label{vdad28}
\end{align} 
where $\varphi^{(1)} (x)$ is the $k_\perp^2$ moment of the TMDA $\psi (x,k_\perp^2)$,
    \begin{align} 
    \varphi^{(1)} (x) \equiv  \int_0^\infty d k_\perp^2 \, 
k_\perp^2 \psi  (x, k_\perp^2   )  \ . 
\label{vdad29}
\end{align} 

Note, that $\psi_1 (x)$  is not necessarily $ x \leftrightarrow \bar x$ symmetric 
even if $\varphi_1 (x)$ is. 
 This is natural, since the fraction $x$ in  $\psi_1  (x,\sigma) $
 corresponds to a ``glued'' field  $g \chi (z_1) \phi (z_1)$,
 while the fraction $\bar x$ is associated with a single field $\phi (z_2)$.
  To see the implications of the  $ x \leftrightarrow \bar x$ symmetry
 on the VDA level, we  apply the equation of motion to the second field
in the matrix element  $\langle p |   \phi(z_1) \phi (z_2)|0 \rangle $, which 
gives  
\begin{align}
 \Phi_1 (\bar x, \sigma) =
i \left [\bar x \frac{\partial}{\partial \bar x}  
 +\sigma  \frac{\partial}{\partial \sigma} \right 
 ]   \sigma \Phi  (x,\sigma)   \ . 
 \label{vdad26a}
\end{align}
Since $\Phi  (x,\sigma) =\Phi  (\bar x,\sigma) $,
this  equation is  consistent with Eq. (\ref{vdad26}).  Still, 
 $\Phi_1  (x,\sigma) \neq \Phi_1  (\bar x,\sigma) $ in general.

 \subsection{Generating hard tail from an initially  soft VDA}
 \label{Hard_tail}
 
 \subsubsection{Scalar fields}

Our next   model involves two  currents 
 carrying momenta 
 $xp$ and $ (1-x)p   \equiv  \bar x p$
  at locations $z$ and $0$, respectively, connected 
 by a  perturbative  propagator
 $D^c (z,m)$, and weighted with a function $\varphi (x)$.  
    \begin{figure}[h]
   \centerline{\includegraphics[width=1.6in]{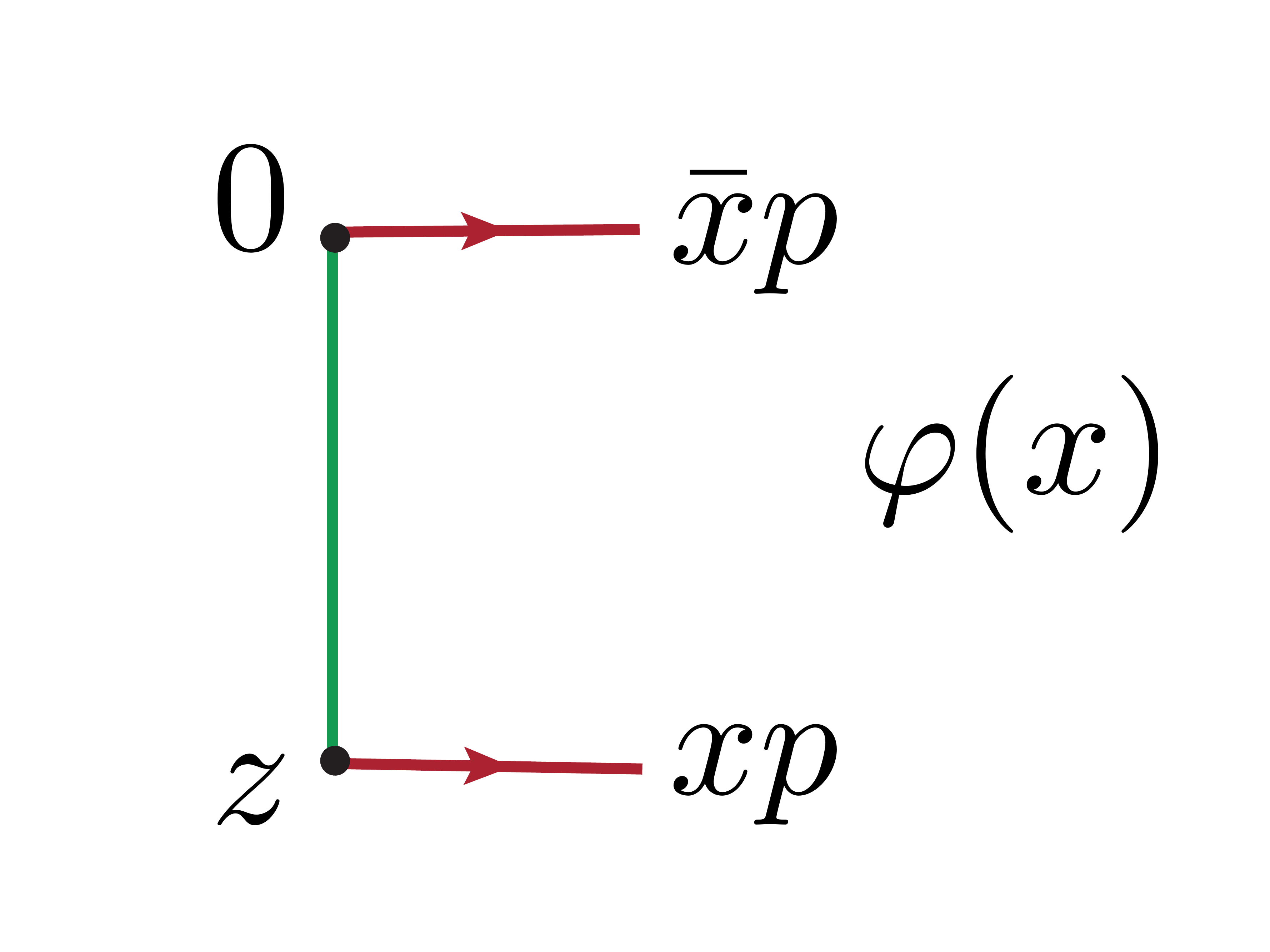}}
   \caption{Modeling VDA by a two-current state.
   \label{twocurrent}}
   \end{figure}
 Then  
 we have
  \begin{align} 
t_1 (z,p) = g^2\, \int_0^1 dx \, e^{i  x(pz)} 
 D^c(z,m) \varphi (x)
     \label{One}
 \end{align}
  where $g$ is the coupling constant.
Using  
\begin{align} 
D^c(z,m) =&  \frac1{(4 \pi)^2}  \int_0^{\infty} e^{-i \sigma z^2 /4 -im^2/ \sigma}  
 {d \sigma}  
\label{alpharD01}
\end{align}
 we obtain 
 \begin{align} 
\Phi (x, \sigma)  =&  \frac{  \varphi (x)}{(4 \pi)^2}  e^{ -im^2/ \sigma}  \  .
\label{One2}
\end{align}
 However, the integral for TMDA, 
  \begin{align} 
\Psi (x, k_\perp)  =&   \frac{  \varphi (x)}{(4 \pi)^2}  
 \int_0^\infty \frac{d\sigma}{\sigma} e^{ -i(k_\perp^2+m^2)/ \sigma} \  ,
\label{One2}
\end{align}
in this case diverges at large $\sigma$.
So, let us take a  nontrivial  primordial VDA $\Phi_0 (x, \sigma_0)$ instead of 
the above model corresponding to  $\varphi (x) \delta (\sigma_0)$. Then
  \begin{align} 
t_1 (z,p) = & g^2\, 
 D^c(z,m) \int_0^1  dx \,  e^{i x (pz)}   \nonumber \\ &  \times 
 \int_0^\infty d\sigma_0 
 e^{-i \sigma _0 z^2 /4}  \Phi_0 (x, \sigma_0)
     \label{OneS}
\end{align}
and 
\begin{align} 
\Phi_1 (x, \sigma)  =&  \frac{g^2}{(4 \pi)^2} \,{\sigma }  \int_0^1 d\beta
e^{ -im^2/ \bar \beta \sigma}  \Phi_0 (x,\beta \sigma) \ , 
\label{OneS2}
\end{align}
which gives 
  \begin{align} 
\psi_1 (x, k_\perp^2)  =&  \frac{g^2}{(4 \pi)^2}   
 \int_{k_\perp^2}^\infty d \kappa^2 \int_0^1 d\beta \, 
 \psi_0(x, \beta (\kappa^2+m^2))
\label{OneS3}
\end{align}
or
  \begin{align} 
\psi_1 (x, k_\perp^2)  =&  \frac{g^2}{(4 \pi)^2}
 \int_{k_\perp^2}^\infty \frac{d \kappa^2}{\kappa^2+m^2} 
 \int_0^{\kappa^2+m^2} d {\kappa'}^2 \psi_0(x, {\kappa'}^2) \  .
\label{OneS3a}
\end{align}

\begin{figure}[h]
   \centerline{\includegraphics[width=1.8in]{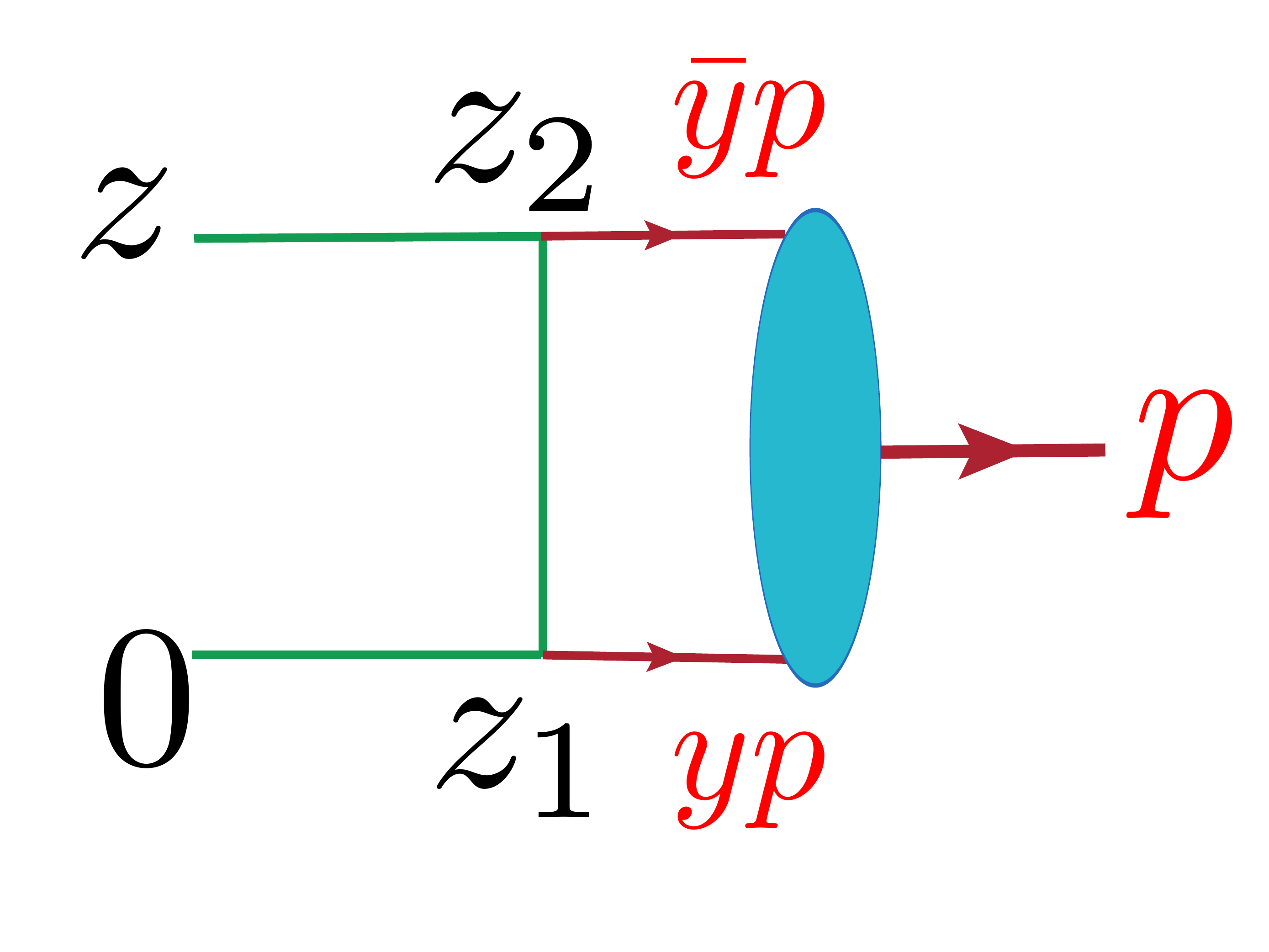}}
   \caption{Getting hard tail from a soft initial distribution. 
   \label{bilocal}}
   \end{figure}

The next step is to  take a model with two perturbative propagators
attached. In other words, consider a model 
 with  vertices  at 
 locations $z_1$ and $z_2$  connected 
 by a  perturbative  propagator
 $D^c (z_1-z_2,m)$ and two perturbative propagators
 connecting these points with points $0$ and $z$  (see Fig. \ref{bilocal}). 
 
 Then we can use the formula (\ref{F1mom2b}) 
 \begin{align}
  \psi_2 (x,k_\perp^2) =& - \frac{1}{k_\perp^2+m^2} \, 
  \frac{\partial}{\partial k_\perp^2} \int_{0}^1 {dy}\,  \nonumber \\ &  \times 
  \psi_1(y, (k_\perp^2+m^2)/V_0(x,y) )
    \ 
\label{F1mom22}
\end{align} 
 with $\psi_1$ given by Eq. (\ref{OneS3}).
 This gives 
     \begin{align}
\psi_2&  (x, k_\perp) =  \frac{g^2}{16\pi^2}  \,  \frac1{( k_\perp^2+m^2 )^2 }   \int_0^1 
dy \, {V_0(x,y)}    \nonumber \\ &  \times  \left [  
\, \int_0^{(k_\perp^2+m^2)/V_0(x,y) }
     \psi_0 (y, { k'_\perp}^2   )  \, d  {k'_\perp}^2  \right ]  \  . 
\label{psi3}
\end{align}

     The part  in square brackets may be  written as 
         \begin{align}
  \Biggl [ \cdots \Biggr ]  =  
  {\varphi_0 (y)}- \int^\infty_{(k_\perp^2+m^2)/V_0(x,y) }
     \psi_0 (y, { k'_\perp}^2   )  \, d  {k'_\perp}^2 
 \ ,
  \end{align} 
where  $\varphi_0 (y)$ is   
the primordial distribution $\Psi_0 (y, k'_\perp)$ integrated
over  all the transverse momentum plane.
 Hence,  for large $k_\perp$, the 
leading $1/k_\perp^4$  term is  determined by 
the DA  $\varphi_0 (y)$ only.
A particular shape  of the \mbox{$k_\perp$-dependence} of  the soft TMDA 
  $\Psi _0(y, k_\perp)$  affects  only 
 the  subleading $\sim [V \otimes \psi_0] (x,k_\perp^2)/k_\perp^2$ term. 
 The form of $k_\perp$ dependence of   $\Psi _0(y, k_\perp)$  is also essential
for the behavior  of  $\Psi^{B_0}  (x, k_\perp)$  
 term at  small $k_\perp$.  In particular,  
 we have 
      \begin{align}
  \Biggl [ \cdots \Biggr ] _{k_\perp=0}  = 
     \psi_0 (y,  k_\perp^2=0   )   
 \ ,
\label {small_k}
  \end{align} 
which gives, e.g., $\varphi_0 (y)/ \Lambda^2$ in the Gaussian model
 (\ref{Gaussian}).

\subsubsection{Spin-1/2 quarks  and scalar gluons}

In case of spin-1/2 quarks interacting $via$ a (pseudo)scalar or vector gluon field 
(in Feynman gauge),
the factor coming from $k$ and $p-k$ legs is given by
    \begin{align}
\frac{ {\rm Tr} \{ \gamma_\alpha\slashed k \slashed p (\slashed p-\slashed k)\}}{k^2 (p-k)^2}
\  .
\label{eq:spi}
\end{align}
Using $\slashed p = (\slashed p-\slashed k)  +\slashed k$, we arrive at  
   \begin{align}
\frac{ {\rm Tr} \{ \gamma_\alpha\slashed k \}}{k^2} +\frac{ {\rm Tr} 
\{ \gamma_\alpha (\slashed p-\slashed k)  \}}{ (p-k)^2}
\  .
\label{eq:spi1}
\end{align}
Representing $\slashed k =  x \slashed p+  (\slashed k - x \slashed p)$
and noticing that $ (\slashed k - x \slashed p)$ results in $ \slashed  z$ 
in coordinate representation,  we may treat  the equation above as 
    \begin{align}
 \left ( \frac{x}{k^2} + \frac{1-x}
{ (p-k)^2}  \right ) \,  {\rm Tr} \{ \gamma_\alpha\slashed p \} +`` {\cal O} (\slashed z)"
\  .
\label{eq:spi2}
\end{align}
Now, for the first term we need to find $\Phi^{B_0}_{1}  (x,1/\alpha) $ satisfying 
 \begin{align}
 \label{PhiBmom2}
&       \int_{0}^{\infty} d \alpha \int_{0}^1 dx\frac1{x} \, 
 \Phi^{B_0}_{1}  (x,1/\alpha) \,  \,  e^{i \alpha (k-xp)^2}  
 \\ &    =  \frac{ \pi^2  g^2}{k^2} \,
   \int_{0}^{\infty} 
    \frac{d \alpha_2}{   \alpha_2}
     \int_{0}^1 dy\, 
  \int_0^1 d\xi \,
 \Phi_0 (y,\xi/\alpha_2) \,   
       e^{ i \alpha_2  ( k-y p)^2
  } 
 \,  .    \nonumber
\end{align} 
This gives (switching $1/\alpha = \sigma$) 
  \begin{align}
 \Phi^{B_0}_{1}  (x,\sigma ) 
=&    \alpha_g  \, 
\int_{x}^1 dy\,  
  \int_0^1 d\xi \,
 \Phi_0 \left (y, \frac{\xi \sigma}{V_1(x,y) }  \right ) \  ,
\label{PhiBmomf2}
\end{align} 
where $V_1(x,y)  =(x/y) \,  \theta (x\leq y)$  and $\alpha_g= g^2/(16\pi^2)$.
The second term in Eq. (\ref{eq:spi2}) gives a similar contribution, with 
$V_1(x,y) \to V_2(x,y) \equiv (\bar x /\bar y) \theta (y\leq x)$. 
  As a result,  the  total contribution for 
 hard TMDA      generated  in case of   spinor quarks  is  given by 
     \begin{align}
  \Psi^{B_0}_{1/2}   (x, k_\perp) =   & \frac{  \alpha_g }{ \pi } \,  
   \int_0^1 dy     \int_0^1 d\xi \,  
     \psi_0 \left  (y, \frac{\xi  k_\perp^2}{ V_0(x,y)} \right )  \ . 
     \label{tailspi}
  \end{align} 
It  differs from the scalar  expression  (\ref{psi3}) just by the absence of the overall 
$1/k_\perp^2$ factor.  One may also   write
       \begin{align}
  \Psi^{B_0}_{1/2}   (x, k_\perp) =   & \frac{  \alpha_g }{ \pi  k_\perp^2} \,  
   \int_0^1 dy \,  { V_0(x,y)}
      \nonumber \\ &  \times 
 \int_0^{k_\perp^2/V_0(x,y) }
     \psi_0 (y, { k'_\perp}^2   )  \, d  {k'_\perp}^2 
     \label{tailspi2B}
  \end{align} 
which corresponds to $1/(k_\perp^2+m^2)$ dependence in the correction (\ref{psi3})   to 
the TMDA. It produces 
 a logarithmically divergent term for DA, corresponding to a logarithmic 
evolution of DA's in such models.

For a soft TMDA $\psi_0 (x, k_\perp^2)$, the integral over all $k_\perp^2$ produces 
the initial distribution amplitude $\varphi_0 (x)$, so we have
       \begin{align}
  \Psi^{B_0}_{1/2}   (x, k_\perp) =   & \frac{  \alpha_g }{ \pi  k_\perp^2} \,  
   \int_0^1 dy \,  { V_0(x,y)}\,  
      \left [ \varphi_0 (y)   - \delta \varphi  \right ]
           \label{tailspi21}
  \end{align} 
where the correction term 
       \begin{align}
\delta \varphi    = 
 \int_{k_\perp^2/V_0(x,y)}^\infty 
     \psi_0 (y, { k'_\perp}^2   )  \, d  {k'_\perp}^2 \ , 
     \label{tailspi3}
  \end{align} 
  quickly vanishes for large $k_\perp^2$.  For an illustration, take 
a factorized Ansatz for the initial TMDA, \mbox{$ \psi_0 (x, k_\perp^2   )  = \varphi_0 (x) \, K(k_\perp^2)$}.
Then 
we have 
      \begin{align}
 \delta \varphi    =     \varphi_0 (y) \int_{k_\perp^2/V_0(x,y)}^\infty 
    K ({ k'_\perp}^2   )  \, d  {k'_\perp}^2\ , 
     \label{tailspi4}
  \end{align} 
  which gives 
      \begin{align}
  \delta \varphi  =     \varphi_0 (y) 
   \,  e^{-k_\perp^2/V_0 (x,y) \Lambda^2} 
     \label{tailspi2G}
  \end{align} 
for a  Gaussian  form $K( k_\perp^2) = 
e^{-k_\perp^2/\Lambda^2}/\Lambda^2$, and 
      \begin{align}
   \delta \varphi  =   \varphi_0 (y)  \,  \frac{2k_\perp K_1 (2 {k_\perp}/{ \Lambda \sqrt{V_0 (x,y)}})
}{ \Lambda \sqrt{V_0 (x,y)}} 
     \label{tailspi2m0}
  \end{align} 
for   the $m=0$  model  when  \mbox{$K ( k_\perp^2)=
2\, {K_0 ( 2 k_\perp / \Lambda) }/{ \Lambda^2 } $}.

 \subsubsection{Evolution in the impact parameter space}

 For initially collinear quarks,
 we have  $\varphi^{\rm conv}  (x, z_\perp)  \sim     K_0 (mz_\perp) \, \delta \varphi (x)$
in the impact parameter space. The  logarithmic divergence for $z_\perp=0$ of this  outcome 
 corresponds to  evolution of the DA.
 In the $B_0$ model, we have  (switching  to $\varphi   (x, z_\perp)   \to \varphi   (x, z_\perp^2)$ in our 
 notations below)    
   \begin{align}
  \varphi^{B_0}   (x, z_\perp^2) =   &   \alpha_g 
   \int_0^1 dy   \,V_0 (x,y) 
        \nonumber \\ &  \times 
   \int_1^\infty  \frac{d \nu}{\nu} \, \varphi_0 
   \Bigl (y, \nu \, z_\perp^2 \,V_0(x,y) \Bigr ) 
     \ . 
     \label{tailY}
  \end{align} 
Substituting  formally 
 $ \varphi_0 (y, z_\perp^2)$ by $\varphi_0 (y)$  
 in   the \mbox{$z_\perp^2\to 0$}  limit, 
we get a   logarithmically divergent integral over $\nu$.  
 However, for a function $ \varphi_0 (y, z_\perp^2)$ that  rapidly decreases 
 when  \mbox{$z_\perp^2 \gtrsim 1/\Lambda^2$,}  
one gets $\ln (z_\perp^2 \Lambda^2)$ as a factor accompanying 
 the convolution of $V_0(x,y)$ and $\varphi_0 (y)$. 
 Hence, the pion  size  cut-off contained in the primordial 
 distribution provides the scale in  $\log (z_\perp^2)$,
 and we may keep the hard quark propagators massless.
 
        \begin{figure}[h]
\centerline{\includegraphics[width=3.1in]{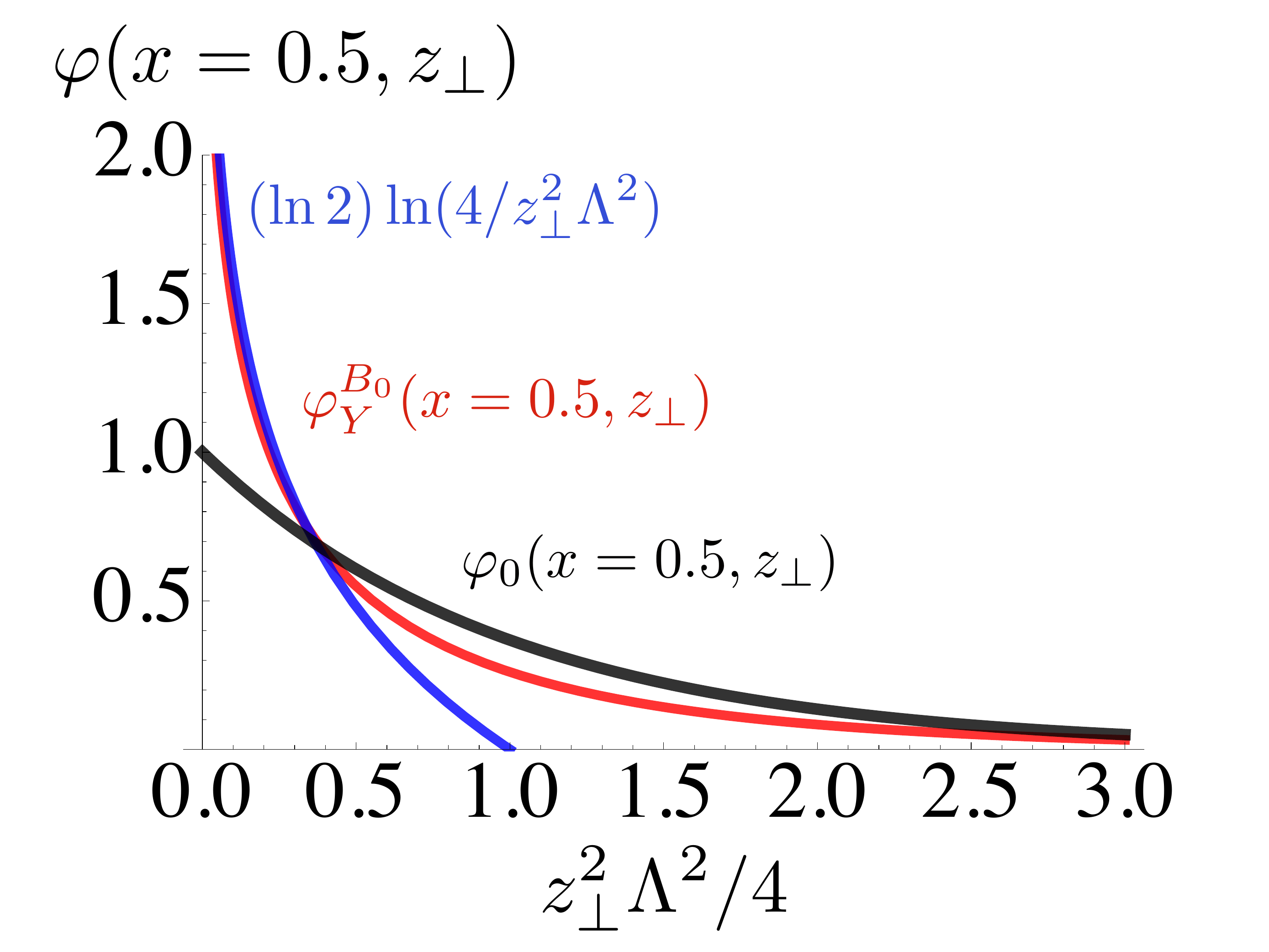} }
   \caption{
Illustration for Gaussian model with flat DA 
$\varphi_0 (x,z_\perp) =\exp [-z_\perp^2 \Lambda^2/4] $ 
   \label{z2_evol}}
   \end{figure}

    \begin{figure}[t]
\centerline{ \includegraphics[width=3.1in]{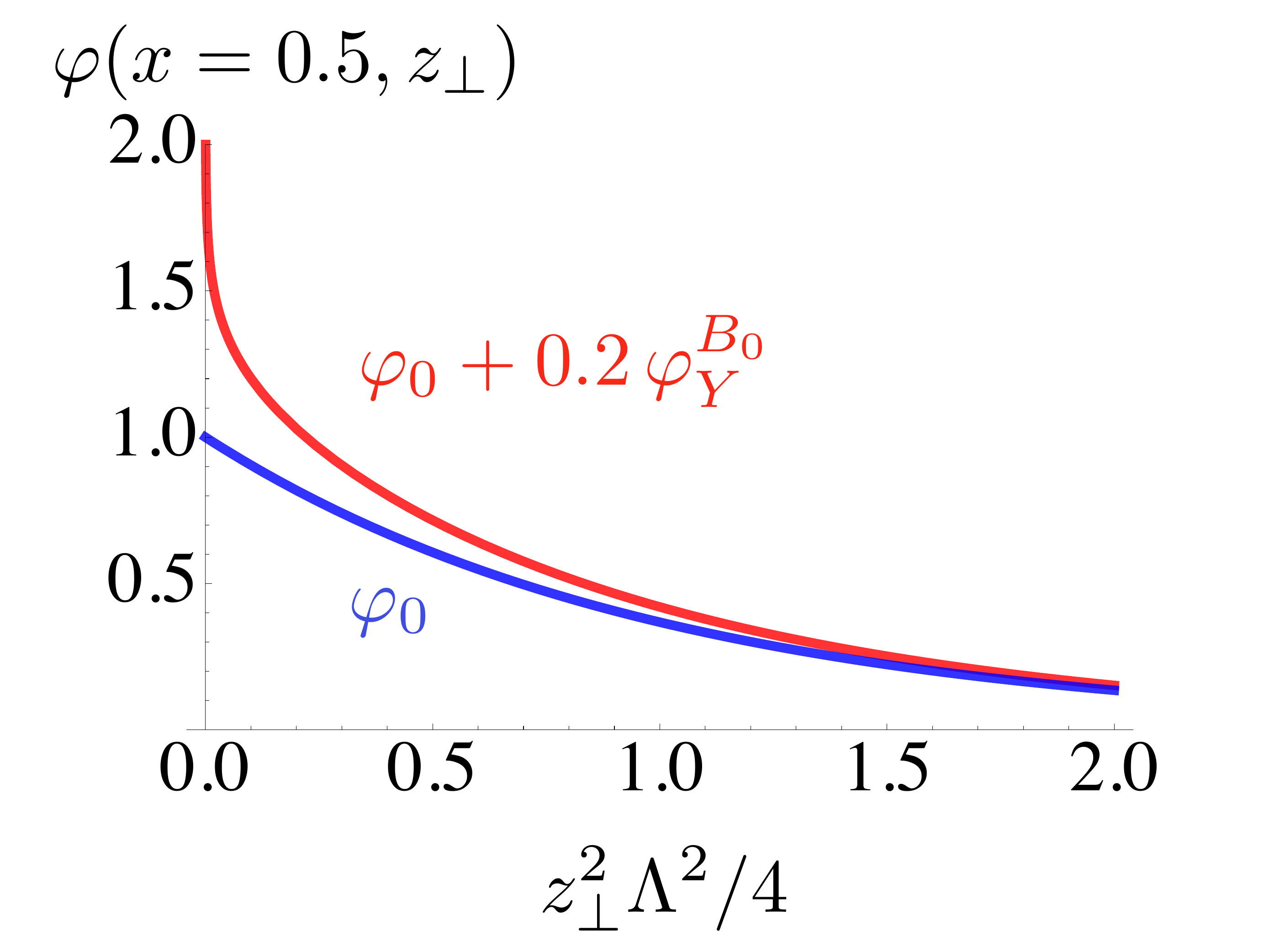} }
   \caption{
Illustration for Gaussian model with flat DA 
$\varphi_0 (x,z_\perp) =\exp [-z_\perp^2 \Lambda^2/4] $ 
   \label{total_z2}}
   \end{figure}

 For scalar gluons, this cut-off also results in a finite value of 
 $ \Psi^{B_0}_Y  $ in the  formal $k_\perp \to 0$ limit:
   \begin{align}
  \Psi^{B_0}_Y   (x, k_\perp=0) =   &\alpha_g   \, 
   \int_0^1 dy       \, 
     \Psi_0 \left  (y,  k_\perp=0 \right )   
     \ . 
     \label{small_k}
  \end{align} 
  Thus, the  \mbox{$\Psi^{\rm conv}_Y  (x, k_\perp) \sim 
1/k_\perp^2$} singularity of the
``collinear model''
\mbox{$\Psi_0 (y, k_\perp) = \varphi_0 (y) \, \delta (k_\perp^2)/\pi$}  
converts  
into a  constant $\sim 1/\Lambda^2  $  in the   Gaussian model. 
Note also that 
 the overall 
 factor in  \mbox{Eq. (\ref{small_k})}  then 
contains  the $x$-independent integral of $\varphi_0 (y)$, i.e. 
$f_\pi$, rather than the  convolution $\delta \varphi (x) $ 
as 
in Eq. (\ref{psi3}). 

It should be emphasized  that 
the VDA  approach provides an  unambiguous prescription      of  
  { generating}
hard-tail terms like $\Psi^{B_0}(x, k_\perp)$    from a soft
primordial distribution 
$\Psi_0 (y, k_\perp)$.     
 
   \subsubsection{Particular choices of the soft $B_0$}

Assuming  a factorized Ansatz   for the initial IDA  $ \varphi_0 (y, z_\perp^2) = \varphi_0 (y) \, Z_0( z_\perp^2 \Lambda^2) $,
we have 
    \begin{align}
  \varphi^{B_0}   (x, z_\perp^2) =     & \alpha_g 
   \int_0^1 dy   \,V_0 (x,y) \, \varphi_0 (y) Z_1 [ \zeta^2 \, V_0(x,y) ]
       \label{tailYZ}
  \end{align} 
   where 
     \begin{align}
       Z_1 (   \zeta^2) \equiv 
   \int_1^\infty  \frac{d \nu}{\nu} \, 
Z_0 ( \nu  \zeta^2 )  \ ,
     \label{tailYZ1}
  \end{align} 
and  we can study the sensitivity of   $\varphi^{B_0}   (x, z_\perp^2) $ to a particular choice 
of the soft factor $Z_0( \zeta^2)$. In particular, in a Gaussian model, 
$Z_{0}^{ \rm exp} ( \zeta^2)= e^{-\zeta^2}$,
we have 
    \begin{align}
      Z_1^{ \rm exp}  (   \zeta^2)  =  \Gamma[0, \zeta^2 ] 
 =
 -\ln (\zeta^2 ) - \gamma_E + {\cal O} (\zeta^2) 
     \ ,
     \label{tailYZexp}
  \end{align} 
where we have explicitly displayed the small-$\zeta^2$ behavior to 
extract the $\sim \ln z_\perp^2$ evolution term, which gives 
    \begin{align}
   \varphi^{B_0}_{\rm exp}  &  (x, z_\perp^2)  =      \alpha_g  \Biggl [
   \ln \left (\frac{e^{-\gamma_E}}{z_\perp^2 \Lambda^2_{\rm exp} } \right ) 
   \int_0^1 dy   \,V_0(x,y) \, \varphi_0 (y) 
       \nonumber \\ & + \int_0^1 dy   \,V_0(x,y) \,    \ln (  V_0(x,y))  \,  \varphi_0 (y) 
  \Biggr ]  + {\cal O} (z_\perp^2 \Lambda^2_{\rm exp} ) 
 \  .
     \label{tailYZexp2}
  \end{align} 
Similarly, for    a slow-decrease Ansatz \mbox{$Z_{\rm slow} ( \zeta^2)= 1/(1+\zeta^2)$,} 
 we have 
    \begin{align}
Z_1^{\rm slow} ( \zeta^2)=\ln (1/\zeta^2+1) =
 -\ln \zeta^2 + {\cal O} (\zeta^2)   \  , 
    \label{tailYZ1slow}
   \end{align}
 so that the major  change for small $z_\perp^2$ is the absence of the $\gamma_E$ term,
  which amounts to a change of  the evolution scale.
  The evolution terms of the two Ans\"atze coincide when $\Lambda^2_{\rm slow} =
  e^{\gamma_E} \Lambda^2_{\rm exp}$.

In fact,   
 the approximate $\ln z_\perp^2 \Lambda^2$ form for the evolution factor  is  inconvenient,
because it  changes sign for \mbox{$z_\perp = 1/\Lambda $}  and 
tends to infinity for large $ z_\perp^2$,  while 
the original expressions for $Z_1 (\zeta^2)$ are positive-definite and  vanish in the
$\zeta \to \infty$  limit.
To avoid this artifact of the small-$\zeta^2$ expansion for $Z_1 (\zeta^2)$,
one may represent, e.g., 
$$Z_1 [\zeta^2V_0 (x,y)] = Z_1 (\zeta^2) + [Z_1 (\zeta^2V_0 (x,y))-Z_1 (\zeta^2)] \  .$$
  Then $ Z_1 (\zeta^2) $ gives a positive-definite  evolution factor with a 
  correct $- \ln (\zeta^2)$ 
  behavior for small $\zeta^2$ and vanishes for large $\zeta^2$. The remainder is 
  finite for $\zeta^2=0$ and also vanishes for large $\zeta^2$.
  Since the logarithmic  $ \ln (\zeta^2)$  part of all $Z_1 (\zeta^2)$'s is universal,
  while the rest  depends on the shape of $Z_0 (\zeta^2)$, 
 it make sense to  declare  one of  $Z_1 (\zeta^2)$'s to be a standard  one,
 i.e. to represent 
 $$Z_1 [\zeta^2V_0 (x,y)] = Z_1^{\rm stand} (\zeta^2) + 
 [Z_1 (\zeta^2V_0 (x,y))-Z_1^{\rm stand}  (\zeta^2)] \  , $$
 thus making the evolution part universal.  Possible choices for $ Z_1^{\rm stand} (\zeta^2) $
are  $ Z_1^{\rm exp} (\zeta^2) = \Gamma [0, \zeta^2]$, and 
 $ Z_1^{\rm slow} (\zeta^2) = \ln [1/\zeta^2 +1]$.  

  \subsubsection{Ultraviolet-related  addition to hard  tail }
  
  Since the axial current is conserved, the full evolution kernel for the pion DA 
  should have  a plus-prescription form with respect to $x$:
  $V_0(x,y) \to [V_0(x,y)]_+$.  In fact, 
  calculating the $\gamma^* \gamma \to \pi^0$ amplitude at $\alpha_g$ order,
  we should include 
the one-loop vertex and 
  self-energy corrections to the hard quark propagator $S^c (z)$. They produce, in particular,
 a   factor of $\ln (z^2 )$ multiplying   $S^c (z)$.
  We may   treat it as  a multiplicative 
  modification of VDA convoluted with the original 
  propagator $S^c (z)$. This corresponds to adding 
  the $\sim \alpha_g \ln (z_\perp^2) \, \varphi_0 (x, z_\perp^2)$
  term 
  to   $\varphi^{B_0}_Y   (x, z_\perp^2)$ as another 
  ${\cal O} (\alpha_g)$ correction to IDA.

  Since this term comes from an ultraviolet divergent
  contribution, we need to  decide which  UV renormalization prescription to use. 
  While the $\sim \ln (z_\perp^2 )$ behavior for small $z_\perp$ is   not  affected 
  by the choice, the  large $z_\perp$ behavior depends on it and  may be adjusted.
  It is convenient   to  take the Bessel function form 
$K_0 (z_\perp \mu)$. Then the correction  vanishes for large 
$z_\perp$, and never changes sign. 
Furthermore, its Fourier transform 
has a simple  $1/(k_\perp^2 + \mu^2)$  form that is finite for 
$k_\perp=0$. Incorporating these considerations we fix the UV-related  correction for  IDA 
to be given by 
   \begin{align}
  \varphi^{B_0,{\rm UV}}_Y   (x, z_\perp^2) =   &-  \alpha_g K_0 (z_\perp \mu) \, 
 \varphi_0 
  (x, \,  z_\perp^2) 
     \ . 
     \label{seY}
  \end{align}

For TMDA, this term gives  
 \begin{align}
  \Psi^{B_0,{\rm UV}}   (x, k_\perp;\mu) =   & -\frac{\alpha_g}{2 \pi} \, 
  \int d^2k'_\perp   \frac{\Psi_0 (x,k'_\perp) }{ (k_\perp-k'_\perp)^2+\mu^2} \,   \  .
     \label{psiSEk}
  \end{align}
    In actual calculations, it is convenient to use the formula
      \begin{align}
  \psi^{B_0,{\rm UV}}   & (x, k^2_\perp;\mu) =    -\frac{\alpha_g}{2} \, 
   \int_0^1 \frac{d\xi}{1-\xi}     \,    
        \nonumber \\ &  \times  
     \psi_0 \Bigl (x, \xi (  k_\perp^2+\mu^2/(1-\xi)) \Bigr ) 
     \ . 
     \label{psiSE}
  \end{align} 
  The leading large-$k_\perp$ term then comes from the $\xi \sim 1/ k_\perp^2$ region of integration.

\begin{figure}[h]
\centerline{\includegraphics[width=2.9in]{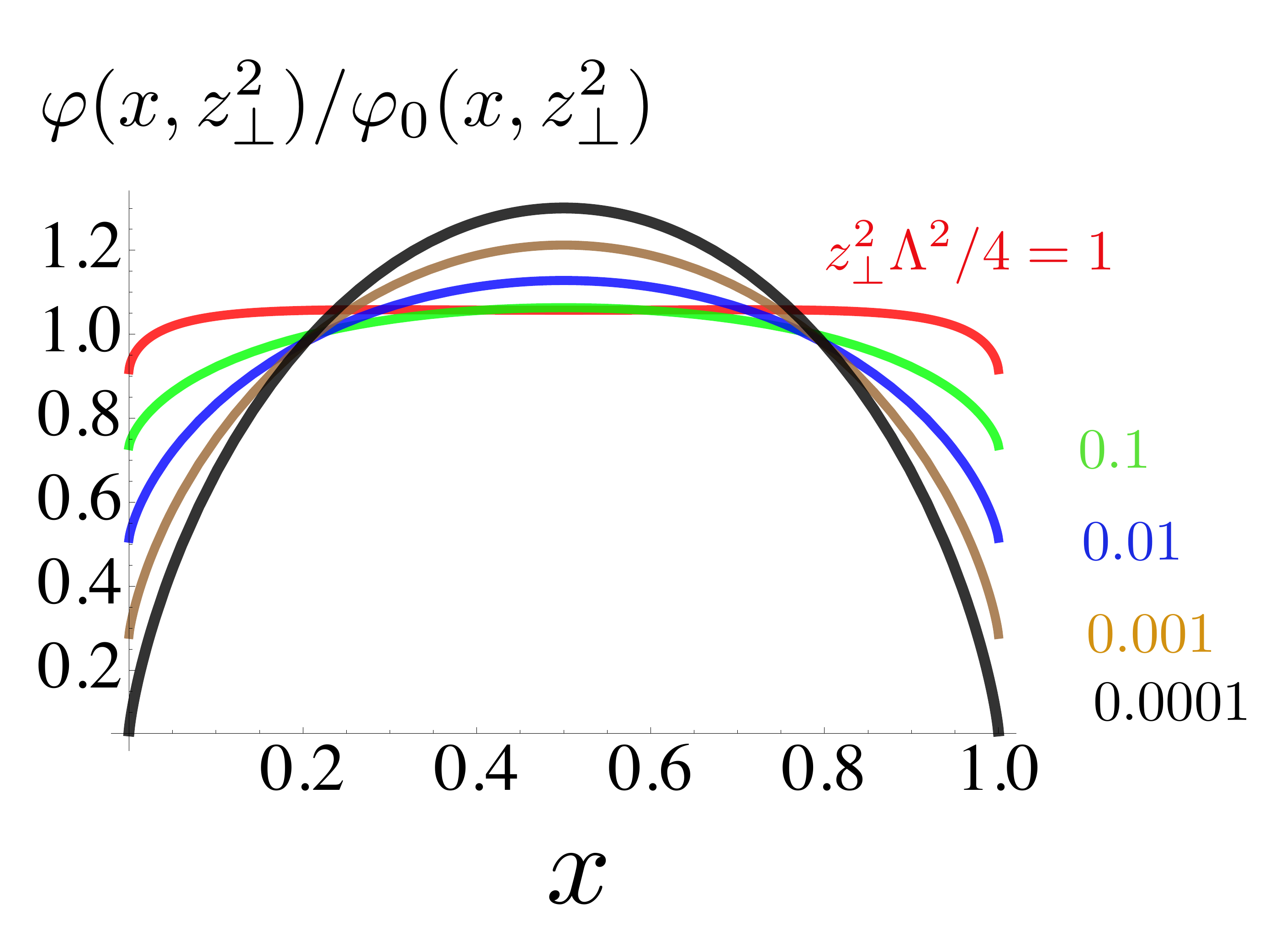} }
   \caption{
 Illustration for Gaussian model with flat DA 
 $\varphi_0 (x,z_\perp^2) =e^{-z_\perp^2/4\Lambda^2}$ and 
$ \alpha_g =0.2$ 
   \label{evol_flat}}
   \end{figure}

Adding (\ref{tailY}) and  self-energy part 
(with $\mu =\Lambda/2$ and Bessel form for  log singularity) gives
   \begin{align}\
   \label{YuHard}
 & \delta \varphi_Y   (x, z_\perp^2) =    \alpha_g 
  \Biggl [  \int_0^1 dy   \,V_0(x,y) 
   \int_1^\infty  \frac{d \nu}{\nu} \, 
    \\ & \times \varphi_0 
   \Bigl (y, \nu \, z_\perp^2 \,V_0(x,y) \Bigr ) 
- K_0(z_\perp \Lambda/2)\,
    \varphi_0   (x, z_\perp^2)  \Biggr ] \ .  
     \nonumber 
  \end{align} 
  In Fig. \ref{evol_flat} we show 
  total IDA $ \varphi   (x, z_\perp^2) = \varphi_0   (x, z_\perp^2)
+  \delta \varphi_Y   (x, z_\perp^2)$ taking for the soft 
IDA a Gaussian model 
for the $z_\perp$-dependence and flat  DA $\varphi_0 (x) =1$.
One can see the change of the  $x$-profile  from 
a flat form for large $z_\perp$ to asymptotic $\sim x\bar x$ 
for small $z_\perp$.

 If the  function $\Psi_0 (x, k'_\perp)$ rapidly decreases with  growing  ${k'}_\perp^2$,
 then the leading contribution for large $ k_\perp^2$ is obtained from 
 the region of small $k'_\perp$ which gives 
   \begin{align}
  \psi^{B_0,{\rm UV}}   (x, k_\perp;\mu) =   & -\frac{\alpha_g}{2} \, 
    \frac{\varphi_0 (x) }{ k_\perp^2} + \ldots     
     \label{psiSEas}
  \end{align} 
for large $ k_\perp^2$. In the formal $k_\perp \to 0$ limit, we have a finite result 
   \begin{align}
  \psi^{B_0,{\rm UV}}   (x, k_\perp^2=0;\mu) =   & -\frac{\alpha_g}{2} \, 
  \int_0^\infty   \frac{\psi_0 (x,{k'}_\perp^2) }{ {k'}_\perp^2+\mu^2} \, d ({k'}_\perp^2)     \  .
     \label{psiSEsmall}
  \end{align}

   \subsubsection{Hard tail contribution to the  transition amplitude}
   
   The integral giving  the transition form factor
  \begin{align}
 F (Q^2) 
= &  
   \int_{0}^1   \frac{dx}{xQ^2}  
       \int_{{k_\perp}^2 \leq  xQ^2}   \Psi (x,  k_\perp   ) 
       \left [ 1- \frac{k_\perp^2} { xQ^2} \right ]     
 \, d^2  k_\perp
 \label{Fspinor40aa}
 \end{align}
 in case of the hard exchange contribution may be written as
  \begin{align}
 F^{B_0} (Q^2) 
= &      \alpha_g  
   \int_{0}^1   \frac{dx}{xQ^2}   \int_0^1 dy   
       \int_0^{ xQ^2} \, {  d  k_\perp ^2   }   \left [ 1- \frac{k_\perp^2} { xQ^2} \right ] \,
  \nonumber \\ &\times  \int_0^1 d \xi  \,        
     \psi_0 \left  (y, \frac{\xi  k_\perp^2}{ V_0(x,y)} \right )  
 \label{Fspinor40ab}
 \end{align}
 or
   \begin{align}
 F^{B_0} (Q^2) 
= &      \alpha_g  
   \int_{0}^1   {dx}  \int_0^1 dy   
       \int_0^{ 1} \, (1-\kappa) \,  {  d  \kappa   } \,
   \nonumber \\ &\times
   \int_0^1 d \xi       
     \psi_0 \left  (y, \frac{\xi  \kappa xQ^2}{ V_0(x,y)} \right )  \ . 
 \label{Fspinor40ac}
 \end{align}
 Denoting $\xi \kappa \equiv \lambda$, we have 
  \begin{align}
 F^{B_0} (Q^2) 
= &      \alpha_g  
   \int_{0}^1   {dx}  \int_0^1 dy   
\,
   \int_0^1 d \lambda  \,  [ \ln (1/\lambda) -1 +\lambda]\,      
         \,
  \nonumber \\ &\times 
     \psi_0 \left  (y, \frac{\lambda xQ^2}{ V_0(x,y)} \right )  \ .
 \label{Fspinor40ad}
 \end{align}
 For large $Q^2$, one can separate here  the 
terms with a power-like  behavior  of  $\ln(Q^2)/Q^2, 1/Q^2$ and $ \Lambda^2/Q^4$
type, 
 and those which have (for a soft $\psi_0$) 
 a faster than an inverse power decrease
 with the increase of $Q^2$.

  \subsection{Hard tail in QCD}
  
\subsubsection{Yukava-type contributions}

  In quantum chromodynamics, working in Feynman gauge,
  the only change in the box diagram and ultraviolet-divergent 
  terms is in the overall factor, namely, one should take
  $\alpha_g = C_F \alpha_s/(2 \pi)$ in Eq. (\ref{YuHard}), with $C_F$ 
  being the color factor.  In particular the $1/k_\perp^2$  hard tail 
  generated  by these ``Yukawa-type'' contributions is accompanied by the
   \begin{align}
V_Y (x, y) = & \frac{\alpha_s}{2 \pi} C_F 
\left [ \frac{x}{y} \theta(x<y) +  \frac{ \bar x}{\bar y }  \theta(x>y) -\frac12 \delta (x-y) \right ]
\nonumber \\ & = \frac{\alpha_s}{2 \pi} C_F [V_0 (x,y)]_+ 
\label{ERBLY}
\end{align} 
  part of the ERBL evolution kernel, 
  with ``+'' denoting the plus-prescription
  with respect to $x$:
     \begin{align}
[V_0 (x, y)]_+ = V_0 (x,y) - \delta (x-y) \int_0^1 dz \, V_0 (z,y) \  . 
\label{plus}
\end{align} 
  For vector gluons, one should also take into account contributions
coming from the gauge link $E(0,z;A)$, which generate 
the remaining part of the QCD ERBL evolution kernel. 

\subsubsection{Link contributions in case of 
collinear initial quarks} 

\begin{figure}[h]
   \centerline{\includegraphics[width=3in]{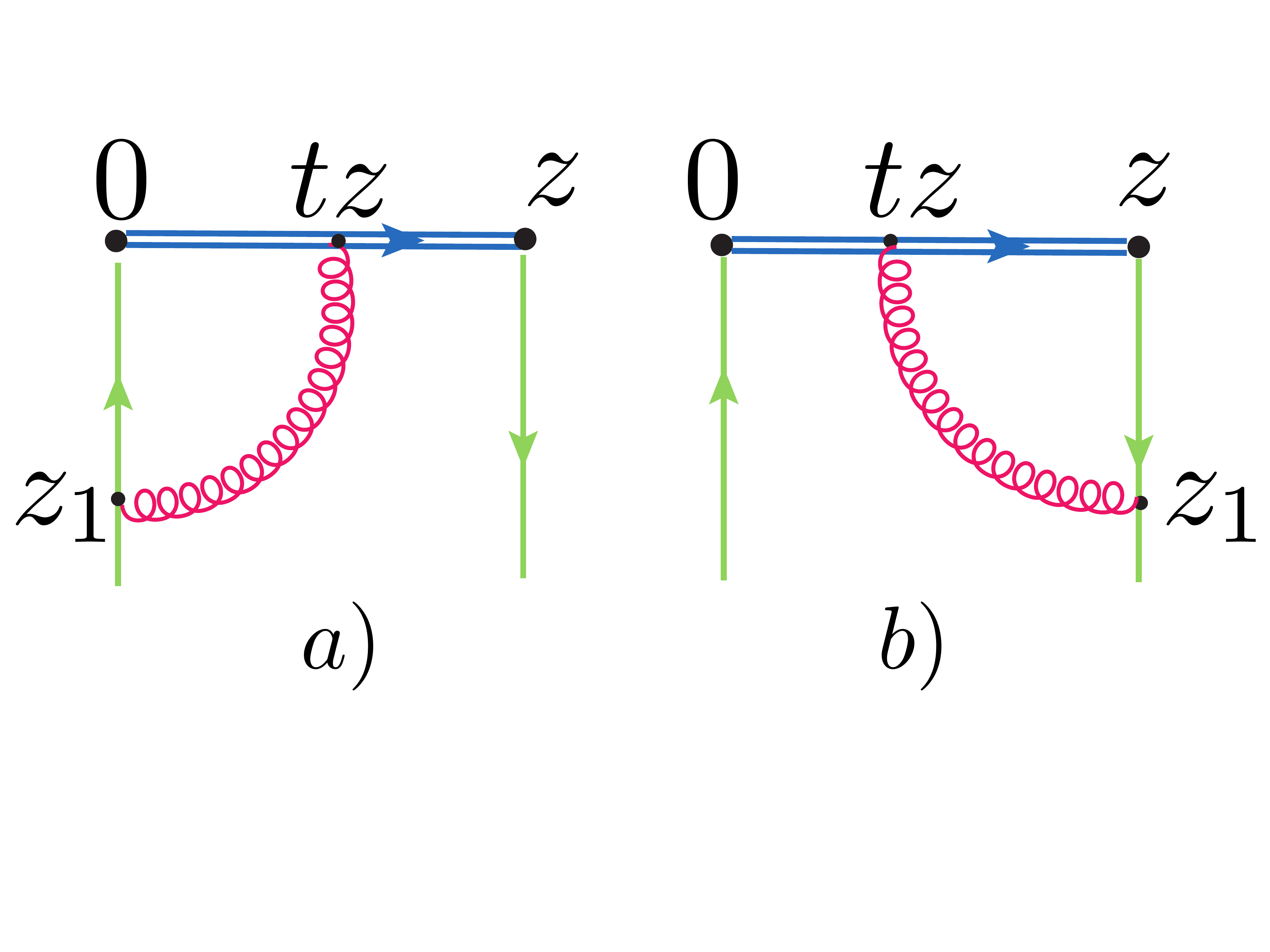}}
        \vspace{-10mm}
   \caption{Insertions of gluons coming out of the gauge link.
   \label{link}}
   \end{figure}

 Let us start with a collinear 
intial state, i.e. take $\Phi_0^{\rm coll} (y, \sigma)= \varphi_0 (y)\, \delta (\sigma)$.
There are two possibilities: gluon may be connected to $yp$ 
(Fig. \ref{link}a) or $(1-y)p$ (Fig. \ref{link}b) 
quark leg. 
Insertion into the $yp$ leg 
 produces  the term
 \begin{align} 
  t_L  (z,p,y) = &  {i} g^2\,C_F \, e^{i \bar y(pz)} \int_0^1 dt \, 
 \int   d^4z_1   \, e^{i y(pz_1)} 
  \nonumber \\ &  \times    S^c(z_1) \slashed z    
  D^c (z_1-tz)
    \label{gauge1}
 \end{align}
If the gluon  is inserted into the $\bar y p$ line, we start with 
  \begin{align} 
 t_R (z,p,y) = & {i}   g^2\,C_F  \int_0^1 dt \, 
  \int   d^4z_1  \, e^{i \bar y(pz_1)} 
   \nonumber \\ &  \times
  \slashed z     S^c(z-z_1) 
  D^c (z_1-tz) \, 
  \  .
    \label{gauge1Pa}
 \end{align}
 Shifting $z_1 \to z_1 +z$, changing $z_1 \to -z_1$,    $ t \to 1-t$ and using that $D^c (z)$ is an even function gives 
    \begin{align} 
t_R (z,p,y) = & i g^2\, C_F   e^{i \bar y(pz)}  \int_0^1 dt \, 
  \int   d^4z_1  \, e^{-i \bar y(pz_1)} 
   \nonumber \\ &  \times
   \slashed z     S^c(z_1) 
  D^c (z_1-tz) \, 
  \  .
    \label{gauge1Pa}
 \end{align}
Thus,  these two cases are  related by symmetry, and  it is sufficient to 
analyze just the first of them, Eq. (\ref{gauge1}). 
 Integrating there  over  $z_1$   gives
 \begin{align} 
  t_L (z,p,y) = &i \frac{ g^2}{16 \pi^2} \, C_F e^{i \bar y(pz)} \int_0^1 dt \, 
 \int_0^\infty \sigma_1 \frac{d\sigma_1 d \sigma_2 }{(\sigma_1 + \sigma_2)^3}
   \nonumber \\ &  \times e^{i (yt (p z)  \sigma_2 -
    \sigma_1  \sigma_2 t^2 z^2/4 )/(\sigma_1 + \sigma_2)} 
  \nonumber \\ &  \times  
    \left  (y\slashed p \slashed z 
   + t \sigma_2 \frac{z^2}{2}  \right )   \ .
    \label{gauge2}
 \end{align}
Switching to $\alpha_i=1/\sigma_i$,   introducing common 
\mbox{$\lambda=\alpha_1+\alpha_2$,}  and then relabelling $\lambda=1/\sigma$,
we obtain 
  \begin{align} 
 t_L & (z,p,y) = i \frac{ g^2}{16 \pi^2} \, C_F \,e^{i \bar y(pz)} \int_0^1 dt \, 
 \int_0^\infty \frac{d\sigma}{\sigma} 
   \nonumber \\ &  \times  \int_0^1 d \beta \, e^{i yt \beta (p z)}  e^{-  i \sigma
    t^2 z^2/4 } 
\left  (y \bar  \beta \slashed p \slashed z 
   + t \sigma  \frac{z^2}{2 } \right )   
    \label{gauge3fins}
 \end{align}
 
Representing $ \slashed p \slashed z = (pz)+   [\slashed p, \slashed z] /2$
and using this result in the expression for transition amplitude we end up with 
 the operator 
\mbox{$\sim \epsilon_{\mu \alpha \beta \nu}  p^\alpha z^\beta  \bar \psi (0)  \gamma^\nu  \ldots  \psi (z)$ }  whose $\langle p | \ldots |0 \rangle$ matrix element  vanishes.
 
The  second term  in Eq. (\ref{gauge3fins}) does not contribute to the hard tail 
since it lacks $z^2$-dependence after $\sigma$-integration. 
 Thus, only the $(pz)$-part of the 
 first term  contributes to the hard tail, which in the coordinate 
 representation is reflected by
 a  logarithmic $\ln z^2$ singularity  resulting from the $\sigma $ integration.
 As discussed earlier,  $\ln z^2$ 
 reflects  the DA evolution.

First,  we are going to get  rid of  the
integration over
            $t $   specific  for the vector gluons.  
            Its calculation is  complicated by the $e^{-  i \sigma t^2 z^2/4} $ factor.
            Let us represent
           \begin{align} 
e^{-  i \sigma  t^2 z^2/4} = e^{-  i \sigma z^2/4}
+ \left [ e^{-  i \sigma  t^2 z^2/4} -  e^{-  i \sigma  z^2/4} \right ] \  .
    \label{gauge4aa}
\end{align} 
The bracketed term here formally vanishes for $z^2=0$,   which means a suppression for
small $z^2$.  As a result,  it  does not  contain $\ln z^2$ terms and hence does not 
contribute to the  hard tail. 
Integrating over $t$ in the part corresponding to the first term, we get 
  \begin{align} 
 t^{(1)}  (z,p,y) =   & \gamma_\mu \, \alpha_g\,  e^{i \bar y(pz)}
 \int_0^\infty    \frac{d \sigma }{ \sigma }\,  e^{ - i \sigma   z^2/4}
\nonumber \\ &  \times
  \int_0^1 d \beta \,  \frac{ \bar  \beta} {\beta} 
\left [ e^{i y \beta (p z)}  - 1 
    \right ]   \  ,
    \label{gauge4a}
 \end{align}
 where $\alpha_g =  C_F \alpha_s/(2 \pi)$.
   This  contribution    generates the evolution corresponding to 
 the $x \leq y$ part of the QCD ERBL  kernel:  
 \begin{align} 
  t^{(1)}   (z,p,y) = &  \gamma_\mu \alpha_g\,   
   \ln (z^2) \int_0^1 dx \, e^{i \bar x(pz)}
V_1 (x,y)
  \ ,
     \label{gauge6a}
  \end{align}
where
 \begin{align} 
V_1 (x,y)=& \int_0^1 d\beta \, \delta (x -y + \beta y) \left ( \frac{\bar \beta}{\beta} \right )_+
    \nonumber \\ &
     = \left ( \frac{ x}{y}\, \frac{\theta (x \leq y)} {y-x} \right )_+
  \ .
     \label{V1}
  \end{align}
   Similarly, attaching the gluon to the $\bar y p$ quark line,
  we get the $x\geq y$ part 
   \begin{align} 
V_2 (x,y) = & \int_0^1 d\beta \, \delta (x -y - \beta \bar 
y) \left ( \frac{\bar \beta}{\beta} \right )_+
    \nonumber \\ &
     = 
     \left ( \frac{ \bar x}{\bar y}\, \frac{\theta (x \geq y)} {x-y} \right )_+
     \label{V2}
  \end{align}
 of the ERBL evolution kernel.
  Thus, in the convolution model we have an evolution  contribution
   \begin{align} 
    \delta \varphi^{(1)}   (x,z^2 ) = & \alpha_g\,   \ln (z^2) \int_0^1 dy \,
  V (x,y) \varphi_0 (y) \ , 
       \label{gauge7a}
    \end{align}
  with the ERBL kernel in its correct ``plus prescription'' form.

\subsubsection
{Noncollinear initial quarks}

Let us now consider the case when we have an initiall soft distribution
described by a  VDA $\Phi_0 (y, \sigma_0)$. 
Again, it is sufficient to consider  the diagram corresponding 
to the gluon insertion into the $yp$ leg. 
The contribution of  the $\bar y p$  diagram  can  be added at the end
using the symmetry considerations.
.

 \paragraph{Derivation of a general formula for VDA.}
 The result of integration over  $z_1$ has the structure similar to that obtained in the collinear quarks case (\ref{gauge2}).   Keeping again        the $\sim (pz)$ term only, 
and  representing the $z^2$-dependent factor 
 as a sum of its $t=1$ value and the rest (which formally  vanishes for $z^2=0$,
 and thus does not contribute to the hard tail), and  
integrating over $t$  in this (``hard'') part, we get 
  \begin{align} 
&  t^{\rm (h)}  (z,p) =  g^2\,    \int_0^1 dy  \, e^{i \bar y(pz)} 
 \int_0^\infty    d\sigma  \,  e^{- i \sigma z^2  /4 } \,  \int_0^1   \frac{d \xi}{\bar \xi}
    \nonumber \\ &  \times  
 \,
\int_0^1   \frac{d \beta}{\beta}
  \left ( e^{  i y  \beta  (p z)} -  e^{ i y  \xi \beta (p z)}  \right )  
 \Phi_0 (y,  \xi \sigma / \bar \beta)  \ .
    \label{gauge6P}
 \end{align}
 In terms of VDA,  we have
   \begin{align} 
&  \Phi^{\rm (h)}  (x, \sigma )  =   \alpha_g \,   \int_0^1 dy 
 \, \int_0^1   \frac{d \xi}{\bar \xi}
\int_0^1   \frac{d \beta}{\beta}  \, \Phi_0 (y,  \xi \sigma / \bar \beta)
  \nonumber \\ &  \times
  \left [ \delta ( x-y +  y  \beta ) -    \delta ( x-y +  y \xi  \beta ) \right ] 
   \ .
    \label{gauge7P}
 \end{align}
 One can see that
    \begin{align} 
\int_0^1 dx \,   \Phi^{\rm h}   (x, \sigma )  =   0 \ , 
    \label{gauge7Pb}
 \end{align}
 i.e. the  hard addition $ \Phi^{\rm h}   (x, \sigma )$ does not 
 change the $x$-integral  of $\Phi (x,\sigma)$.

\paragraph{Structure of hard TMDA.} 
The integrand  in Eq. (\ref{gauge7P}) contains 
$1/\beta$ and $1/ \bar \xi$ factors resulting in potential singularities
for $\beta=0$ and  $\xi=1$. However, the two $\delta$-functions
 in Eq. (\ref{gauge7P})   coincide in these limits, 
 and as a result  there are no divergencies.
Still,  the most 
natural way to proceed with  integrations  in Eq. (\ref{gauge7P})
is to use  the $\delta$-functions 
to eliminate the   integral over
  $\beta$.  
Then the common $1/\beta$ factor in both terms 
  results in contributions proportional to the  factor $1/(y-x)$
  singular for $y=x$.  Of course, the singularity cancels 
  because it is accompanied by a difference of two VDAs
  coinciding for $y=x$. However,
  each term    produces  a divergent contribution. 
To  be able to  keep  these terms     separately,
  we choose to regularize the  original $1/\beta$ 
    singularity. To this end, we impose  a cut-off for the lower limit of  the 
    $\beta$-integral at  some finite 
$\epsilon >0$ value.  Converting  the result  of $\beta$-integration
into an expression for TMDA, we obtain   
     \begin{align}
  \psi^{\rm h}    (x, & k_\perp^2)  =  g^2\,   \int_0^1  \frac{dy}{y-x} 
 \, \int_0^1   \frac{d \xi}{\bar \xi} \, 
 \Biggl \{
   \theta \Bigl ( {y} (1- \epsilon)  \geq {x}  \Bigr )    \nonumber \\ & \times
  \left [ \psi_0 \left (y,  \frac{\xi k_\perp^2 }{ x/y} \right ) 
   -   \psi_0 \left (y,  \frac{\xi^2 k_\perp^2 }{x/y -\bar  \xi } \right ) 
   \theta \Bigl ( \bar \xi \leq \frac{x}{y} \Bigr ) 
  \right ]  \nonumber \\ &   - 
     \theta \Bigl ( {y} (1- \epsilon) \leq {x}  \Bigr )   \psi_0 \left (y,  \frac{\xi^2 k_\perp^2 }{x/y -\bar  \xi } \right ) 
      \nonumber \\ & \times
  \theta \Bigl ( 1-\frac{x}{y}  \leq\xi \leq (1-\frac{x}{y} )/\epsilon    \Bigr )    \Biggr \}
 \ .
    \label{gauge10P}
\end{align} 
 
\subparagraph{Consider first 
 the $y\leq x/(1-\epsilon)  $  part: } 
      \begin{align}
  \psi_<^{\rm h}    (x, k_\perp^2) =& -  \alpha_g\,  \,   \int_x^{x/(1-\epsilon) }  \frac{dy}{y-x} \,   
    \nonumber \\ & \times
 \, \int_0^1   \frac{d \xi}{\bar \xi} \,    
    \psi_0 \left (y,  \frac{\xi^2 k_\perp^2 }{x/y -\bar  \xi } \right ) 
      \nonumber \\ & \times
  \theta \Bigl ( 1-\frac{x}{y}  \leq\xi \leq\frac1{\epsilon} (1-\frac{x}{y} )    \Bigr )   
 \ .
    \label{gauge12Pa}
\end{align} 
 Note that the interval of integration over 
 $y$ shrinks to zero  when $\epsilon \to 0$.
 Nevertheless, the   integral itself  has a finite value in this limit. 
 
 Since the   region  of integration is specified by  \mbox{$0 \leq 1-x/y \leq  \epsilon $,}
 we write   
$1-x/y = \epsilon z $ with  a new variable  $z$ satisfying 
$0 \leq z \leq 1$.
 Furthermore,  the limits on $\xi$ correspond to    $\xi = (1-x/y)\,  \zeta$
 with $1\leq \zeta \leq 1/ \epsilon$. 
 Thus, we denote  $\xi = \nu z$ with $\epsilon \leq \nu \leq 1$ to get 
      \begin{align}
  \psi_<^{\rm h}    \left (x
, k_\perp^2 \right ) = &  -  \alpha_g\,   \,   \int_0^1     \frac{  dz}{ 1- \epsilon z }  \,  
 \, \int^1_{\epsilon}    \frac{d \nu}{1- z \nu  }      \nonumber \\ & \times  \,  
    \psi_0 \left ( \frac{x}{1- \epsilon z},  \frac{\nu^2  }{\nu  -\epsilon  } z k_\perp^2 \right ) 
 \ .
    \label{gauge12Pd}
\end{align} 
Taking  the $\epsilon \to 0$ limit  and introducing $\tau = \nu z$, we obtain 
  \begin{align}
&  \psi_<^{\rm h}    (x, k_\perp^2) =  \alpha_g\,  
 \, \int^1_{0}  d \tau \,   \frac{  \ln \tau}{1-\tau}  \,    
    \psi_0 \left ( {x} ,  \tau  k_\perp^2  \right )  \  . 
    \label{gauge12Pd}
\end{align}
There is no change in the longitudinal momentum  fraction $x$ in this term,
i.e. this contribution may be written as a $\delta (y-x)$ term under the $y$-integral.
Note also that this contribution  is finite for $k_\perp=0$   if the 
initial TMDA $\psi (x, k_\perp^2)$ is finite for  $k_\perp=0$.
Namely,
 \begin{align}
  \psi_<^{\rm h}    (x, k_\perp^2=0) =  &\alpha_g \,     
 \, \int^1_{0}  d \tau \,   \frac{  \ln \tau}{1-\tau}  \,    
    \psi_0 \left ( {x} ,    k_\perp^2=0  \right )
      \nonumber \\ &
      = - \alpha_g \, \frac{\pi^2}{6} \,    \psi_0 \left ( {x} ,    k_\perp^2=0  \right ) \  . 
    \label{gauge12Pe}
\end{align}

\subparagraph{Sudakov logarithm.}
In the opposite limit of    large $k_\perp^2$,  
the
$\tau$-integral for  $\psi_<^{\rm h}    ( x,  k_\perp^2 )$ 
is dominated by small values   $\tau \sim \Lambda^2 /k_\perp^2$,
where $\Lambda^2$ is the scale characterizing 
the decrease of TMDA with $k_\perp^2$. 
Approximating   the $\tau$-integral  as
       \begin{align}   
  \int^{ \Lambda^2 /k_\perp^2}_{0}  & d \tau \,    \ln \tau   \,    
    \psi_0 \left ( {x} ,  \tau  k_\perp^2  \right ) 
     \nonumber \\ & 
     = \frac1{k_\perp^2}
      \int^{\Lambda^2} _{0}  d \kappa^2  \,     \ln \frac{\kappa^2}{k_\perp^2 }\,    
    \psi_0 \left ( {x} ,  \kappa^2  \right ) 
    \nonumber \\ & = - 
         \frac1{k_\perp^2} \ln  \left (\frac{k_\perp^2}{\Lambda^2} \right ) 
      \int^{\infty }_{0}  d \kappa^2  \,        
    \psi_0 \left ( {x} ,  \kappa^2  \right ) 
     + \ldots   
    \nonumber   \\ &  = -  \frac{1}{k_\perp^2  }  \, \varphi_0 (x) \, 
\ln  \left (\frac{k_\perp^2}{\Lambda^2} \right ) + \ldots \  .
    \end{align} 
we see that 
the term 
 $\psi_<^{\rm h}    (x, k_\perp^2) $ behaves like 
$-  (\ln  k_\perp^2 )/ k_\perp^2  $, i.e., has an extra
$\ln k_\perp^2$ on top of expected $1/  k_\perp^2$
hard tail factor.  After  integration  over $k_\perp^2$  till $xQ^2$ 
in the transition amplitude, 
such a  term 
produces a negative 
doubly-logarithmic contribution $\sim -  \ln^2 (xQ^2/\Lambda^2)$,
i.e. a Sudakov
logarithm. 
However, it is known that there should be no such  logarithms in the total result
for  transition form factor. 

\subparagraph{So, let us  consider the $y (1-\epsilon)\geq x $  part:  } 
      \begin{align}
  \psi_> ^{\rm h}   (x, k_\perp^2) & =  \alpha_g \,   \int_{x /(1- \epsilon )}^1  \frac{dy}{y-x} \,  
 \, \int_0^1   \frac{d \xi}{\bar \xi} \, 
  \left [ \psi_0 \left (y,  \frac{\xi k_\perp^2 }{ x/y} \right ) 
  \right.   \nonumber \\ &  \left. 
   -   \psi_0 \left (y,  \frac{\xi^2 k_\perp^2 }{x/y -\bar   \xi } \right ) 
   \theta \Bigl ( \bar \xi \leq \frac{x}{y} \Bigr ) 
  \right ]    
 \ .
    \label{gauge11P}
\end{align} 
Since  the two terms in the square brackets coincide when $y=x$,
 the integral over $y$  converges even if we take \mbox{$\epsilon =0$.}  
Another potential divergence is due to the $1/\bar \xi$ factor in the $\xi$-integral.
Again, the two terms  in the square brackets coincide when $\xi =1$,
thus  we can safely put \mbox{$\epsilon =0$.}

In particular, for small $k_\perp^2$, assuming that $\psi (y, k_\perp^2=0)$ is finite, we get
      \begin{align}
  \psi_> ^{\rm h}   (x, k_\perp^2=0) & =  \alpha_g \,   \int_{x }^1  \frac{dy}{y-x} \,  
     \ln  \left  ( \frac{y}{x} \right  ) \,  \psi_0 \left (y,  k_\perp^2=0 \right )   
 \  . 
    \label{gauge11P}
\end{align} 

However, we also need  to see that the $y (1-\epsilon)\geq x $  part  
contains a term that cancels the Sudakov contribution that we obtained for 
 the $y (1-\epsilon)\geq x $ part. 

For  large $k_\perp^2$, the leading 
 $\sim 1/k_\perp^2$  behavior  is obtained  from integration over the regions where 
the second argument of $\psi_0$ vanishes,
i.e. $\xi \to 0$ for both  terms in Eq. (\ref{gauge11P}). 
In this limit, the singularity of the integrand  for $\xi =1$ is irrelevant, so to
 avoid formal complications related to this singularity, we write
$1/ \bar \xi = 1 + \xi/\bar \xi$ 
and observe that 
 the $\xi/\bar \xi$ factor produces a   suppression   in the $\xi \to 0$ region
 resulting in extra powers of $1/k_\perp^2$ for large $k_\perp^2$, i.e. the terms 
accompanied by it   do not contribute to the leading behavior.  
Thus we have
     \begin{align}
   \psi_> ^{\rm h}  & (x, k_\perp^2)  =  \alpha_g \,   \int_{x /(1- \epsilon )}^1  \frac{dy}{y-x} \,  
 \, \int_0^1 d \xi  \,  
  \left [ \psi_0 \left (y,  \frac{\xi k_\perp^2 }{ x/y} \right ) 
   \right.    \nonumber \\ & \left. 
   -   \psi_0 \left (y,  \frac{\xi^2 k_\perp^2 }{x/y -\bar   \xi } \right ) 
   \theta \Bigl ( \xi \geq 1- \frac{x}{y} \Bigr ) 
  \right ]    + \ldots 
 \ .
    \label{gauge11Pb}
\end{align} 
We keep finite $\epsilon$ here because we want again to treat separately 
the two terms of this expression. 
The first term looks similar to what we had in the  nongauge case,
and can be written as 
     \begin{align}
   \psi_> ^{\rm h,1}  & (x, k_\perp^2)  =  \frac{\alpha_g }{k_\perp^2  }   \,  \int_{x /(1- \epsilon )}^1  \frac{dy}{y-x} \,  V_0 (x,y)
 \,
 \nonumber \\ &  \times  \int_0^{ k_\perp^2 /V_0 (x,y)}  d \kappa^2  \,  
 \psi_0 \left (y,  \kappa^2  \right ) 
 \ . 
    \label{gauge11Pc}
\end{align} 
At large $k_\perp^2$,  this term gives 
     \begin{align}
   \psi_> ^{\rm h,1}  & (x, k_\perp^2)  =  \frac{\alpha_g }{k_\perp^2  }  \,    \int_{x /(1- \epsilon )}^1  \frac{dy}{y-x} \,  V_0 (x,y)
 \varphi_0 \left (y \right ) +\ldots 
 \ , 
    \label{gauge11Pd}
\end{align} 
involving the  expected part of the ERBL evolution kernel.
The $y$-integral is singular at the lower limit producing $\ln (1/\epsilon)$ term, namely
     \begin{align}
   \psi_> ^{\rm h,1}  & (x, k_\perp^2)  =  \frac{\alpha_g }{k_\perp^2  }  \,  
   \ln \frac1{\epsilon}  \int_0^{ k_\perp^2 }  d \kappa^2  \,  
 \psi_0 \left (y,  \kappa^2  \right )  \nonumber \\ & 
 +{\rm regular  \  part  } 
 \   . 
    \label{gauge11Pe}
\end{align} 
This term  should be compensated by the second term in Eq. (\ref{gauge11Pb}).
Introducing $t \equiv  (1-x/y)$, the latter  may be written as
    \begin{align}
   \psi_> ^{\rm h,2}  & (x, k_\perp^2)  =  -\alpha_g \,   \int_{\epsilon }^{1-x}  \frac{dt}{t(1-t)} \,  
  \nonumber \\ &  \times    \, \int_{t}^1 d \xi  \,   
  \psi_0 \left (\frac{x}{1-t} ,  \frac{\xi k_\perp^2 }{1- t/  \xi } \right ) 
 \  . 
    \label{gauge11Pg22}
\end{align} 
Its singular part comes from   the $t\to 0$ region of  integration of the $t$-integral. 
Getting  it, we can take $t=0$ in the $\xi$-integral. The resulting singular part 
is evidently opposite 
to that in Eq. (\ref{gauge11Pe}).  
To get a more accurate estimate, let us neglect $t$ only when it stays  in $1-t$ combination,
but keep it in other places. Namely, consider
    \begin{align}
   \psi_> ^{\rm h,2,approx}  & (x, k_\perp^2)  =  -\alpha_g \,   \int_{\epsilon }^{1-x}  \frac{dt}{t} \,  
  \nonumber \\ &  \times    \, \int_{t}^1 d \xi  \,   
  \psi_0 \left ({x},  \frac{\xi k_\perp^2 }{1- t/  \xi } \right ) 
 \  . 
    \label{gauge11Pg221}
\end{align} 
Representing $\xi = t (1+u)$ we can write 
    \begin{align}
 \psi_> ^{\rm h,2,approx}  & (x, k_\perp^2)  =  -\alpha_g \,   \int_{\epsilon }^{1-x}  {dt} \,  
   \int_0^{1/t-1}   {d u}  \nonumber \\ &  \times 
 \,   \,  
  \psi_0 \left (x\, ,  t \, k_\perp^2 \,  \frac{(1+u)^2}{u}  \right ) 
 \  . 
    \label{gauge11Pg22}
\end{align} 
Note that $(1+u)^2/u$ is always larger than 4, which means the leading large 
large-$k_\perp^2$ power behavior comes  from integration 
over small $t\sim \Lambda^2/k_\perp^2$,
where $\Lambda^2$ is  a scale characterizing the fall-off of TMDA. 
Thus  we can substitute  the upper  limit of  integration  over $u$ by infinity 
without changing the leading large-$k_\perp^2$ power behavior of the integral 
and  write 
    \begin{align}
   \psi_> ^{\rm h,2,approx}  & (x, k_\perp^2)  =  -\alpha_g \,   \int_{\epsilon }^{\Lambda^2/k_\perp^2} {dt} \,  
   \int^{\infty} _{0}  du   \nonumber \\ &  \times 
 \,   \,  
  \psi_0 \left (x, \, t \, k_\perp^2\,  \frac{(1+u)^2}{u}   \right ) + \ldots
 \  . 
    \label{gauge11Pg22a}
\end{align} 
For  large $u$, we can approximate $(1+u)^2/u$ by $u$ to get 
    \begin{align}
   \psi&_> ^{\rm h,2,approx}   (x, k_\perp^2)  =  -\frac{\alpha_g }{k_\perp^2}\,   \int_{\epsilon }^{\Lambda^2/k_\perp^2} \frac{dt}{t}  \,  
    \nonumber \\ &  \times 
 \,    \int^{\infty} _{0} d \kappa^2  \,  
  \psi_0 \left (x, \, \kappa^2\,   \right ) + \ldots\nonumber \\ &
=  -\frac{\alpha_g }{k_\perp^2}\,\varphi_0 (x) \, \left [  \ln \frac1{\epsilon} - \ln  \left (\frac{k_\perp^2}{\Lambda^2}
\right )  \right ] + \ldots 
 \  . 
    \label{gauge11Pg22abc}
\end{align}

Thus,  we conclude
that the $  \psi_> ^{\rm h,2}   (x, k_\perp^2) $ part contains the $\ln 1/ \epsilon $ term
canceling the singularity of the $  \psi_> ^{\rm h,1}   (x, k_\perp^2) $ part.
In other words, it supplies the subtraction term
  \begin{align}
 &  \psi ^{\rm h,subtr}   (x, k_\perp^2)       =  - 
     \frac{\alpha_g }{  k_\perp^2 } 
       \,  
  \varphi_0 (x) \,  \left [ \ln \frac1{\epsilon}  \right ]
 \ 
    \label{gauge11Pg22}
\end{align} 
  generating  the plus prescription  for the part of the ERBL kernel
displayed in 
 Eq. (\ref{gauge11Pd}).

The $  \psi_> ^{\rm h,2}   (x, k_\perp^2) $ part  
also contains  the $\left ( \ln  {k_\perp^2}/{\Lambda^2} \right )/k_\perp^2$ contribution 
 that cancels a similar logarithm contained in the
  $  \psi_< ^{\rm h}   (x, k_\perp^2) $ term, 
  thus 
 guaranteeing that in $\alpha_g $ order
there will be no $\ln^2 Q^2$ Sudakov double logarithms 
in the transition form factor.

 \subparagraph{Summarizing,}  after properly taking the $\beta$-integral
in the original hard contribution (\ref{gauge7P}), the resulting hard addition to 
the TMDA is given by a   sum of the $\psi_<^{\rm h}    (x, k_\perp^2) $ term (\ref{gauge12Pd})
and the $\psi_>^{\rm h}    (x, k_\perp^2) $ term (\ref{gauge11P})
in which one can safely take $\epsilon=0$. 
Furthermore, the ``unwanted''  $  \left ( \ln{k_\perp^2}/{\Lambda^2} \right )/k_\perp^2$
Sudakov logarithms  present in both terms cancel, while the 
remaining $\sim 1/k_\perp^2$ terms generate evolution governed 
by the relevant part of the ERBL kernel with a correct plus-prescription.


 \subparagraph{Full result for the link contribution.}   Finally,  we should  add 
 the contribution of the diagram with   the insertion into the $\bar yp$ leg 
 (which 
is obtained by changing  \mbox{$y \to \bar y$}  and $x \to \bar x$) 
to get hard contribution due to gluons coming from the gauge link 
        \begin{align}
&  \psi^{\rm h,total}_{\rm link}    (x, k_\perp^2) =  \alpha_g \,   \int_x^1  \frac{dy}{y-x} 
 \, \int_0^1   \frac{d \xi}{\bar \xi} \, 
   \nonumber \\ & \times
\left   [ \psi_0 \left (y,  \frac{\xi k_\perp^2 }{ x/y} \right ) 
   -   \psi_0 \left (y,  \frac{\xi^2 k_\perp^2 }{x/y -\bar  \xi } \right ) 
   \theta \Bigl ( \bar \xi \leq \frac{x}{y} \Bigr ) 
  \right ]  \nonumber \\ & 
 + \alpha_g \,     
 \,   \int^1_{0}  d \tau \,  \frac{  \ln \tau}{1-\tau}  \,    
    \psi_0 \left ( {x} ,  \tau  k_\perp^2  \right ) 
     \nonumber \\ &
      + \Biggr \{y \to \bar y \ , \ x \to \bar x  \Biggr  \}  \  .
    \label{gauge12Pf}
\end{align} 

We  can also transform this result into  the  impact parameter space
        \begin{align}
&  \varphi^{\rm h,total}_{\rm link}    (x, z_\perp^2) =  \alpha_g \, 
\int_1^\infty    \frac{d \nu}{\nu -1} \, 
\Biggl \{ -      
    \varphi_0 \left ( {x} ,  \nu \, z_\perp^2   \right ) \, \ln \nu \,
     \nonumber \\ &+
  \int_x^1  \frac{dy}{y-x} \frac{x}{y}  
 \, 
  \Biggl  [\varphi_0 \left (y,  \nu \, { z_\perp^2\, \frac{x}{y}  } \right ) 
   -    \theta \Bigl (y  \leq x \frac{\nu }{\nu -1} \Bigr )   
     \nonumber \\ &  \times 
   \varphi_0 \left (y,  (\nu {x}/{y}  -\nu +1)\nu  z_\perp^2  \right ) 
    \Biggr ]   \Biggr \} 
         \nonumber \\ &
  + \alpha_g \,
\int_1^\infty    {d \nu}  \int_x^1  \frac{dy}{y} \theta \Bigl (y  \leq x \frac{\nu }{\nu -1} \Bigr )  
  \nonumber \\ &  \times 
   \varphi_0 \left (y,  (\nu {x}/{y}  -\nu +1)\nu  z_\perp^2  \right ) 
      + \Bigl  \{y \to \bar y \ , \ x \to \bar x  \Bigr \}  \  .
    \label{gauge12Pimpb}
\end{align}

  
 \subparagraph{Large $k_\perp^2$.} 
 As we have seen, for large $k_\perp^2$  this expression contains all the 
 relevant 
 terms of the 
 $\sim 1/k_\perp^2$ hard tail. It also contains contributions 
 that have a large-$k_\perp^2$ behavior similar to that of the soft TMDA
 $\psi_0 (x, k_\perp^2)$. 
 
\begin{figure}[h]
\centerline{\includegraphics[width=2.9in]{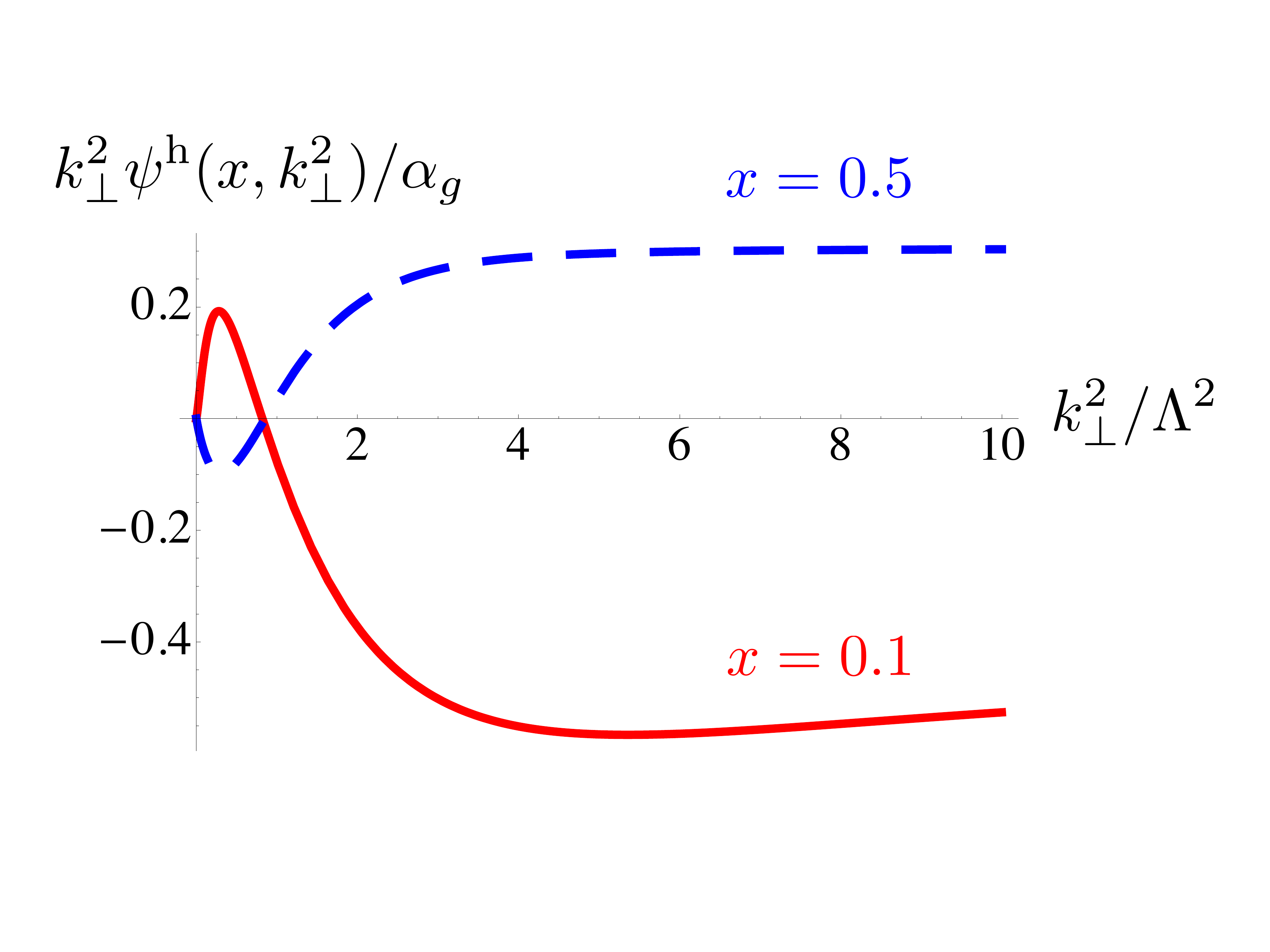} }
\vspace{-1cm} 
   \caption{
 $k_\perp$-dependence of the hard tail for a Gaussian model with flat DA 
for $x=0.1$ (solid line, red online) and $x=0.5$ (dashed line, blue online).
   \label{tail_flat}}
   \end{figure}

 For a Gaussian Ansatz with a flat profile, $ \psi_0^{\rm F} \left (y,  k_\perp^2 \right ) =e^{-k_\perp^2/\Lambda^2}/\Lambda^2$,    
 Fig. \ref{tail_flat}    illustrates 
 the $k_\perp$-dependence of the hard addition
 $\psi^{\rm h, total}_{\rm link} (x,k_\perp^2)$ for $x=0.5$ and $x=0.1$.
One can see that the product $k_\perp^2 \psi^{\rm h}$  flattens 
 for $k_\perp^2 \gtrsim 4 \Lambda^2$,   clearly demonstrating the presence 
 of   a $\sim 1/k_\perp^2$ contribution in  this case.
 
\begin{figure}[h]
\centerline{\includegraphics[width=2.9in]{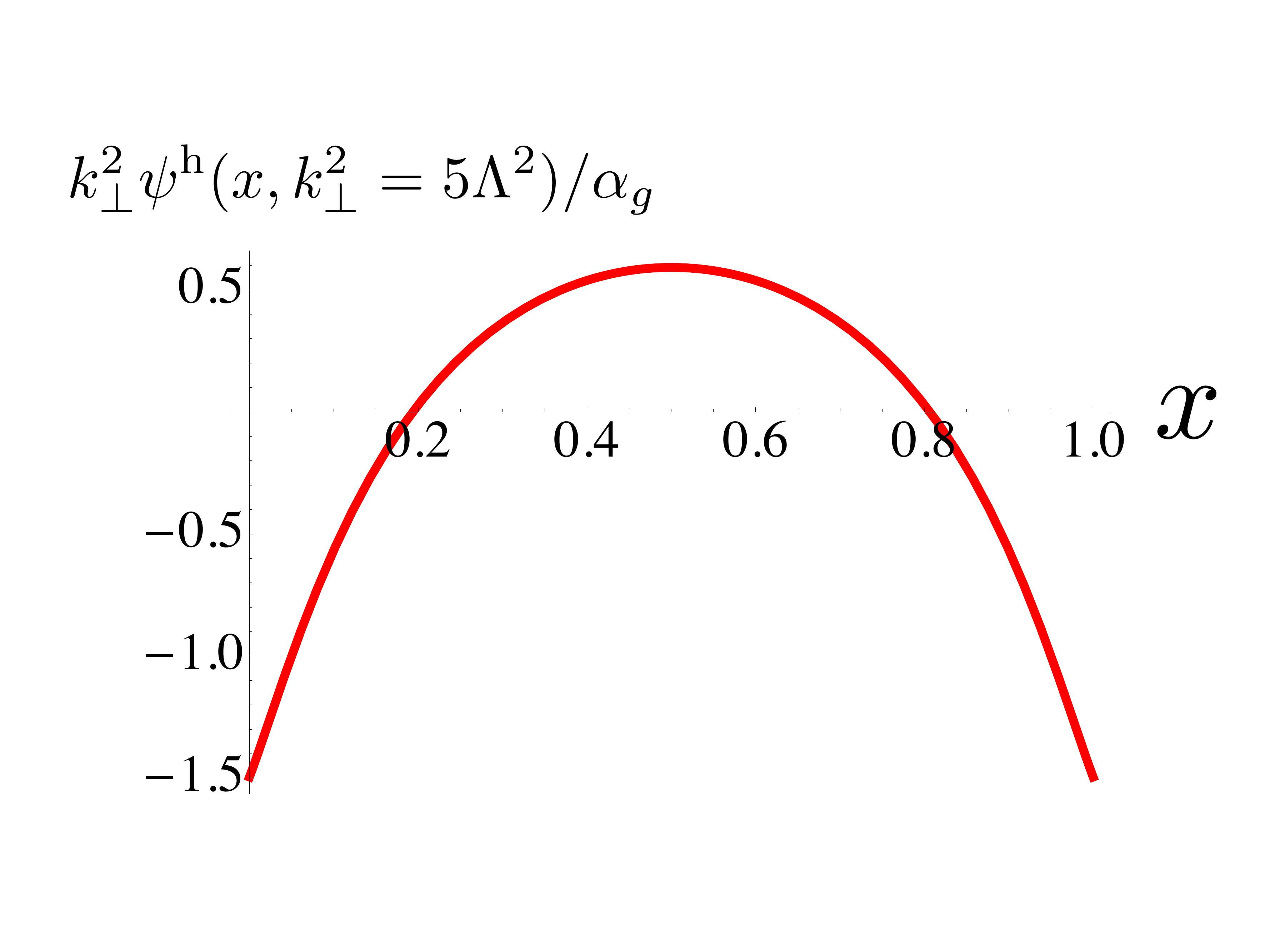} }
\vspace{-5mm}
   \caption{
 $x$-dependence of the hard tail for a Gaussian model with flat DA 
for $k_\perp^2 = 5 \Lambda^2$.  
   \label{tail_flat_x}}
   \end{figure}

 For sufficiently large $k_\perp^2$, the hard correction 
 is positive for the middle point $x=0.5$ and negative for 
 $x=0.1$.   In Fig. \ref{tail_flat_x} we show the $x$-dependence of 
  $\psi^{\rm h, total}_{\rm link} (x,k_\perp^2)$  for a particular value
  $k_\perp^2 = 5 \Lambda^2$ and initially flat $x$-distribution.
  Clearly, the hard correction tends to make the combined  distribution   narrower.
 
\begin{figure}[h]
\centerline{\includegraphics[width=2.9in]{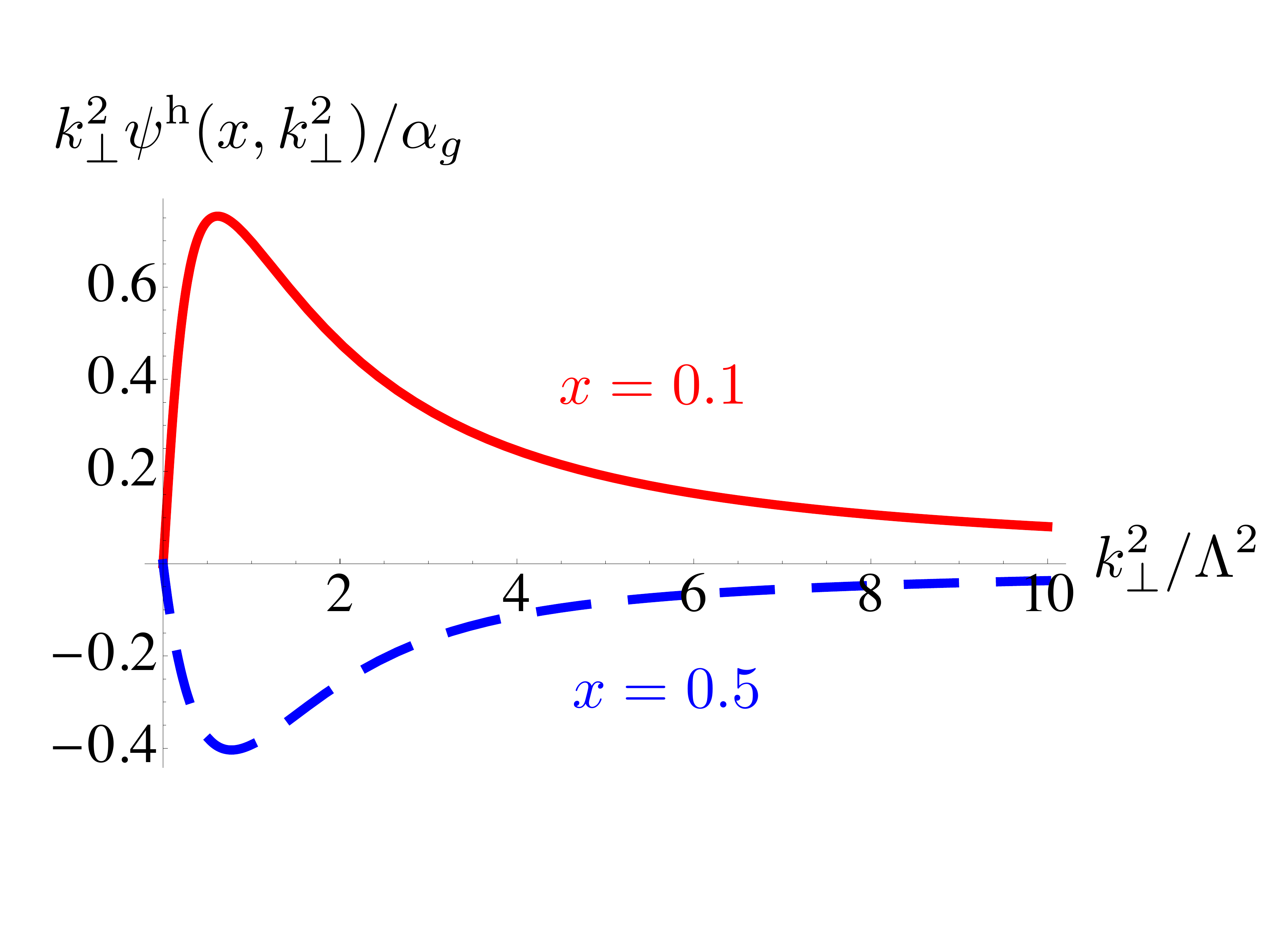} }
\vspace{-1cm} 
   \caption{
 $k_\perp$-dependence of the hard tail for a Gaussian model with flat DA 
for $x=0.1$ (solid line, red online) and $x=0.5$ (dashed line, blue online).
   \label{tail_as}}
   \end{figure}

 The situation changes when one takes an initially asymptotic profile,
 e.g. $ \psi_0^{\rm as} \left (y,  k_\perp^2 \right ) =6 y \bar y e^{-k_\perp^2/\Lambda^2}/\Lambda^2$.
 As one can see from  Fig.\ref{tail_as}, instead of flattening to a nonzero value, the product $k_\perp^2 \psi^{\rm h}$ 
 continues to decrease to zero for large $k_\perp^2$  thus demonstrating the 
 absence of the $\sim 1/k_\perp^2$ hard tail  in  this case, because the 
 convolution of the ERBL kernel with the asymptotic profile vanishes.

 
 \subparagraph{Small $k_\perp$. }  For  $k_\perp^2 \to 0$,
 Eq. (\ref{gauge12Pf})   has a finite
 limit, namely 
     \begin{align}
    \psi & ^{\rm h,total}_{\rm link}       (x, k_\perp^2=0) =  \alpha_g \,==    \int_{0}^1 dy  \left \{ \left [  \frac{\theta(y>x)}{y-x} \,  
     \ln  \left  ( \frac{y}{x} \right  )\right ]_+ \,     \right. \nonumber \\ & \left.+
         \left [ \frac{\theta(y<x)}{x-y} \,  
     \ln  \left  ( \frac{1-y}{1-x} \right  )\right ]_+ \, \right \} \psi_0 \left (y,  k_\perp^2=0 \right )   
 \ . 
    \label{gauge11PP}
\end{align} 
For a Gaussian Ansatz with a flat profile, $ \psi_0^{\rm F} \left (y,  k_\perp^2=0 \right ) =1/\Lambda^2$, 
this gives 
    \begin{align}
  \psi & ^{\rm h, total, F}_{\rm link}       (x, k_\perp^2=0) =  \frac{\alpha_g }{2\Lambda^2}
   \left [ \ln^2 \frac{ \bar x}{x}- \frac{\pi^2}{3} \right ]\,
 \ .
    \label{gauge11PFT}
\end{align} 
 Similarly, for a  Gaussian Ansatz 
 with asymptotic profile, $ \psi_0^{\rm as}
  \left (y,  k_\perp^2=0 \right ) =6 x (1-x)/\Lambda^2$,  we have 
  \begin{align}
  \psi  ^{\rm h,total, as}_{\rm link}    (x, k_\perp^2=0) & =  \frac{ 3\alpha_g }{\Lambda^2}
   \left [-\frac32  + 5  \, x \bar x + (\bar x - x) \ln \frac{ \bar x}{x} 
   \right. \nonumber \\ & \left. +
  x \bar x  \left [ \ln^2 \frac{ \bar x}{x}- \frac{\pi^2}{3} \right ]\right ]\,
 \ .
    \label{gauge11PFT}
\end{align} 
In both cases, the correction decreases TMDA
 in  the middle of the $x$-interval, and enhances it at the 
 ends. One can check that the integral of 
 $ \psi^{\rm h,total}   (x, k_\perp^2=0) $ over $x$ 
 in these two cases gives zero.  In fact, 
as follows from  Eq. (\ref{gauge7Pb}),  integrating  
 $ \psi^{\rm h,total} _{\rm link}   (x, k_\perp^2) $ over $x$ gives zero for 
 any initial  $\psi_0 \left ( {x} ,   k_\perp^2  \right ) $  and for all $k_\perp$.
 
 One may ask if it makes sense to consider a perturbatively obtained 
 hard term in the small
 $k_\perp$ region. To this end,   we remind that 
 we already use a model soft TMDA in this region, to
 which our  hard term gives just a small correction. 
 So, using the ``soft+hard''  combination is simply 
 another model for the small-$k_\perp$ region.  
 We think  that an  approach in which  the hard term 
 is naturally finite for small $k_\perp$ 
 has advantages compared to a usual practice when
 the explosion of the $1/k_\perp^2$  hard term  for small $k_\perp$ 
 is stopped  by an arbitrarily chosen cut-off.
 


 \section{Summary and outlook}
\label{Summary}

 In the present paper, we  outlined a 
 new approach to transverse momentum dependence 
 in hard processes.  Its starting point, just like in the OPE formalism,
  is the use 
 of coordinate representation. At handbag level,  the structure  
 of a hadron with momentum $p$  is described  
 by   a matrix element 
 of the bilocal operator ${\cal O} (0,z)$,   
 treated as a function of $(pz)$ and $z^2$.  It is parametrized 
 through a    {\it virtuality distribution}  $\Phi (x, \sigma)$,  in which 
 the variable $x$ is Fourier-conjugate to $(pz)$,  and has the usual 
 meaning of a  parton  momentum fraction.  Another parameter, $\sigma$,  is 
conjugate to $z^2$ through an analog  of   Laplace transform.
 
 Projecting ${\cal O} (0,z)$ onto  a spacelike interval with   $z_+=0$,  
 we   introduce {\it transverse momentum distributions} 
 $\Psi (x, k_\perp)$  and show that they can be written  
 in terms of  virtuality distributions $\Phi (x, \sigma)$. 
This fact opens  the possibility 
 to  convert the results of covariant calculations,  
written in terms  of $\Phi (x, \sigma)$,  into expressions 
involving  $\Psi (x, k_\perp)$. 
This procedure  being a crucial feature of our approach,   is illustrated  
 in the present paper  
by its application to  hard exclusive transition process
 $\gamma^* \gamma \to \pi^0$  at the handbag level
 (which is analogous to the 2-body Fock state approximation).
   Starting with scalar toy models, we then extend the analysis  
   onto the case of spin-1/2 quarks  and vector  gluons.

We studied  a few simple models for soft VDAs/TMDAs,
and  then used  them  for comparison of VDA results with experimental 
(BaBar and BELLE)  data 
on  the pion transition form factor. 
 
A  natural  next step
is going beyond the handbag approximation. 
 In QCD, an important feature is that  quark-gluon interactions 
generate a hard $\sim 1/k_\perp^2$ tail for TMDAs. 
To demonstrate the capabilities  of the VDA approach 
in this direction, 
we  describe 
the basic elements of generating hard tails from soft primordial 
TMDAs.  

Another direction is to include the   contribution 
due to a 3-body  quark-gluon TMDA.   In the OPE
it  starts
with a $\bar q G q$ operator  related to $\bar q D_\perp^2 q$ 
operator that  appears in our treatment of 
the handbag contribution as an explicit $k_\perp^2/Q^2$ correction 
to the leading term.  

Among other  possible  directions  for  future studies is  the use  of the 
VDA approach for a systematic study 
of quark virtuality corrections to the
one-gluon contribution for  the pion electromagnetic form factor $F_\pi (Q^2)$.
As we have seen, for the transition form factor such corrections
strongly reduce its magnitude in the region of 
 moderately large
$Q^2$. In case of $F_\pi (Q^2)$, one should expect 
even more drastic reduction, since two TMDAs are involved.

One more  direction, also suggested by the pion form factor analysis,
is a study of TMDAs corresponding to pseudoscalar
$\bar \psi \gamma_5 \psi$  and tensor $\bar \psi \sigma^{\mu \nu} \psi$ 
bilocal operators. It was argued \cite{Ahmad:2008hp,Goloskokov:2011rd}
  that such  chiral-odd projections 
may play  an important role 
in  understanding JLab data on 
 the deeply virtual electroproduction of pions \cite{Bedlinskiy:2014tvi}.
It should be empasized that  ``purely collinear'' pQCD formulas
in these cases are known to 
produce diverging  integrals like 
$\varphi (x)/x^2$. Within the TMDA formalism,
transverse momentum effects are expected to 
regulate such singularities, which necessitates 
the extension of the TMDA approach onto distributions
related to 
chiral-odd bilocal operators.


\section*{Acknowledgements}

I thank I. Balitsky, V.M. Braun, G. A.  Miller,  A.H. Mueller, 
A. Prokudin, A. Tarasov  and  C. Weiss for discussions.
This work is supported by Jefferson Science Associates,
 LLC under  U.S. DOE Contract \#DE-AC05-06OR23177
 and by U.S. DOE Grant \#DE-FG02-97ER41028.

 \bibliographystyle{apsrev4-1.bst}

\bibliography{tmda_1012.bib}

\end{document}